\documentclass[fleqn,usenatbib]{mnras}

\usepackage{newtxtext,newtxmath}

\usepackage[T1]{fontenc}

\DeclareRobustCommand{\VAN}[3]{#2}
\let\VANthebibliography\thebibliography
\def\thebibliography{\DeclareRobustCommand{\VAN}[3]{##3}\VANthebibliography}

\usepackage{graphicx}	
\usepackage{amsmath}	

\title[Numerical MHD simulations of solar flares]{Numerical MHD simulations of solar flares and their associated small-scale structures}

\author[Gonz\'alez-Serv\'in \& Gonz\'alez-Avil\'es]{
Mauricio Gonz\'alez-Serv\'in $^{1}$ and
J. J. Gonz\'alez-Avil\'es $^{2}$\thanks{E-mail: jgonzaleza@enesmorelia.unam.mx} 
\\
$^{1}$ Universidad Nacional Aut\'onoma de M\'exico. Instituto de Radioastronom\'ia y Astrof\'isica. 58089. Morelia, Michoac\'an, M\'exico. \\
$^{2}$ Universidad Nacional Aut\'onoma de M\'exico. Escuela Nacional de Estudios Superiores, Unidad Morelia. 58190. Morelia, Michoac\'an, M\'exico.
}

\pubyear{2023}

\begin{document}
\label{firstpage}
\pagerange{\pageref{firstpage}--\pageref{lastpage}}
\maketitle

\begin{abstract}
Using numerical simulations, we study the formation and dynamics of solar flares in a local region of the solar atmosphere. The magnetohydrodynamics (MHD) equations describe the dynamic evolution of flares, including space-dependent and anomalous magnetic resistivity and highly anisotropic thermal conduction on a 2.5 D slice. We adopt an initial solar atmospheric model in magnetohydrostatic equilibrium, with a magnetic configuration consisting of a vertical current sheet, which helps trigger the magnetic reconnection process. Specifically, we study three scenarios, two with only resistivity and the third with resistivity plus thermal conduction. The main results of the numerical simulations show differences in the global morphology of the flares, including the post-flare loops and the current sheet in three cases. In particular, localized resistivity produces more substructure around the post-flare loops that could be related to the Ritchmyer-Meshkov Instability (RMI). Furthermore, in the scenario of anomalous resistivity, we identify the formation of a plasmoid and a jet at coronal heights. On the other hand, in the scenario with resistivity plus thermal conduction, the post-flare loops are smooth, and no apparent substructures develop. Besides, in the $z-$ component of the current density for the Res+TC case, we observe the development of multiple magnetic islands generated due to the Tearing instability in the non-linear regime.
\end{abstract}

\begin{keywords}
(magnetohydrodynamics) MHD -- Sun: flares -- Sun: corona -- methods: numerical
\end{keywords}



\section{Introduction}
\label{S-Introduction} 

Solar flares are the most energetic explosions in the heliosphere, and they power space weather events that can affect the magnetic environment of our Earth. Solar flares efficiently convert magnetic energy into internal and kinetic energy, producing emissions in different wavelengths. One of the standard models of the solar flares has been discussed in \citealp{2002plap.book.....T,2010hssr.book..159F,2014masu.book.....P}, where the authors extended the cartoon view of magnetic reconnection using observational facts and, therefore, identify processes such as fast evaporation upflows, X-ray loops, and hard X-ray (HRX) source regions \citep{Masuda_et_al_1994, Su_et_al_2013}.   
One of the most relevant 2D MHD simulations of a solar flare based on a magnetic reconnection model with anisotropic thermal conduction was performed by \cite{Yokoyama&Shibata_2001}, who studied the thermal evolution in the post-flare loops to determine the flare temperature. Similar MHD simulations of magnetic reconnection have been performed to study coronal mass ejections and associated giant arcades \citep{Shiota_et_al_2005}, and chromospheric evaporation jets \citep{Miyagoshi&Yokoyama_2004}. More recently, \cite{Ruan_et_al_2020} presented a self-consistent solar MHD flare model that demonstrated the observationally suggested relationship between flux swept out by the hard X-ray footpoint regions and the actual reconnection rate at the X-point, which is a significant unknown in flaring scenarios.  
In solar flares, various plasma flows above the post-reconnection flare arcade have been reported by \cite{Liu_et_al_2013, Lin_et_al_2005, Warren_et_al_2018}. In particular, flare current sheets have rapidly descended, dark finger-like features commonly referred to as supra-arcade downflows (SADs), often observed in extreme ultraviolet (EUV) soft X-ray images. The slow speeds of SADs have been discussed in numerical models containing the bursty jet \citep{2015ApJ...807....6C} or Rayleigh-Taylor-type (RT) instabilities in reconnection downflow regions \citep{Guo_et_al_2014}. Other features observed in post-flare loops are the formation of magnetic islands at the top of the loops and along the current sheet (CS). These magnetic islands are believed to be closely related to fine structures CS in the wake of the eruption \citep{Shen_et_al_2011, Lin_et_al_2005}. Moreover, the magnetic islands' coalescence can further produce secondary CSs and islands \citep{Barta_et_al_2011}. More recently, \cite{Wang_et_al_2021} performed 2.5D MHD simulations to study the dynamics of magnetic reconnection in the solar CS, and they suggest that annihilation of magnetic islands at the flare loop top, which is not included in the standard flare model, plays a non-negligible role in releasing magnetic energy to heat flare plasma and accelerate particles. 

In this paper, we study the formation and evolution of a solar flare, including the post-flare loops, the CS, and the development of associated substructures such as shocks,  MHD instabilities, and small-scale magnetic islands, including the effects of localized and anomalous resistivity, and highly anisotropic thermal conduction on the formation and dynamics of a solar flare following the main ideas of the classical model developed by \cite{Yokoyama&Shibata_2001}. In particular, we base our simulations on three cases: i) Res case, where we only consider the effects of a localized resistivity; ii) anomalous resistivity case; and iii) Res+TC, where we include the effect of the highly anisotropic thermal conduction additional to the localized resistivity. We briefly describe the results of case ii) corresponding to the anomalous resistivity profile.

We organize the paper as follows. First, in section \ref{Model_methods}, we describe the system of MHD equations, the solar atmospheric model, the magnetic field configuration, the resistivity and thermal conduction models, and the numerical methods used to solve the MHD equations. Next, section \ref{results_num_simulations} contains the results of the two study cases, including the analysis of MHD instabilities, shock development, and magnetic island formation. Finally, we draw our concluding remarks in section \ref{Conclusion}.  

We organize the paper as follows. First, in section \ref{Model_methods}, we describe the system of MHD equations, the solar atmospheric model, the magnetic field configuration, the resistivity and thermal conduction models, and the numerical methods used to solve the MHD equations. Next, section \ref{results_num_simulations} contains the results of the two study cases, including the analysis of MHD instabilities, shock development, and magnetic island formation. Finally, we draw our concluding remarks in section \ref{Conclusion}.  

\section{Model and numerical methods}
\label{Model_methods}

\subsection{The system of MHD equations}
\label{MHD_equations}

We consider a gravitationally stratified solar atmosphere described by a plasma obeying the resistive MHD equations with anisotropic thermal conduction to model the plasma that constitutes the solar flare. In particular, we write the equations in a conservative dimensionless form that helps apply the numerical methods easily:

\begin{align}
\frac{\partial\varrho}{\partial t} + \nabla\cdot(\varrho{\bf v}) &= 0, \label{density}\\
\frac{\partial(\varrho{\bf v})}{\partial t} + \nabla\cdot(\varrho{\bf v}{\bf v}-{\bf B}{\bf B} + p_{t}{\bf I}) &= \varrho {\bf g},  \label{momentum} \\
\frac{\partial E}{\partial t} +\nabla\cdot((E+p_{t}){\bf v}-{\bf B}({\bf v}\cdot{\bf B})) &= {\varrho}{\bf v}\cdot{\bf g} \nonumber\\
-\nabla\cdot(\eta{\bf J}\times{\bf B})+\nabla\cdot{\bf F_{c}}, \label{energy} \\ 
\frac{\partial{\bf B}}{\partial t} +\nabla\cdot({\bf v}{\bf B} -{\bf B}{\bf v}) &= -\nabla\times({\bf \eta J}), \label{evolB} \\
\nabla\cdot{\bf B} &= 0, \label{divergenceB}
\end{align}

where $\varrho$ is the mass density, ${\bf v}$ represents the velocity field, ${\bf B}$ is the magnetic field, $T$ is the temperature, $\eta$ is the magnetic resistivity, and ${\bf F_{c}}$ is the thermal conduction flux.

Besides, $p_{t}=p+{\bf B}^{2}/2$ is the total (thermal + magnetic) pressure, ${\bf I}$ is the unit matrix, and $E$ is the total energy density, that is, the sum of the internal, kinetic, and magnetic energy densities, 

\begin{equation}
E = \frac{p}{\gamma-1} + \frac{\varrho{\bf v}^{2}}{2} + \frac{{\bf B}^{2}}{2}. \label{total_energy}
\end{equation}

\noindent For the fluid, we consider the adiabatic index $\gamma=5/3$. The system of equations (\ref{density})-(\ref{divergenceB}) is closed with the ideal gas law

\begin{equation}
p = \frac{k_{B}}{\bar{m}}\varrho T, \label{eos}  
\end{equation}

\noindent where $T$ is the temperature of the plasma, $\bar{m}=\mu m_{H}$ is the particle mass specified by a mean molecular weight value $\mu=$0.6, for a fully ionized gas, and $m_{H}$ is hydrogen's mass, and $k_{B}$ is Boltzmann's constant. The gravitational source term on the right-hand side of equations (\ref{momentum}) and (\ref{energy}) is given by ${\bf g}=[0,-g]$ with magnitude $g=2.74\times10^{4}$ cm s$^{-2}$, which represents an average over the solar surface. In this paper, we use a 2.5D model, meaning all the state variables depend on $x$ and $y$, where $x$ is a horizontal coordinate, and $y$ represents height. Besides, a nontrivial magnetic field $z-$component $B_{z}$ varies with $x$ is allowed. We solve the 2.5D version of the resistive MHD equations with thermal conduction on the $xy$ plane with resolution $\Delta x$ and $\Delta y$. 

\subsubsection{Model of the solar atmosphere}

\begin{figure*}
\centering
    \includegraphics[width=6.5cm, height=5cm]{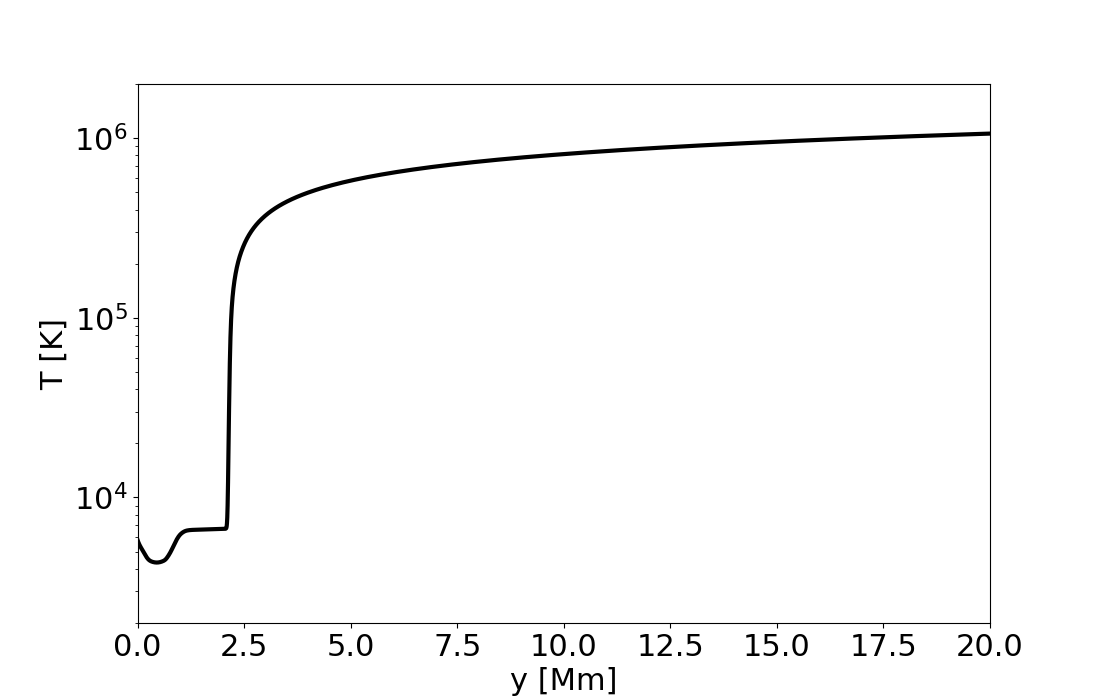}
     \includegraphics[width=6.5cm, height=5cm]{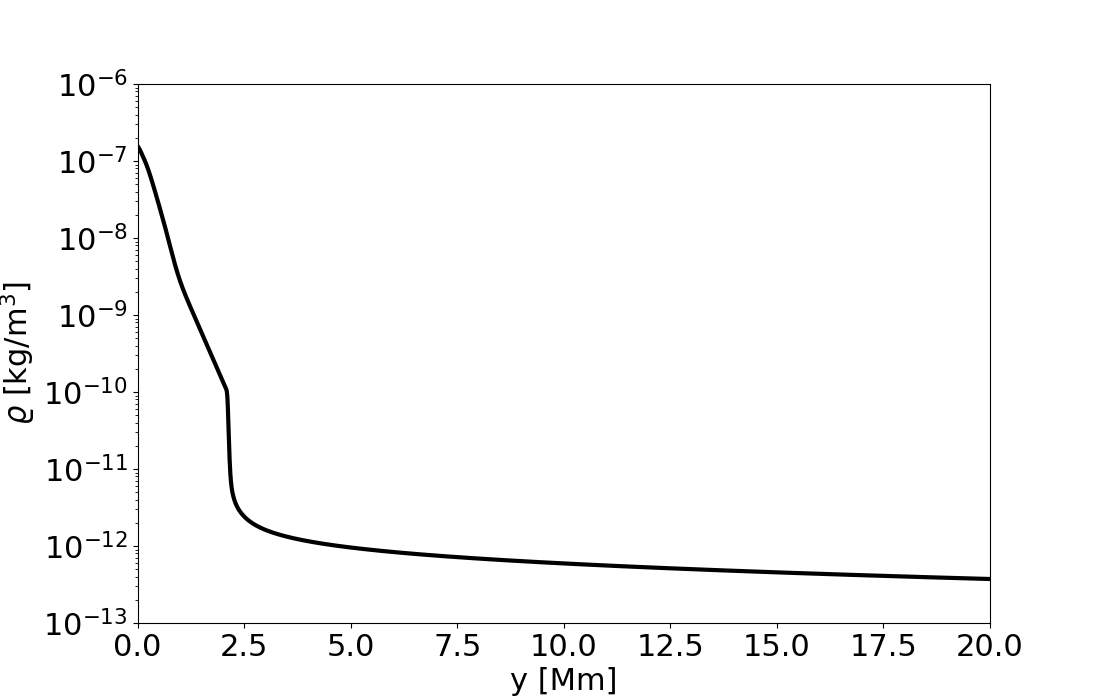}
    \caption{Equilibrium temperature in Kelvin (left) and mass density in gr cm$^{-3}$ (right) as functions of height $y$.}
    \label{fig:atmosphere_model}
\end{figure*}

At the initial time of the simulations, we assume the solar atmosphere is in hydrostatic equilibrium. In particular, we choose the C7 semiempirical model to describe the temperature field of the chromosphere-transition region \citep{Avrett&Loeser_2008}, smoothly extended to the solar corona as in \cite{Gonzalez_Aviles_et_al_2017}. In Fig. \ref{fig:atmosphere_model}, we display the equilibrium temperature and mass density profiles as functions of height $y$, where the steep gradient at the transition region $y\sim2.1$ Mm is discernible.      

\subsubsection{The magnetic field}
\label{magnetic_field}

\begin{figure*}
    \centering
    \includegraphics[width=5.0cm, height=5.5cm]{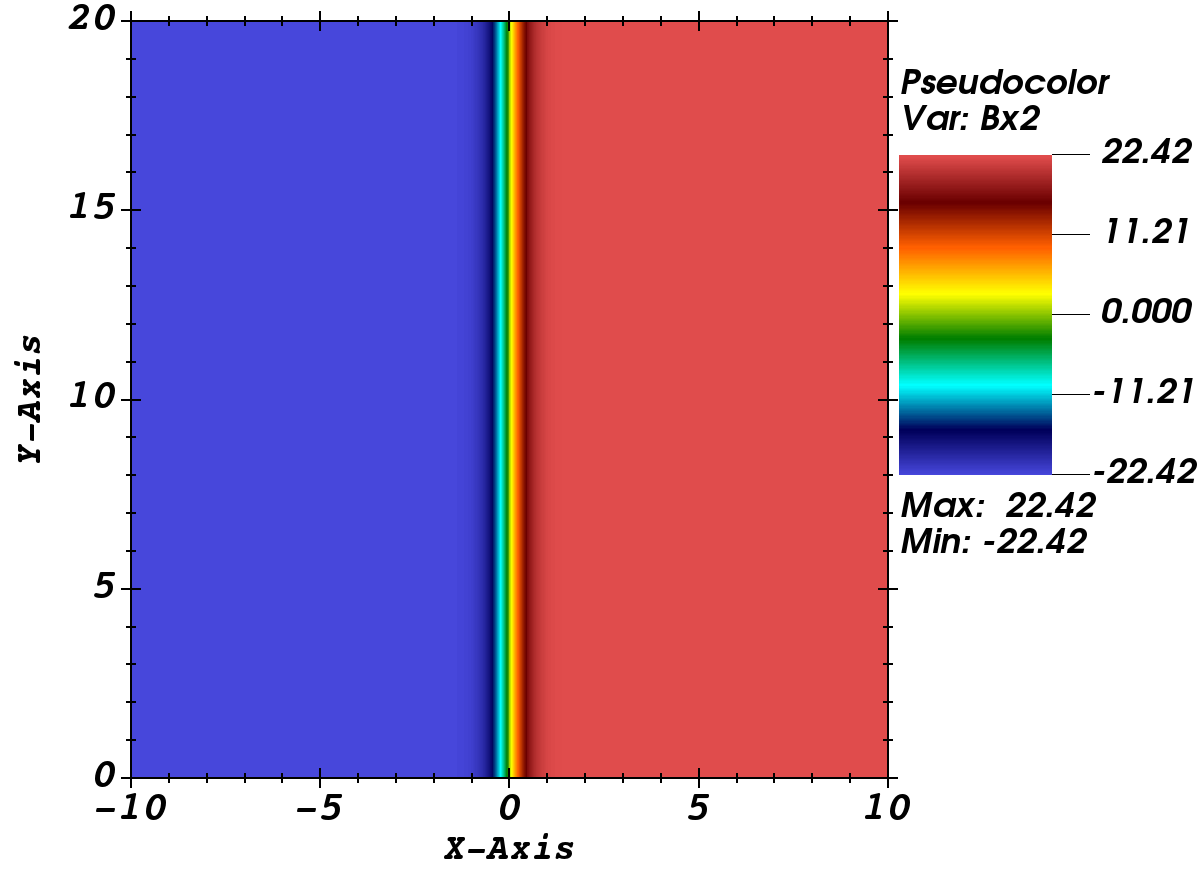}
     \includegraphics[width=5.0cm, height=5.5cm]{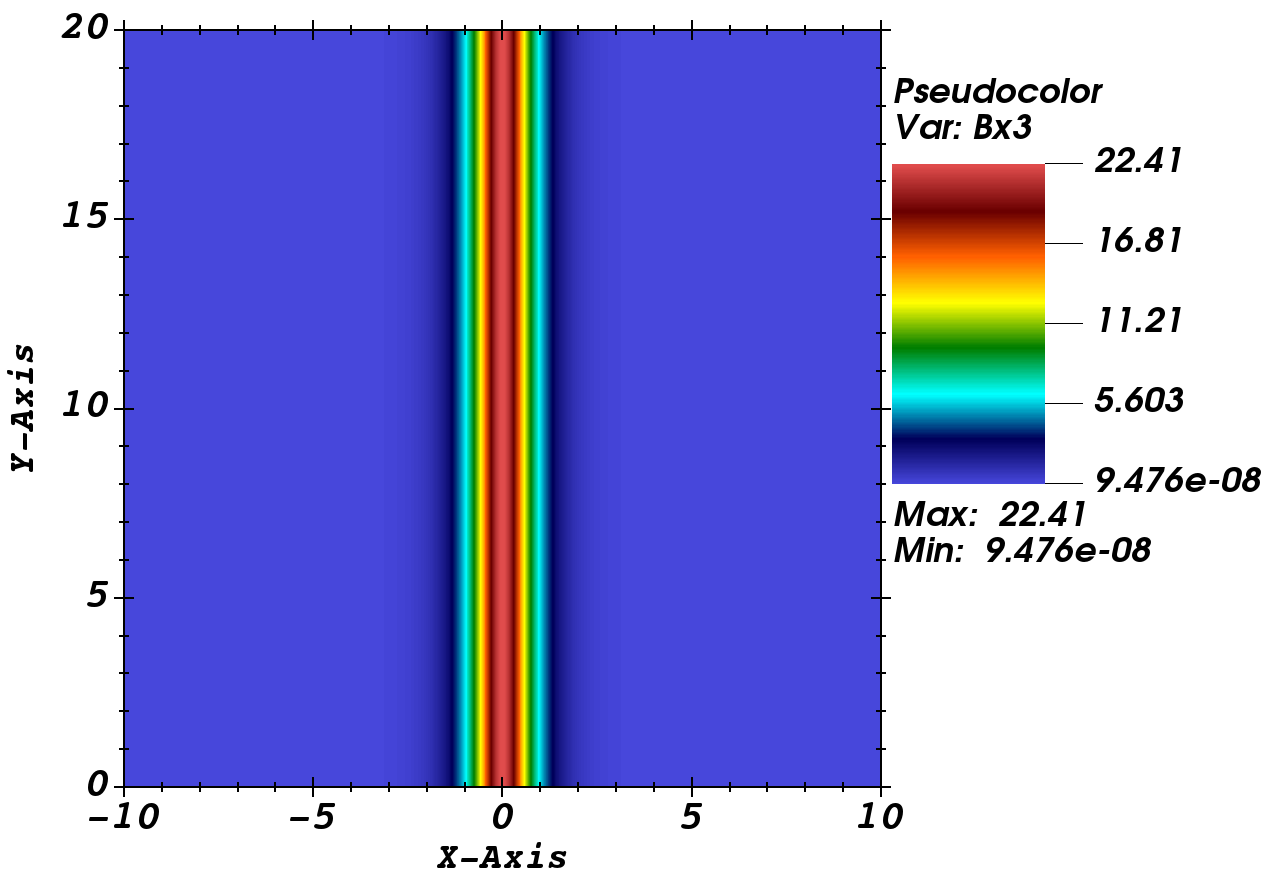}
     \includegraphics[width=5.0cm, height=5.5cm]{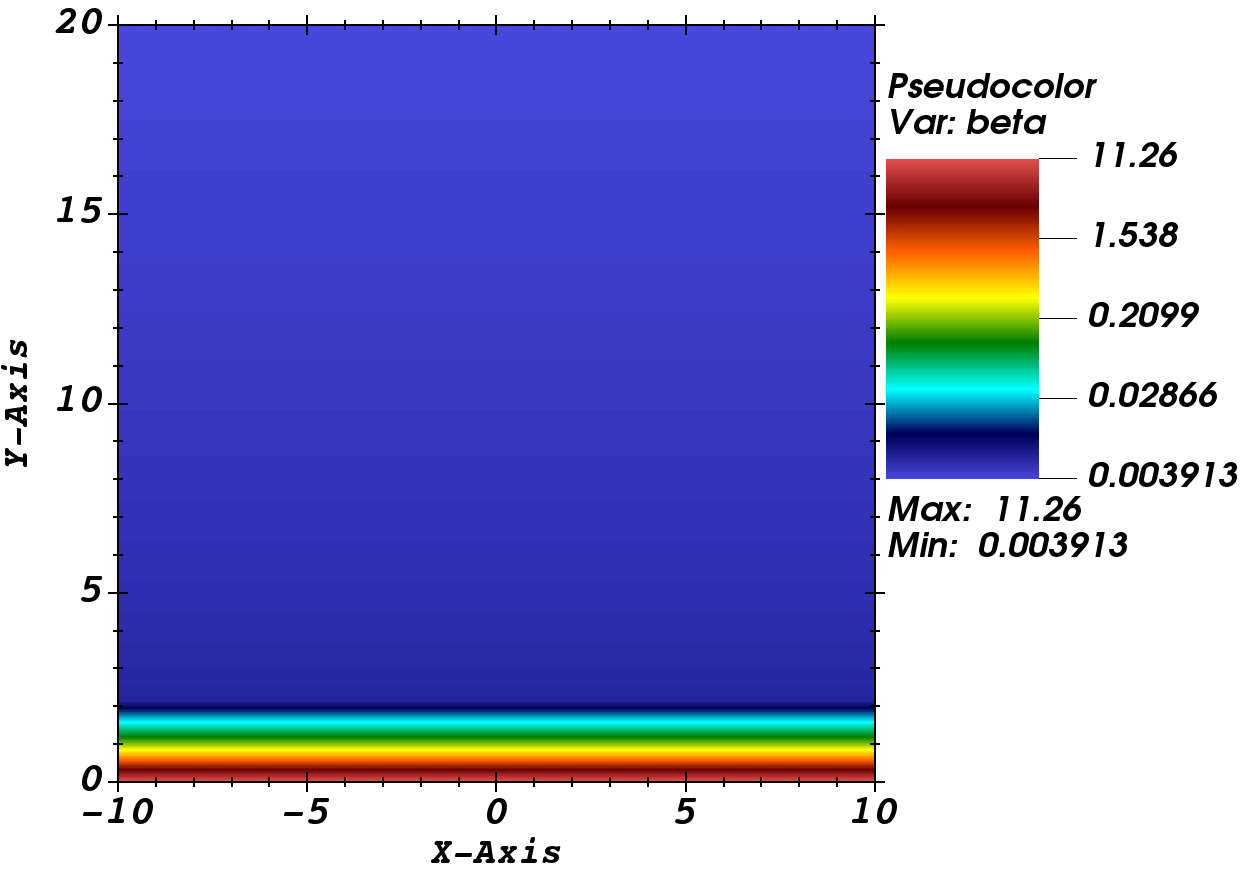}
    \caption{Magnetic field components $B_{y}$ (left) and $B_{z}$ (middle) in Gauss, and plasma $\beta$ (right) at the initial time of the simulations.}
    \label{fig:magneticfield}
\end{figure*}

Initially, we consider a force-free magnetic field  ($(\nabla\times{\bf B})\times{\bf B}={\bf 0}$) similar to that in \cite{Yokoyama&Shibata_2001, Ruan_et_al_2020}, which essentially represents a vertical CS defined as follows:

\begin{eqnarray}
B_{x} &=& 0, \\ \label{Bx} 
B_{y} &=& B_{0}\tanh(x/w), \\ \label{By}
B_{z} &=& B_{0}/\cosh(x/w), \label{Bz}
\end{eqnarray}

here $B_{0}=22.42$ G is the magnetic field strength and $w=5\times10^{7}$ cm is the width of the CS. We show the 2D maps of the $B_{y}$ and $B_{z}$ components in Fig. \ref{fig:magneticfield}, where it is notable that both components only vary in $x$ and are constant along the vertical direction $y$. However, initially, there is a non-negligible $ z-$ component of the current density, which produces a CS that helps trigger the magnetic reconnection at late times in the simulation. Additionally, we plot the plasma $\beta$, which is a parameter that estimates the ratio between gas pressure and magnetic pressure, i.e.,

\begin{equation}
\beta(x,y) = p(y)/B^{2}/2.     
\end{equation}

Here, the pressure $p(y)$ is given by the hydrostatic equilibrium, and $B^{2}=(B_{y}^{2}+B_{z}^{2})$. We display the plasma $\beta$ at the initial time in the right of Fig. \ref{fig:magneticfield}, where it is evident that $\beta$ is about $10$ in the bottom of the photosphere $y=0$ Mm, while in the solar corona ($y>2.1$ Mm) is approximately $10^{-3}$. This parameter will help us explain the dominance of pressures in regions where the magnetic islands form, as described in Section \ref{results_num_simulations}. Moreover, the solar atmospheric model, which is in hydrostatic equilibrium, and the magnetic field configuration, which is force-free, satisfying the magnetohydrostatic equation, ${\bf J}\times{\bf B} - \nabla p - \rho g = {\bf 0}$. Therefore, the whole system achieves the magnetohydrostatic equilibrium at the initial time of the simulations.    

\subsubsection{Spatially dependent and anomalous resistivity profiles}
\label{resistivity_profiles}

To perturb the atmosphere in magnetohydrostatic equilibrium and trigger the magnetic reconnection, we set a spatial dependent function for the resistivity, $\eta(x,y)$. In particular, we use a similar localized resistivity function as in \cite{Takasao_et_al_2015}, which is defined as follows:

\begin{equation}
\eta(x,y) = \eta_{0}\exp\left[{-(r/w_{\eta})^{2}}\right],  
\label{localized_res}
\end{equation}

where $\eta_{0}=10^{16}$ cm$^{2}$ s$^{-1}$ is the strength of the resistivity,  $r=\sqrt{x^{2}-(y-h_{p})^{2}}$, with $h_{p}=6\times10^{8}$ cm (6 Mm) is the vertical location of the Gaussian profile, and $w_{\eta}=1\times10^{8}$ cm (1 Mm) is its width. This profile is time-independent, but it can induce a fast and quasi-steady magnetic reconnection with a single X-point \citep[see, e.g.,][]{Ugai_1992}.

We also include an anomalous resistivity function. Microscopic instabilities physically cause this resistivity. Specifically, we adopt the following profile:

\begin{equation}
\eta(x,y) = \begin{cases}
            \text{min}(\alpha_\eta(\nu_d/\nu_c)-1,\eta_{min}), \ \text{if} \ t<t_\eta \\
             0, \ \text{if} \ t>t_\eta
             \end{cases}
             \label{anomalous_res}
\end{equation}

where, $\alpha_\eta = 1\times10^{-4}$, $\eta_{min}=$0.1, and $t_{\eta}= 10$ s. In the expression above, $\nu_d(x,y) = {\bf J}/(eN_{e})$ represents an instantaneous distribution that quantifies the relative ion-electron drift velocity, where ${\bf J}$ is the current density magnitude, $e$ is the electron charge and $N_{e}$ is the electron number density. Profile of equation \ref{anomalous_res} is similar to the used in \cite{Ruan_et_al_2020}; however, we do not couple with the localized profile of equation \ref{localized_res} and consider a small threshold drift velocity $\nu_{c}$.

\subsubsection{Anisotropic thermal conduction}
\label{thermal_conduction_profile}

In the equation (\ref{energy}), the thermal conduction flux ${\bf F_{c}}$ defines a flux-limited expression that smoothly varies between the classical and saturated thermal conduction regimes $F_{class}$ and $F_{sat}$, respectively, that is:

\begin{equation}
{\bf F_{c}} = \frac{F_{sat}}{F_{sat}+|{\bf F}_{class}|}{\bf F}_{class}.     
\label{Conduction_flux}    
\end{equation}

The above equation represents highly anisotropic thermal conduction suppressed mainly in the direction transverse to the magnetic field lines. For example, in this paper, we denote $\hat{{\bf b}}={\bf B}/|{\bf B}|$ as the unit vector in the direction of the magnetic field; therefore, the classical thermal conduction flux is as follows:

\begin{align}
{\bf F}_{class} &= \kappa_{\parallel}\hat{{\bf b}}(\hat{{\bf b}}\cdot\nabla T) + \kappa_{\perp}[\nabla T - \hat{{\bf b}}(\hat{{\bf b}}\cdot\nabla T)], \\
|{\bf F}_{class}| &= \sqrt{({\bf\hat{b}}\cdot\nabla T)^{2}(\kappa_{\parallel}^{2}-\kappa_{\perp}^{2})+\kappa_{\perp}^{2}\nabla T^{2}}
\end{align}

where the subscripts $\parallel$ and $\perp$ denote, respectively, the parallel and normal components to the magnetic field, $T$ is the temperature, $\kappa_{\parallel}$ and $\kappa_{\perp}$ are the thermal conduction coefficients along and across the field. In most astrophysical scenarios, such as in the Sun, $\kappa_{\perp}/\kappa_{\parallel} = 2\times10^{-31}n_{H}^2/(T^{3}|{\bf B}|^{2})$ is negligible \citep{Priest_2014}, and the thermal conduction is mainly along the field lines. In this paper, we chose $\kappa_{\parallel}=8\times10^{-7}$ erg s$^{-1}$ cm$^{-1}$ K$^{-1}$, which is a typical value in the solar corona \citep{Ruan_et_al_2020}. Saturated effects are accounted for, making the flux independent of $\nabla T$ for substantial temperature gradients, such as the transition region to solar corona in the solar atmosphere, which is a novel ingredient of this paper compared to the similar works where include the thermal conduction. In this limit, the flux magnitude approaches $F_{sat}=5\phi\rho c_{iso}^{3}$, where $c_{iso}$ is the isothermal speed of sound and $\phi<1$ is a free parameter that helps to account for the uncertain numerical coefficient in $F_{sat}$ \citep{Giuliani_1984}, which we take equal to 0.3 in this paper. The equations \ref{density}-\ref{divergenceB} were normalized by the scale factors in CGS units as shown in table \ref{table1:normalization_units}, which are typical scales of the solar atmosphere. The CGS units are used throughout the manuscript to be consistent with the normalization factors.   

\begin{table*}
\normalsize
\centering
    \begin{tabular}{|c|c|c|c|}
    \hline
    Variable & Quantity & Unit & Value \\
    \hline    
    $x$,$y$ & Length & $l_{0}$ & $10^{8}$ cm \\
    $\varrho$ & Density & $\varrho_{0}$ & $10^{-15}$ gr cm$^{-3}$ \\
     ${\bf v}$ & Velocity & $v_{0}$ & $10^{8}$ cm s$^{-1}$ \\
     $t$ & Time & $t_{0}=l_{0}/v_{0}$ & 1 s \\
     $p$ & Gas pressure & $p_{0}=\varrho_{0}v_{0}^{2}$ &  10 gr cm$^{-1}$ s$^{-2}$ (dyn cm$^{-2}$)\\
    ${\bf B}$ & Magnetic field & $B_{0}=\sqrt{4\pi\varrho_{0}v_{0}^{2}}$ & 11.21 G \\
    ${\bf J}$ & Current density & $J_{0}= cB_{0}/l_{0}$ & 3360.5 gr$^{1/2}$ cm$^{-1/2}$ s$^{-2}$ (statA cm$^{-2}$) \\
    $\eta$ & Resistivity & $\eta_{0}=v_{0}l_{0}$ & $10^{16}$ cm$^{2}$ s$^{-1}$ \\
    $\kappa$ & Thermal conductivity & $\kappa_{0}=\mu m_{u}/\varrho_{0}v_{0}l_{0}k_{B}$ & $7.265\times10^{-10}$ erg s$^{-1}$ cm$^{-1}$ K$^{-1}$  \\
      \hline
    \end{tabular}
    \caption{Normalization scale factor in CGS units}
    \label{table1:normalization_units}
\end{table*}


\subsection{Numerical methods}
\label{subsection:numerica_methods}

We numerically solve the equations (\ref{density})-(\ref{divergenceB}) using the PLUTO code \citep{Mignone_et_al_2007}. In all simulations, we set the Courant-Friedrichs-Levy (CFL) number equal to 0.4 and choose the second-order total variation diminishing (TVD) Runge-Kutta time integrator. Additionally, we use the Harten-Lax-van Leer discontinuities (HLLD) approximate Riemann solver \citep{Miyoshi&Kusano_2005} in combination with a linear reconstructor and the minmod limiter.

The numerical evolution of the MHD equations can lead to the violation of the divergence-free constraint equation given by (\ref{divergenceB}), developing unphysical results like the presence of a net magnetic charge. We use the extended generalized Lagrange multiplier method \citep{Dedner_et_al_2002} to control the constraint violation's growth. This method is robust in highly magnetized regions or low plasma beta scenarios, as implied in the solar corona on the left panel of Fig. \ref{fig:magneticfield}. PLUTO evolves the magnetic resistivity and thermal conduction separately from advection through operator splitting, using the super-time-stepping technique \citep{Alexiades_et_al_1996}. This method is robust enough to handle the highly parabolic (diffusion) behavior of space-dependent resistivity and anisotropic thermal conduction in the MHD equations adopted in this paper. Our simulations are carried out in a domain  $x\in[-10,10]$, $y\in[0,20]$, in units of Mm ($10^{8}$ cm), covered by 800$\times$1000 grid cells. Here, $y=0$ Mm represents the bottom of the photosphere. We used the following boundary conditions: outflow boundary conditions at $x_{min}=-10$ Mm, $x_{max}=10$ Mm, and $y_{max}=20$ Mm. In contrast, at $y_{min}=0$ Mm, we employ asymmetric boundary conditions in the $y-$component of the magnetic field $B_{y}$, outflow in the $z-$component of the magnetic field $B_{z}$, mass density and pressure, and zero boundary conditions in the component of the velocity field. These boundary conditions are similar to those used in \cite{Yokoyama&Shibata_2001}. 

\section{Results of numerical simulations}
\label{results_num_simulations}

In this section, we show the most representative results of the numerical simulations performed for the following three scenarios: (i) resistivity case (Res), ii) anomalous resistivity case, and (iii) resistivity plus thermal conduction case (Res+TC). We only include the most representative results for the case of the anomalous resistivity profile. 


\subsection{Resistivity case}
\label{Res_case}

\begin{figure*}
    \centering
    \centerline{\Large \bf   
      \hspace{0.18\textwidth}  \color{black}{\Large{Res}}
       \hspace{0.28\textwidth}  \color{black}{\Large{Res+TC}}
         \hfill}
          \centerline{\Large \bf   
      \hspace{0.26\textwidth}  \color{black}{(a)}
       \hspace{0.31\textwidth}  \color{black}{(b)}
         \hfill}
     \includegraphics[width=6.0cm, height=5.0cm]{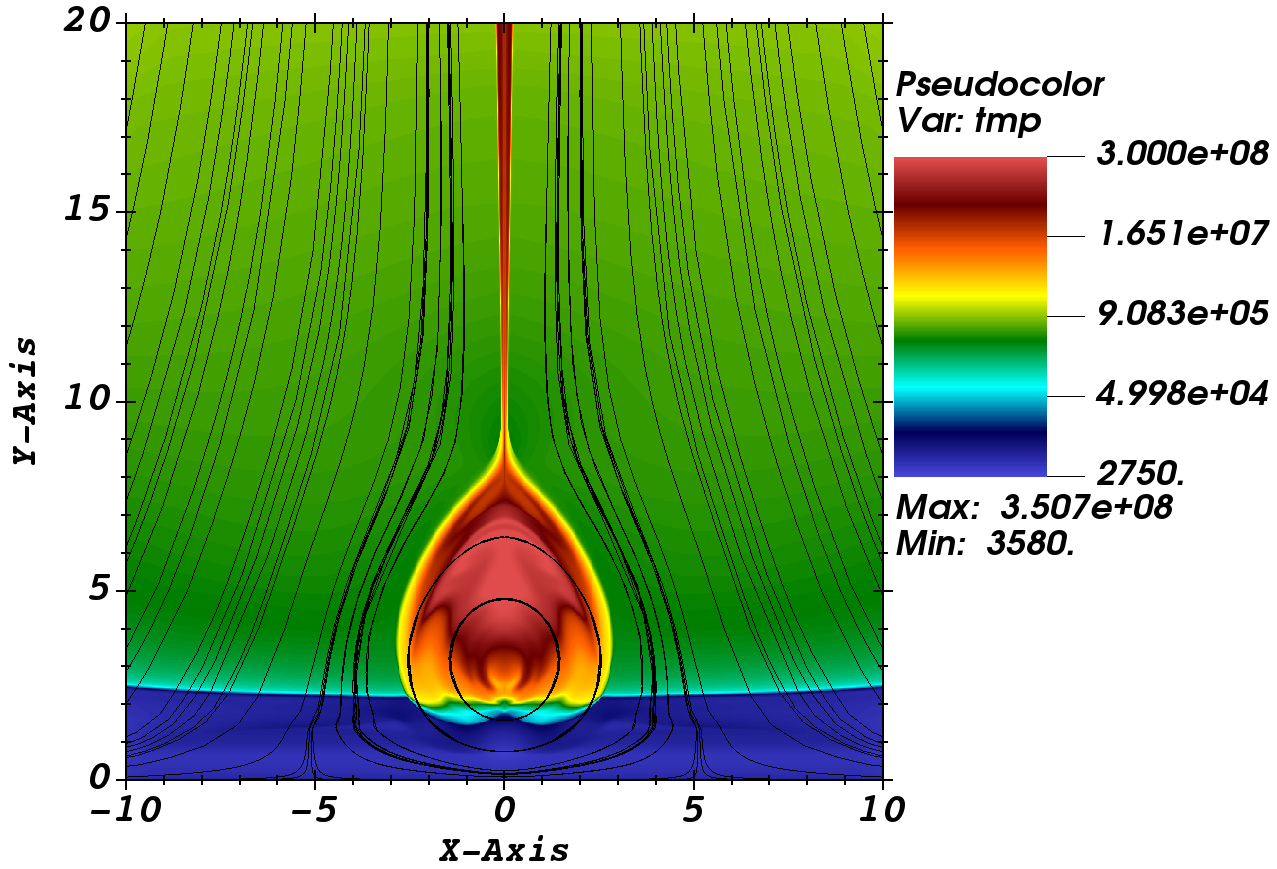}
     \includegraphics[width=6.0cm, height=5.0cm]{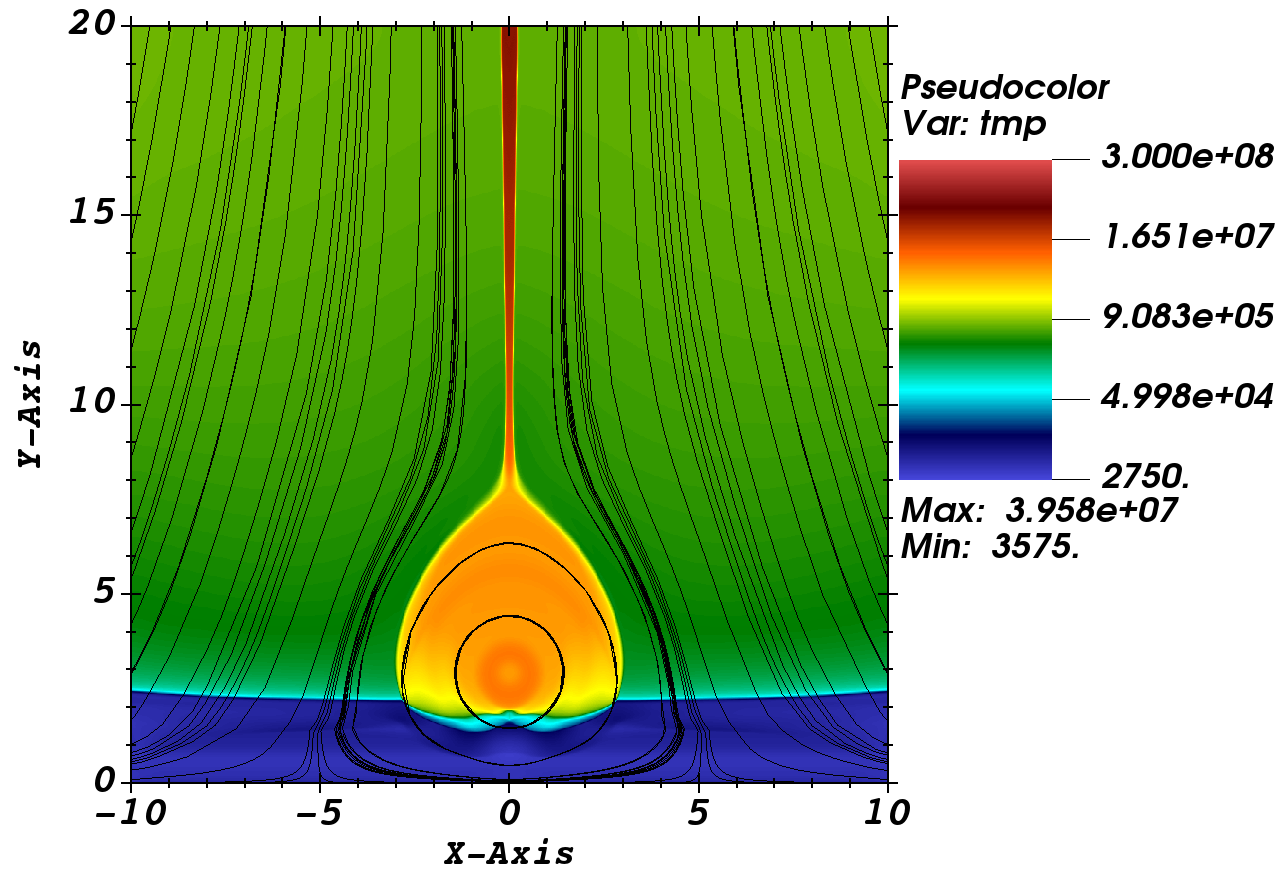}
     \centerline{\Large \bf   
      \hspace{0.275\textwidth}  \color{black}{(c)}
       \hspace{0.295\textwidth}  \color{black}{(d)}
         \hfill}
     \includegraphics[width=6.0cm, height=5.0cm]{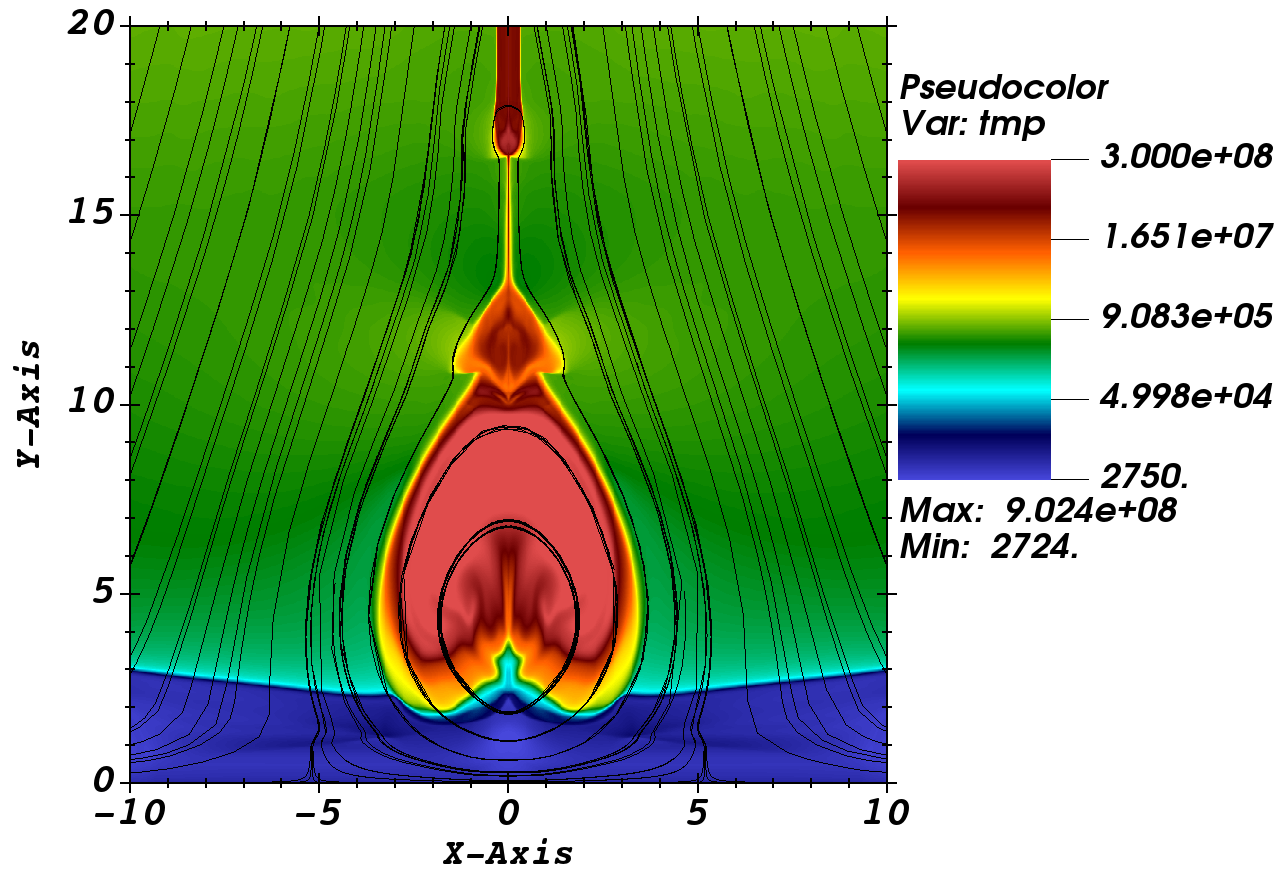}     
     \includegraphics[width=6.0cm, height=5.0cm]{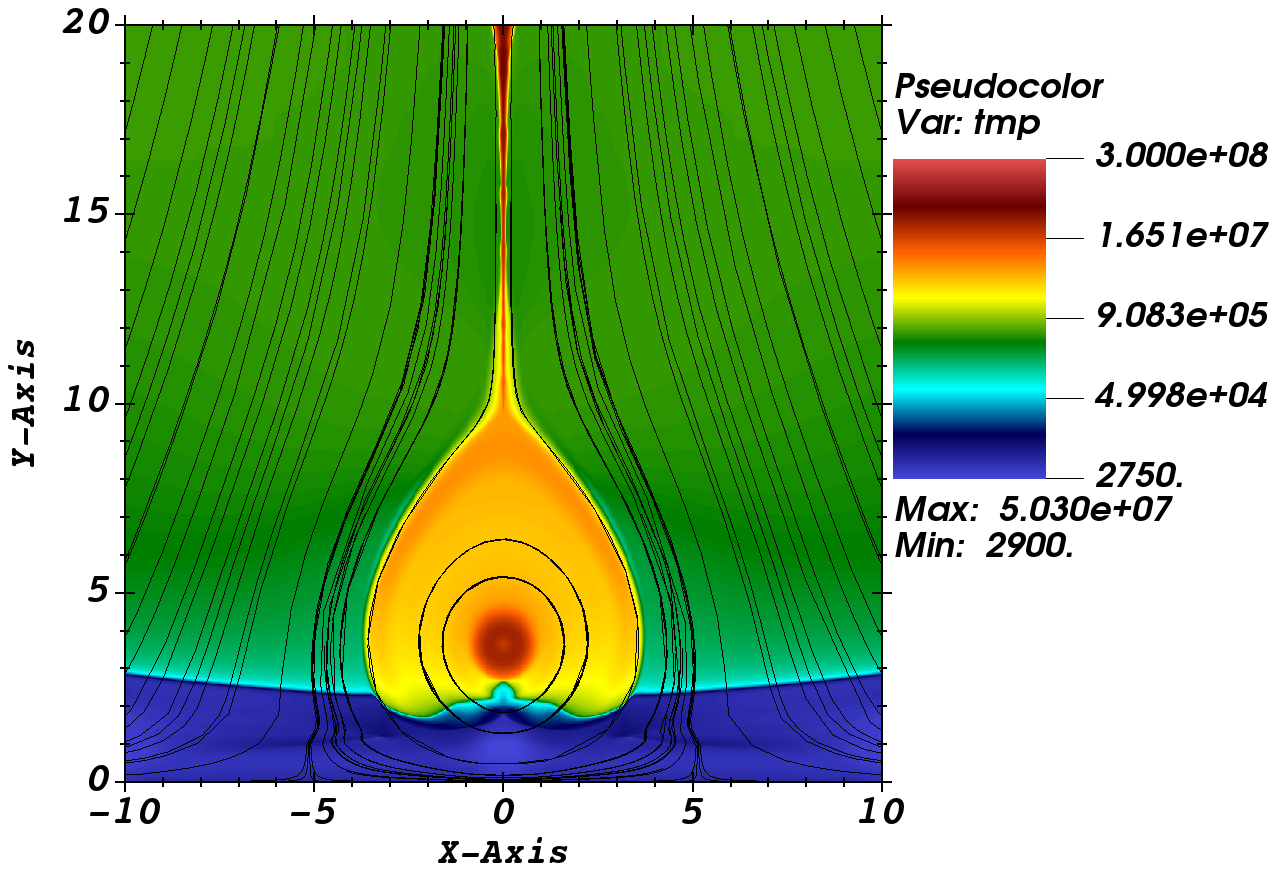}
     \centerline{\Large \bf   
      \hspace{0.275\textwidth}  \color{black}{(e)}
       \hspace{0.295\textwidth}  \color{black}{(f)}
         \hfill}
      \includegraphics[width=6.0cm, height=5.0cm]{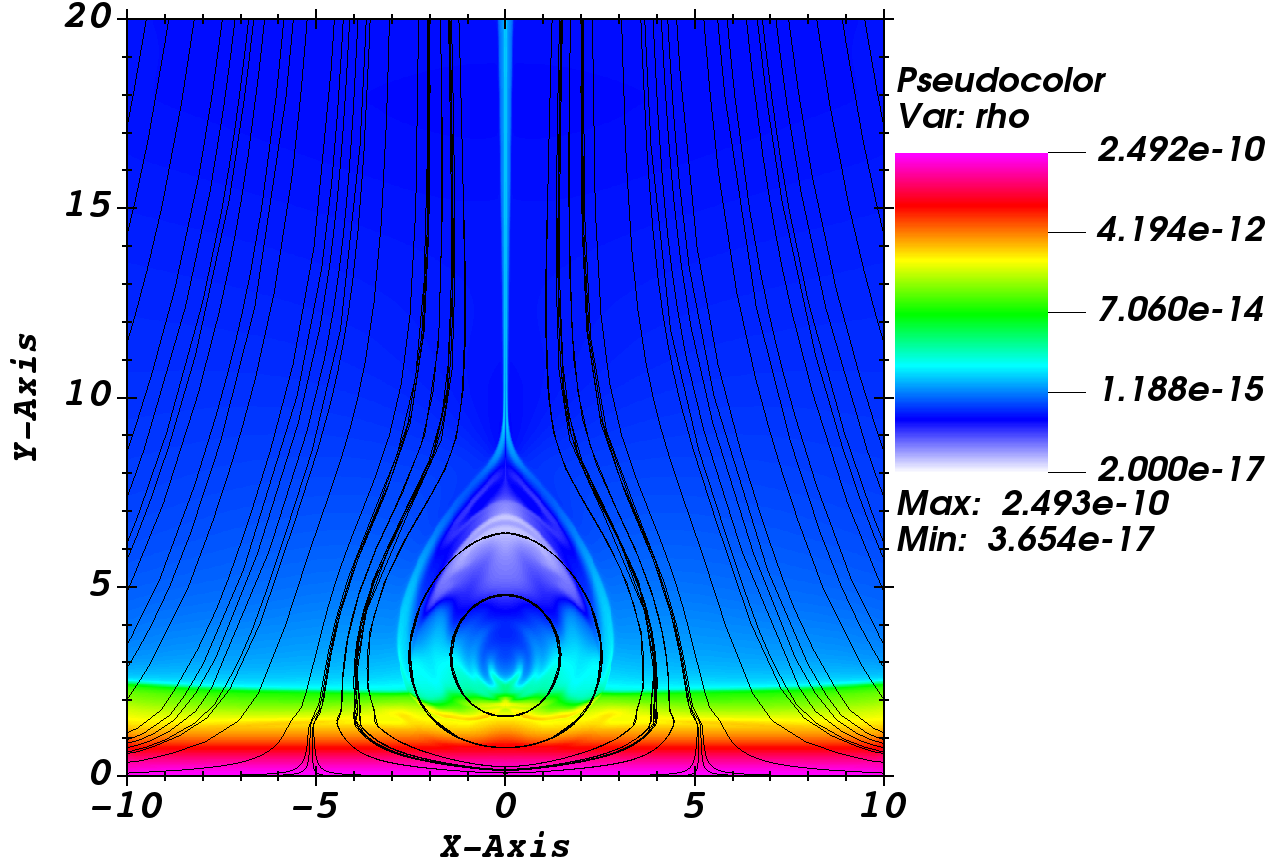} 
      \includegraphics[width=6.0cm, height=5.0cm]{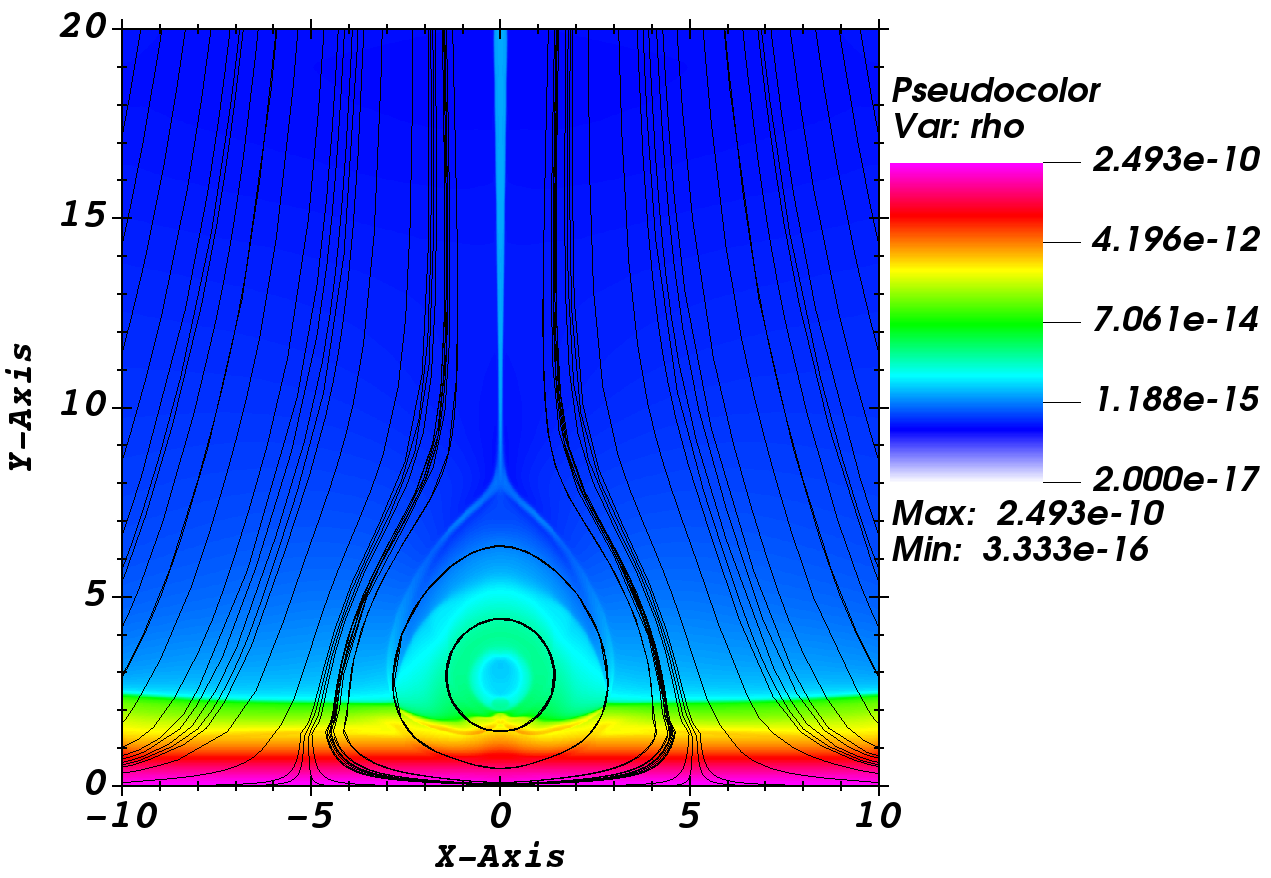}
       \centerline{\Large \bf   
      \hspace{0.275\textwidth}  \color{black}{(g)}
       \hspace{0.295\textwidth}  \color{black}{(h)}
         \hfill}
       \includegraphics[width=6.0cm, height=5.0cm]{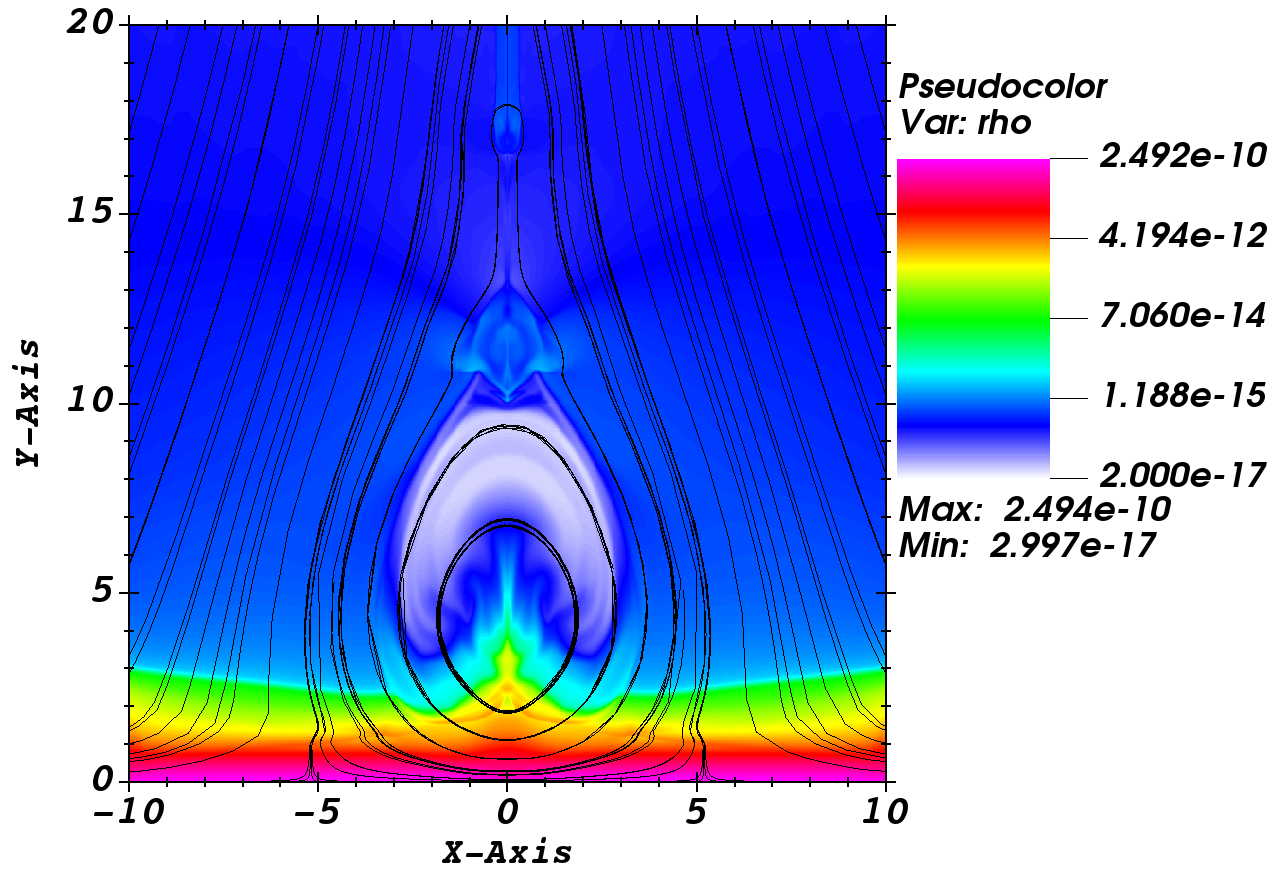}     
       \includegraphics[width=6.0cm, height=5.0cm]{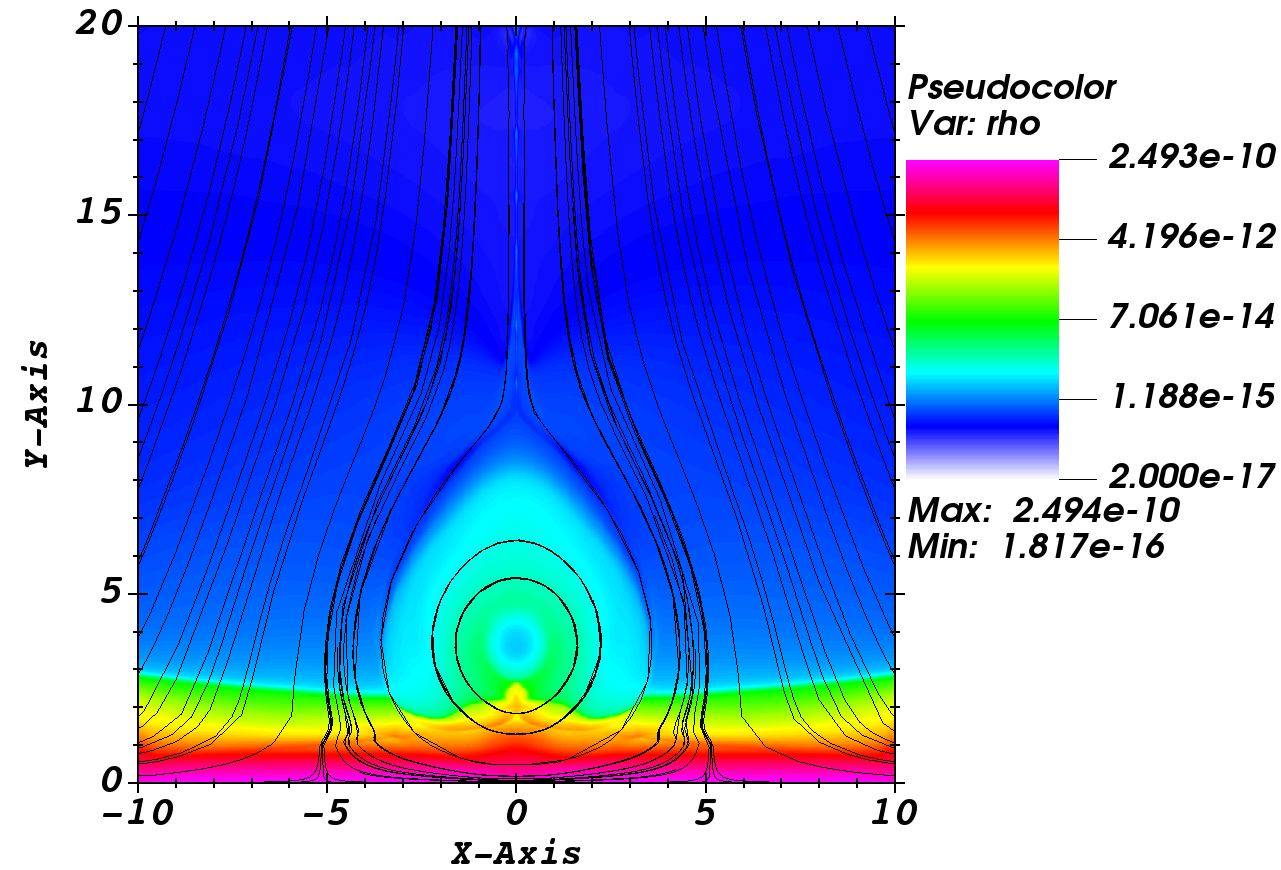}
       \caption{Temperature maps in kelvin overlap with the magnetic field lines at $t=25$ s (panels (a) and (b)) and $t=50$ s (panels (c) and (d)). In panels (e), (f), we show the mass density in gr cm$^{-3}$ for the time $t=25$ s, while in panels (g) and (h), we display mass density at $t=50$ s.}
    \label{fig:temperature_mass_density}
\end{figure*}

In Fig. \ref{fig:temperature_mass_density}, we show the temperature in Kelvin and mass density in gr cm$^{-3}$ with the magnetic field lines for the Res case. For example, in panel (a), we show the temperature at $t=25$ s. There, we note the following plasma structures that are consistent with the standard flare model \citep{Yokoyama&Shibata_2001}: i) formation of a hot flare loop ($\sim 10^{8}$ K) due to the magnetic reconnection triggered by the localized resistivity, ii) the reconnection outflow collides with the flare loop, and iii) production of evaporation flows of about $10^{6}$ K in the flare loop. In panel (c), the temperature at $t=50$ s shows the formation of a small hot region of about $10^{6}$ K at the top of the loops, and it is also evident that inside the loop region, the chromospheric evaporation develops. In panel (e), we show the mass density at $t=25$ s, consistently proving the behavior demonstrated on temperature maps. In particular, chromospheric evaporation appears in dense regions of about $10^{-14}$ gr cm$^{-3}$. Finally, in panel (g), the mass density at $t=50$ s shows how dense material ($\sim 10^{-12}$) moves upwards while it forms dense region ($\sim10^{-17}$ gr cm$^{-3}$) around it, and it is also visible the small region (but hot) at the top of it, as shown by the temperature maps.

\begin{figure*}
    \centering
    \centerline{\Large \bf   
      \hspace{0.32\textwidth}  \color{black}{\Large{Anomalous resistivity case}}
         \hfill}
          \centerline{\Large \bf   
      \hspace{0.27\textwidth}  \color{black}{(a)}
       \hspace{0.3\textwidth}  \color{black}{(b)}
         \hfill}
     \includegraphics[width=6.0cm, height=5.0cm]{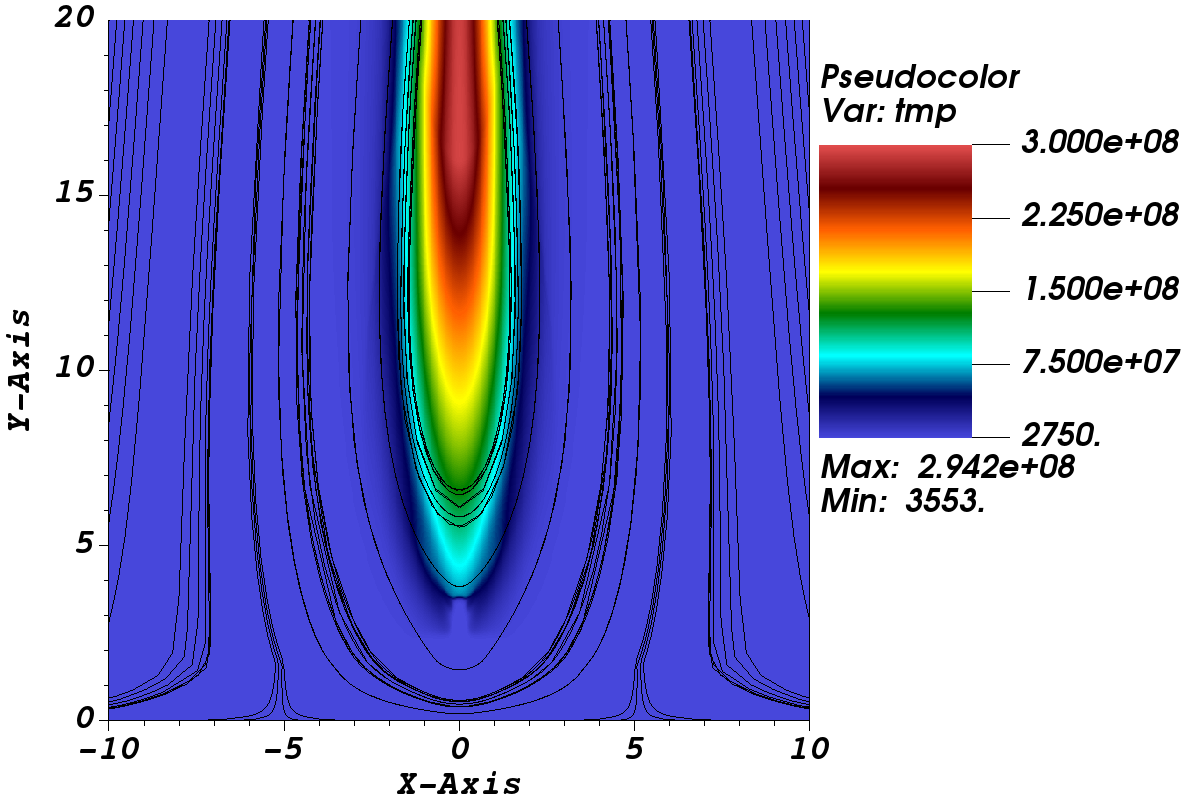}
     \includegraphics[width=6.0cm, height=5.0cm]{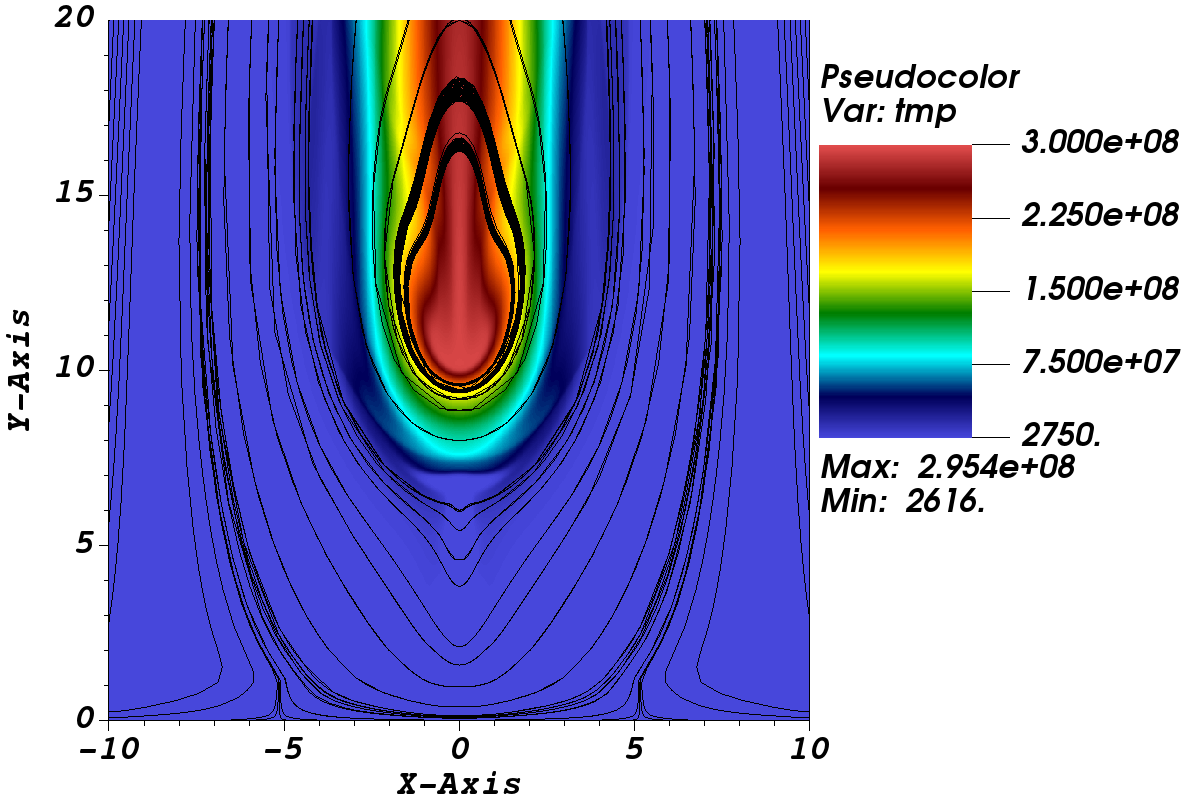}     
     \centerline{\Large \bf   
      \hspace{0.27\textwidth}  \color{black}{(c)}
       \hspace{0.3\textwidth}  \color{black}{(d)}
         \hfill}
      \includegraphics[width=6.0cm, height=5.0cm]{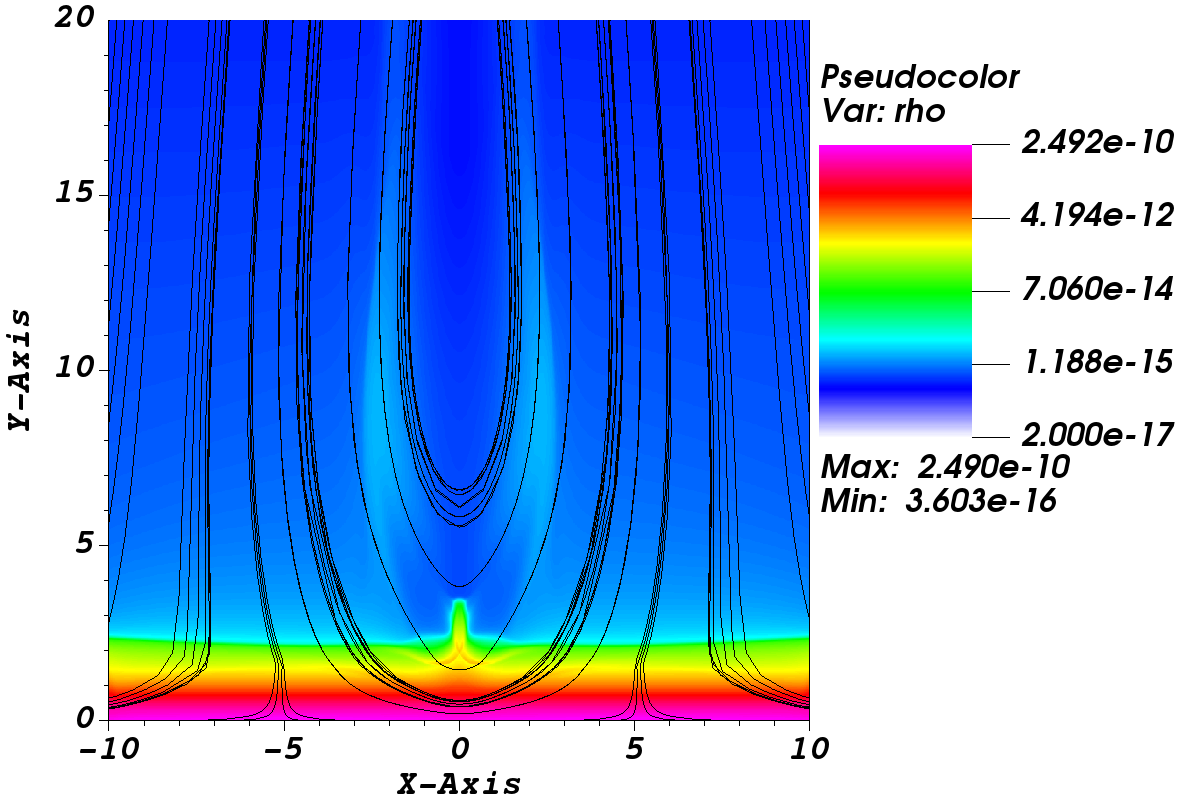}     
       \includegraphics[width=6.0cm, height=5.0cm]{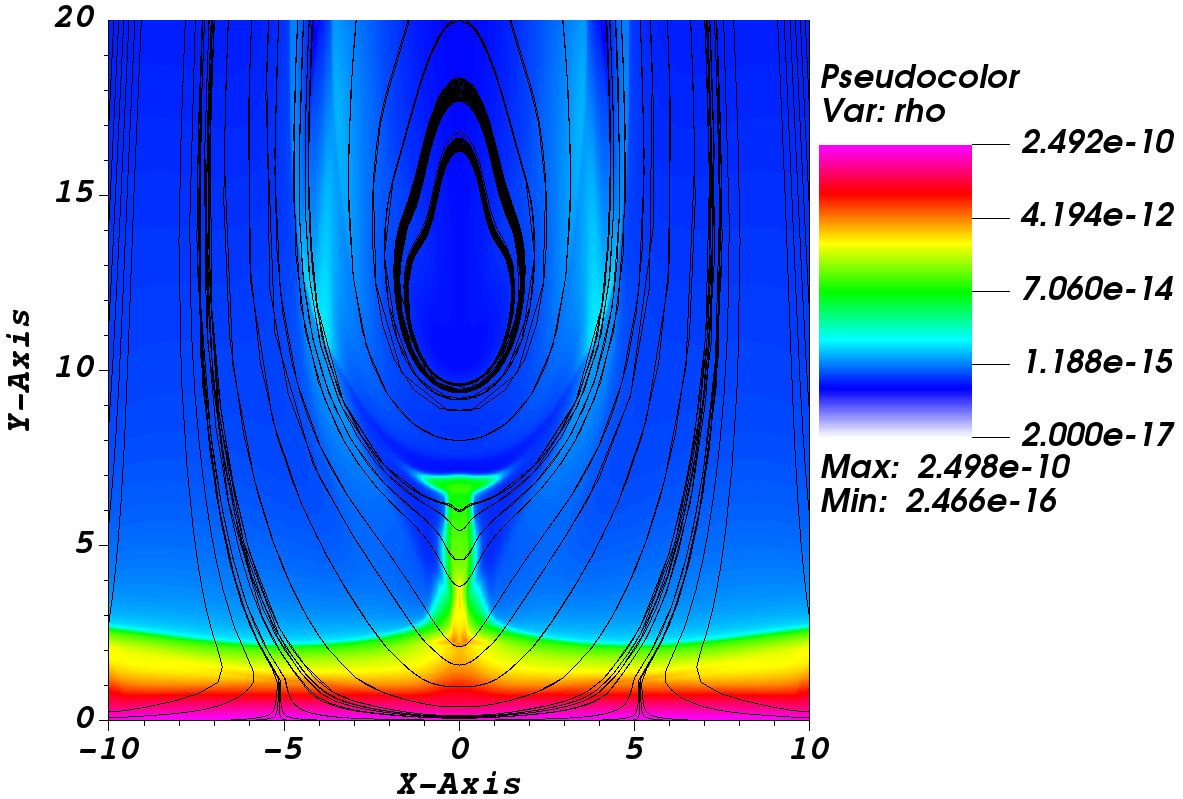}     
       \caption{Temperature maps in kelvin and mass density in gr cm$^{-3}$ overlay with the magnetic field lines at $t=25$ s (panels (a) and (c)) and at $t=50$ s (panels (b) and (d)) for the anomalous resistivity case.}
    \label{fig:temperature_mass_density_anomalous_res}
\end{figure*}

In Fig. \ref{fig:temperature_mass_density_anomalous_res}, we show snapshots of the temperature in kelvin and mass density in gr cm$^{-3}$ at times $t=25, 50$ s for the anomalous resistivity case. In this scenario, we only turn on the resistivity given by the equation (\ref{anomalous_res}) and let the system evolve. In panels (a) and (b), we display the temperature, where it is discernible that any post-flare structure is formed, especially since we do not identify the post-flare loop structures and the development of a thin CS along the vertical direction. Instead, we can see the formation of a plasma structure that becomes wider and moves upwards while the magnetic file lines reconnect. This kind of plasma feature could be related to a plasmoid. In panels (c) and (d), we note that in the mass density, there are no clues of the formation of post-flare loops, but interestingly, we identify the development of a small jet (3 Mm in length) moving upwards from the transition region. This feature is interesting since it could be related to the magnetic reconnection process at microscopic scales, which is triggered by anomalous resistivity at coronal heights, which indicates a relevant result despite the limitations of the MHD to capture the microscopic mechanism associated with the magnetic reconnection process. Finally, it is discernible that these plots are different from the plots of panels (a), (c), (e), and (g) shown in Fig. \ref{fig:temperature_mass_density} for the Res case. 

\begin{figure*}
    \centering
    \centerline{\Large \bf   
      \hspace{0.18\textwidth}  \color{black}{\Large{Res}}
       \hspace{0.25\textwidth}  \color{black}{\Large{Res+TC}}
         \hfill}
           \centerline{\Large \bf   
      \hspace{0.28\textwidth}  \color{black}{(a)}
       \hspace{0.295\textwidth}  \color{black}{(b)}
         \hfill}
    \includegraphics[width=6cm, height=5.0cm]{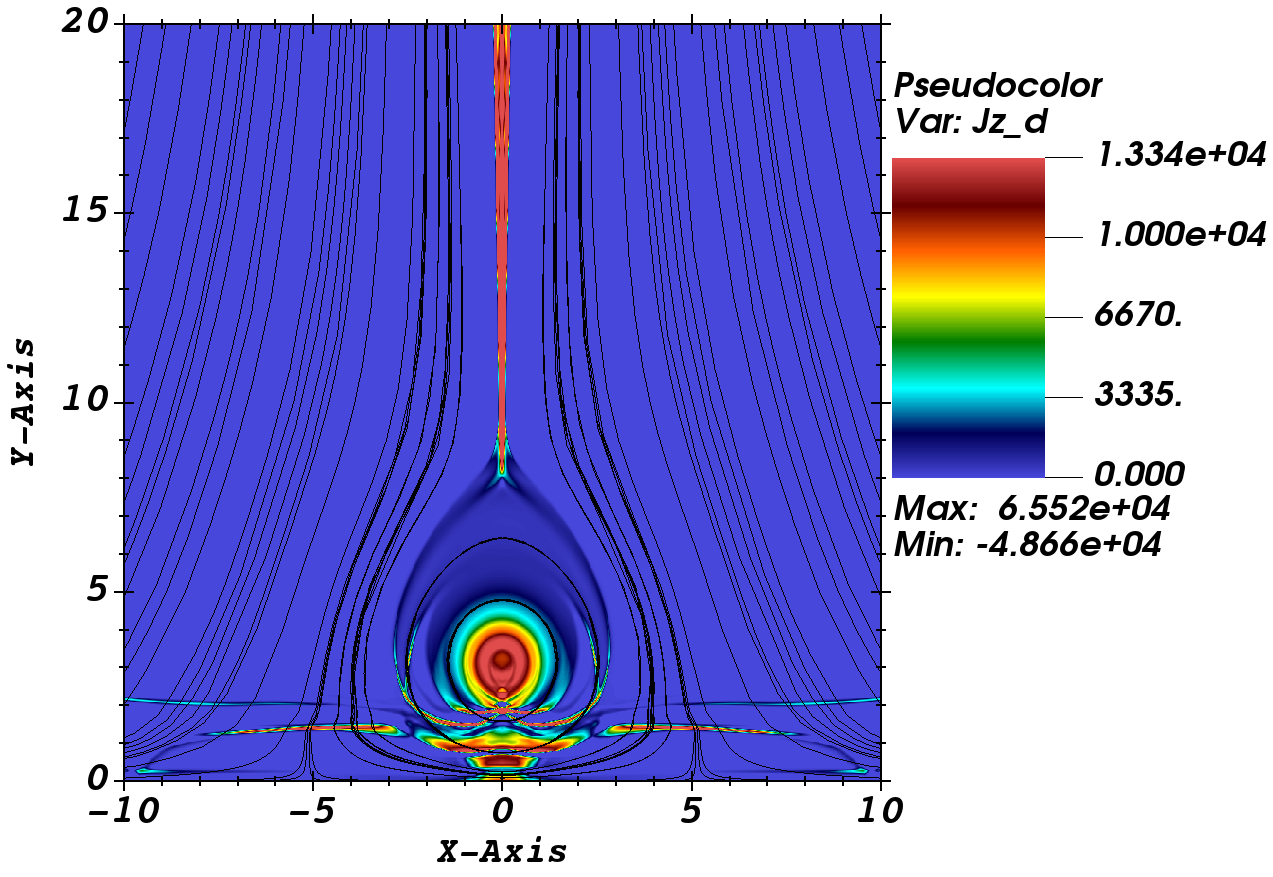}
     \includegraphics[width=6cm, height=5.0cm]{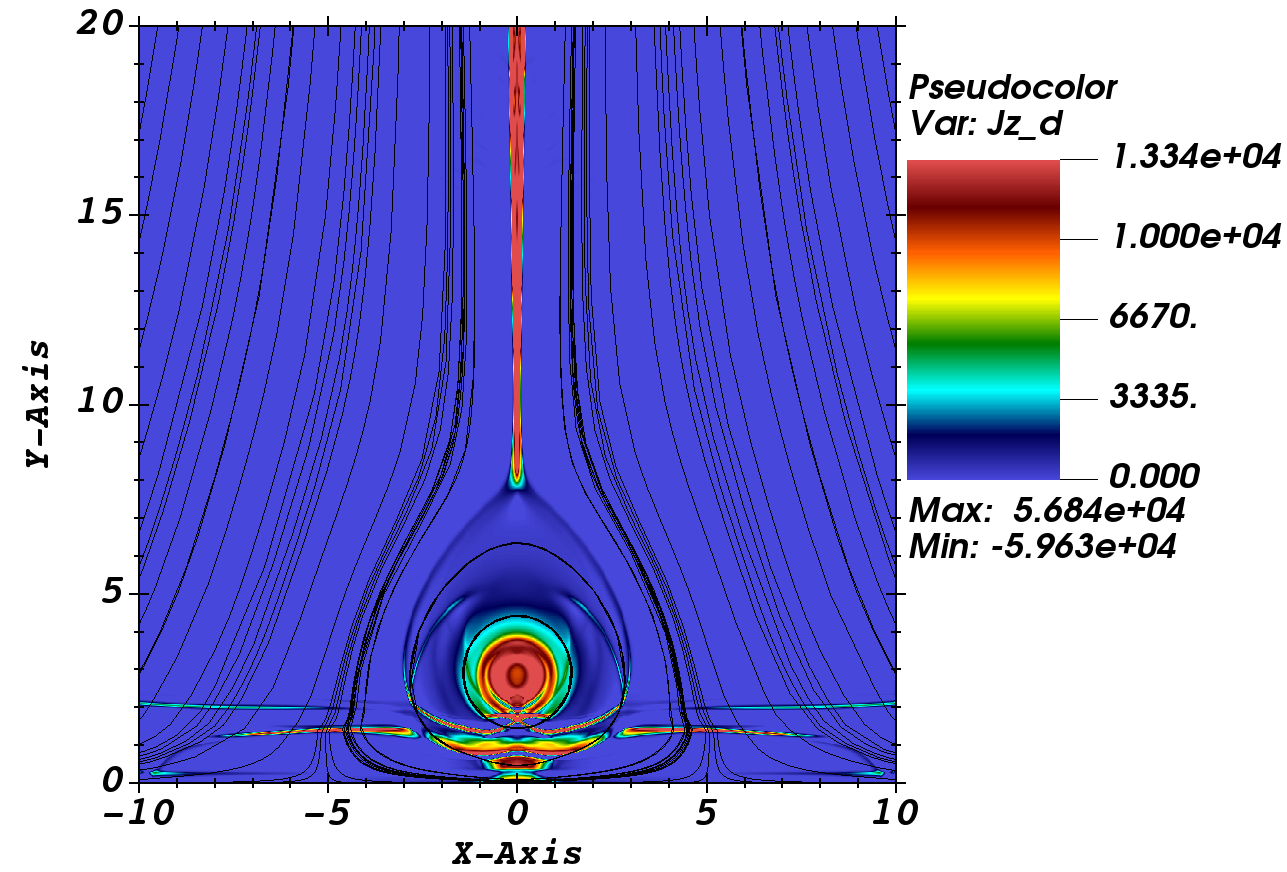}
      \centerline{\Large \bf   
      \hspace{0.28\textwidth}  \color{black}{(c)}
       \hspace{0.295\textwidth}  \color{black}{(d)}
         \hfill}
     \includegraphics[width=6cm, height=5.0cm]{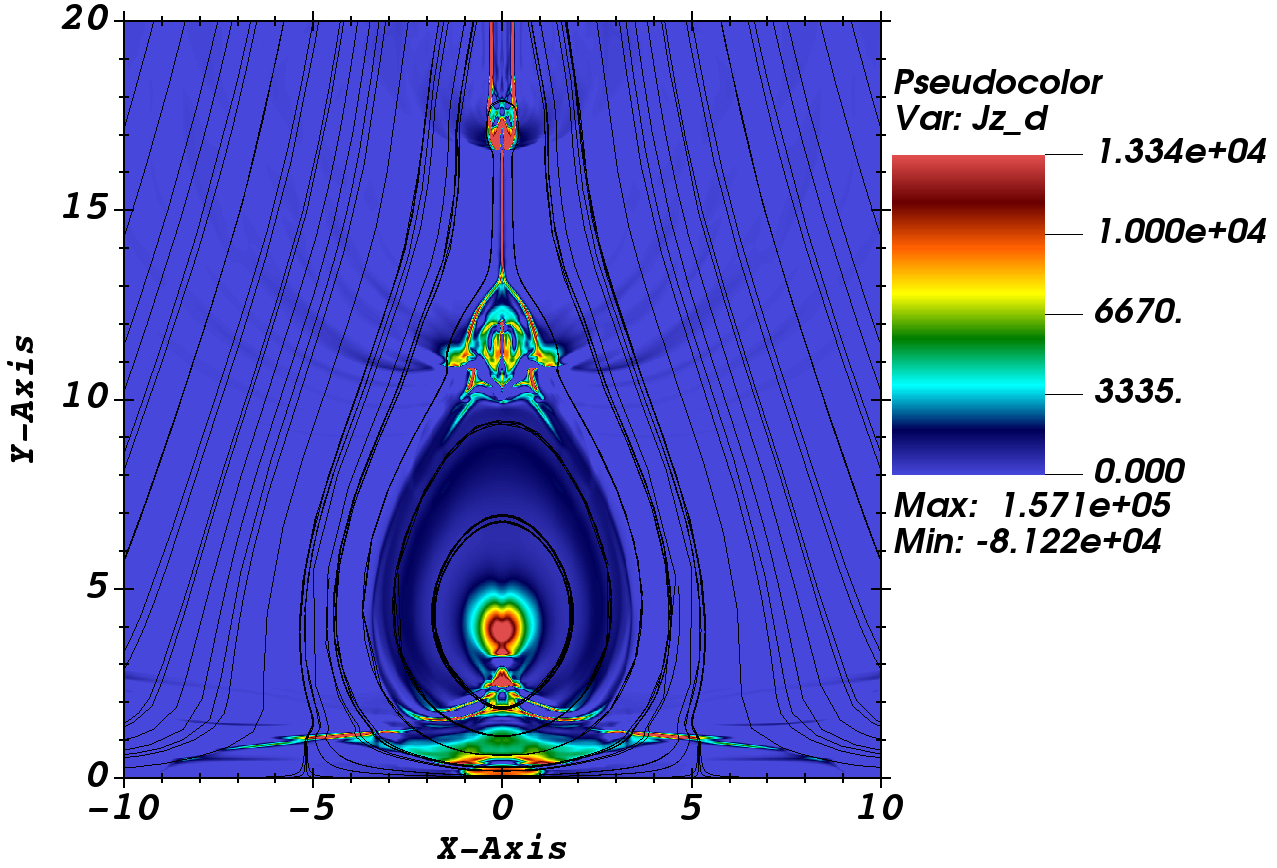}
     \includegraphics[width=6cm, height=5.0cm]{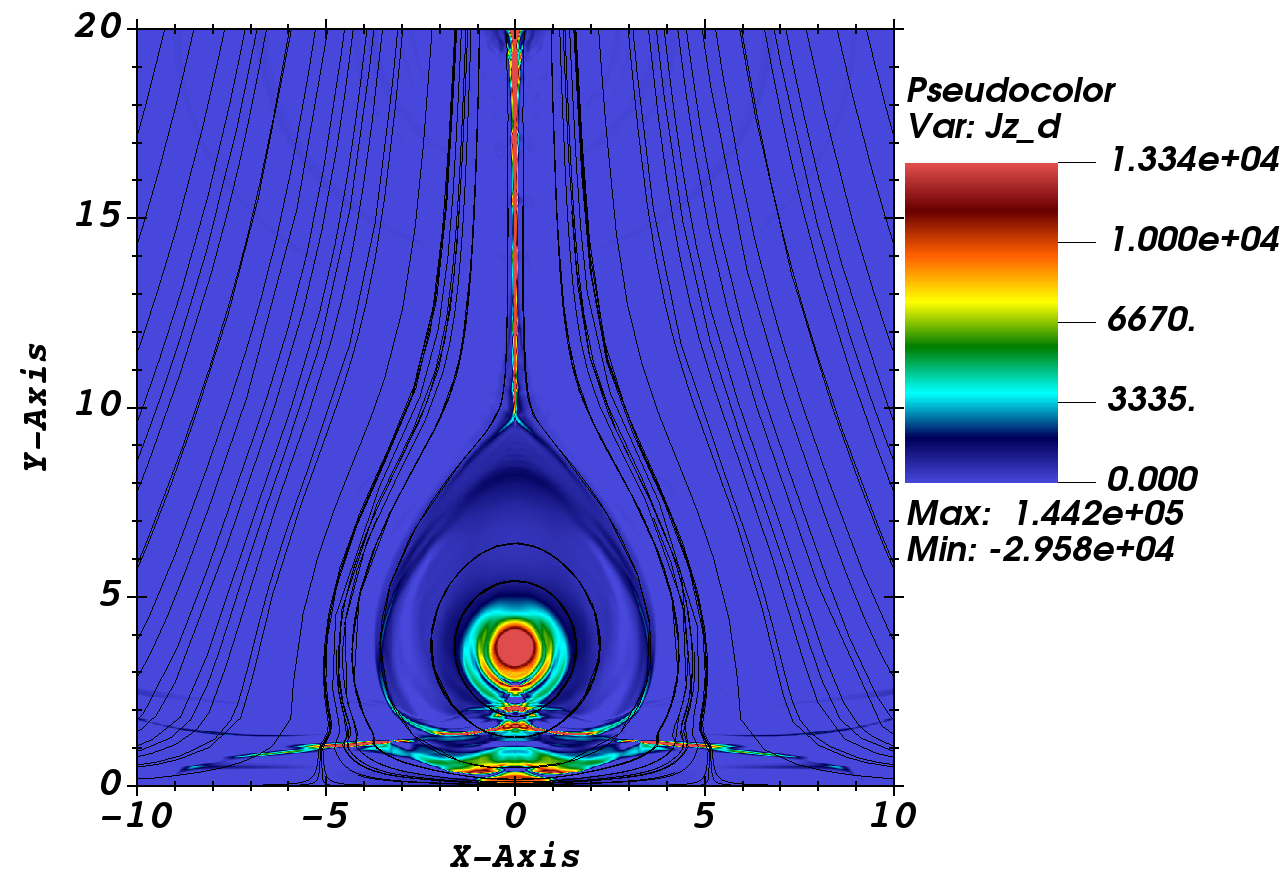}
      \centerline{\Large \bf   
      \hspace{0.28\textwidth}  \color{black}{(e)}
       \hspace{0.3\textwidth}  \color{black}{(f)}
         \hfill}
     \includegraphics[width=6cm, height=5.0cm]{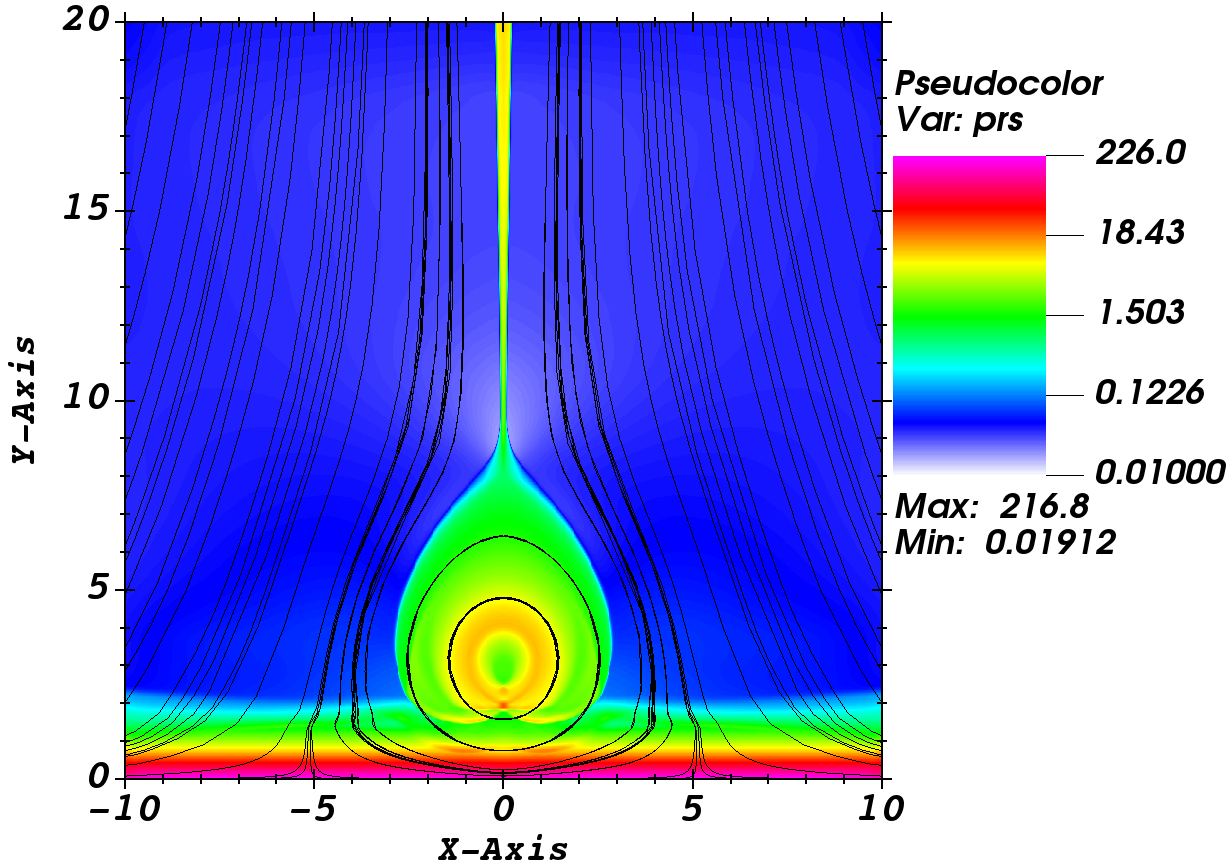}
     \includegraphics[width=6cm, height=5.0cm]{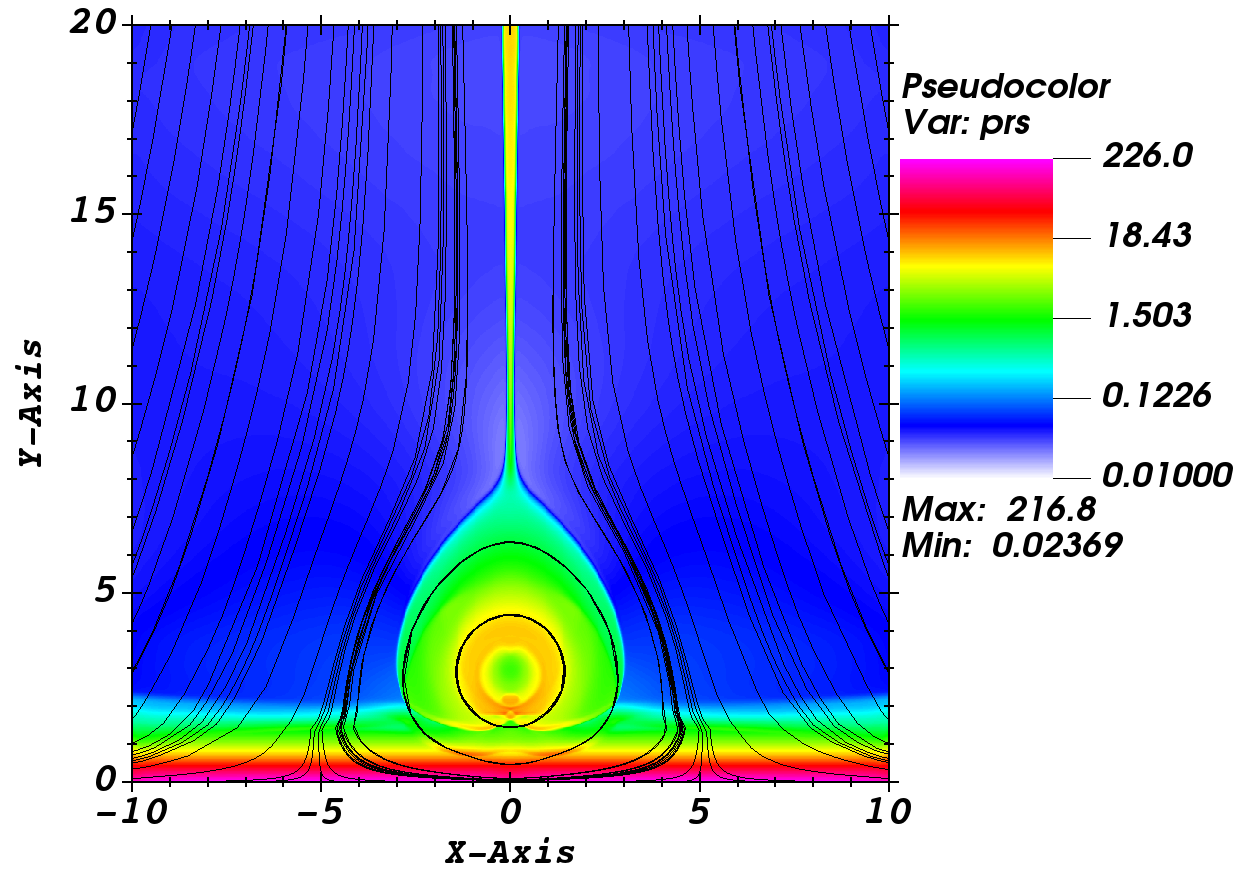}
      \centerline{\Large \bf   
      \hspace{0.28\textwidth}  \color{black}{(g)}
       \hspace{0.3\textwidth}  \color{black}{(h)}
         \hfill}
     \includegraphics[width=6cm, height=5.0cm]{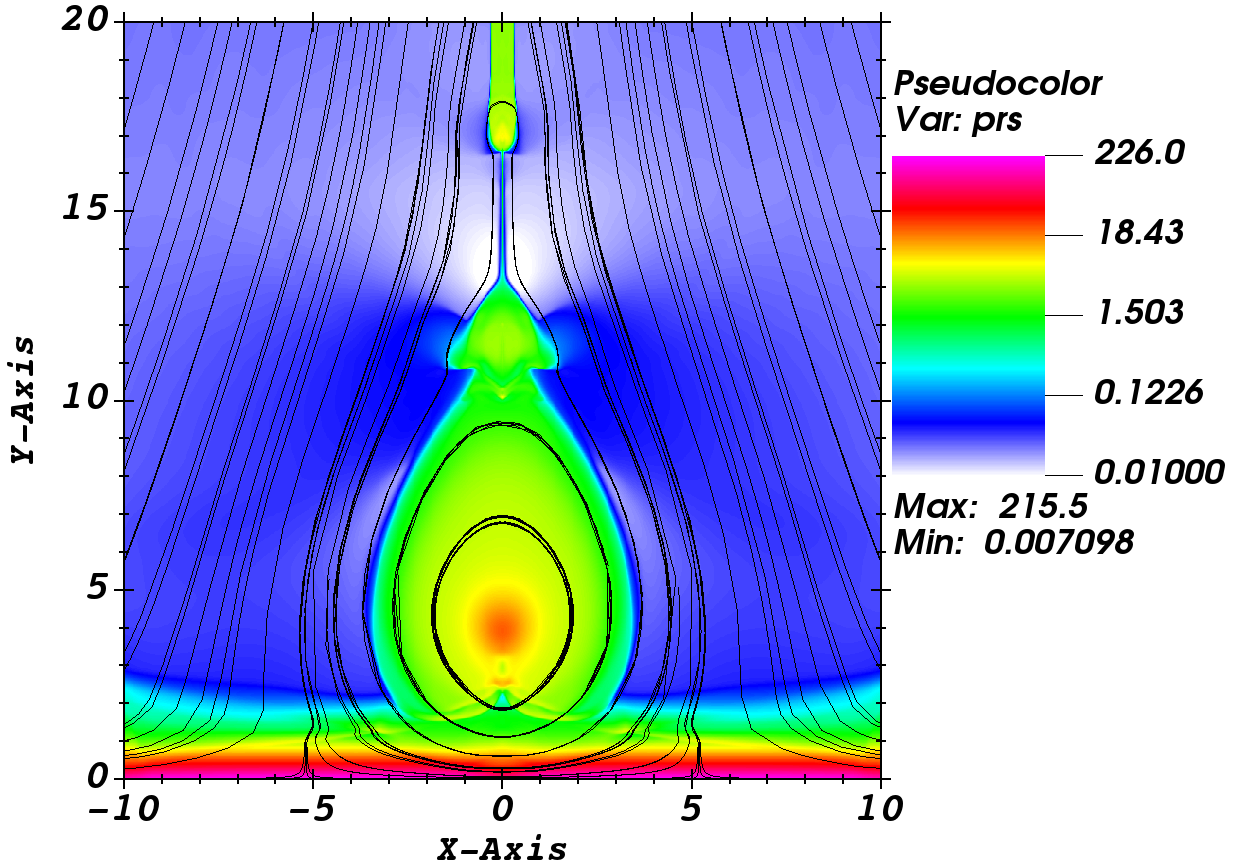}
     \includegraphics[width=6cm, height=5.0cm]{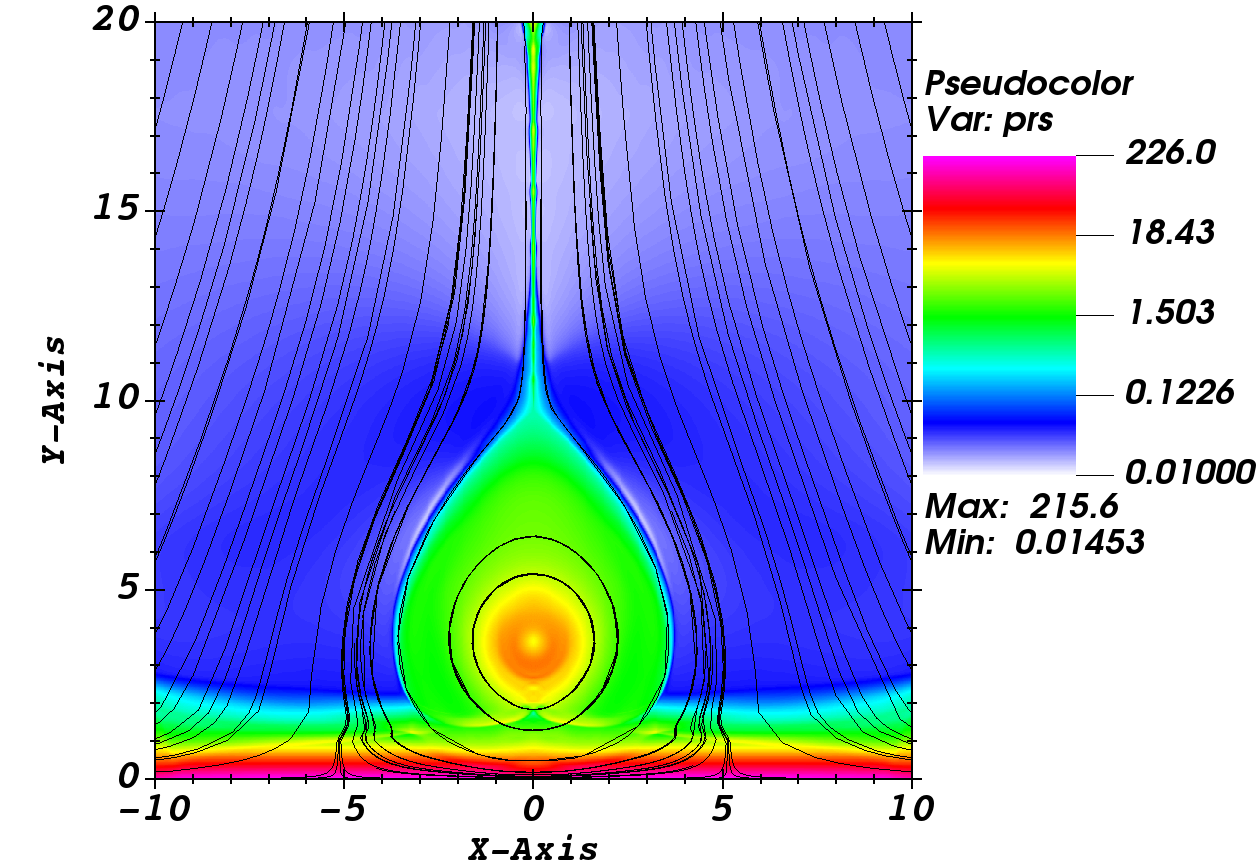}
     \caption{The $z$ component of current density in statA cm$^{-2}$ overlap with the magnetic field lines for the time $t=25$ s (panels (a) and (b)) and $t=50$ s (panels (c) and (d)). In panels (e) (f), we show plasma pressure in dyn cm$^{-2}$ for the time $t=25$ s, while in panels (g) and (h), we display mass density in gr cm$^{-3}$ for $t=50$ s, for the Res case (right panels) and the Res+TC (left panels).}
    \label{fig:current_density_gas_pressure}
\end{figure*}

In Fig. \ref{fig:current_density_gas_pressure}, we show the $z-$component of current density $J_{z}$ in statA cm$^{-2}$ and plasma pressure in dyn cm$^{-2}$ with the magnetic field lines overlap for the two simulation cases. Specifically, in panels (a), (c), (e), and (g), we display the results for the Res case at times $t=25$ s and $t=50$ s. For instance, in panel (a) at $t=25$ s, we identify substructures in $J_{z}$. In particular, there are regions with high current density inside the post-flare loops, and a thin region of high $J_{z}$ is visible that is part of the CS. While, in panel (c), at $t=50$ s, we observe substructures at about $y\approx 12$ Mm and $y\approx 17$ Mm. In panel (e), at $t=25$ s, we identify regions of high plasma pressure ($\sim 18$ dyn cm$^{-2}$) inside the post-flare loops, and $t=50$ s (g), in the plasma pressure is also evident the structure shown in $J_{z}$ of panel (c), in this particular region, the plasma pressure is around 1.5 dyn cm$^{-2}$.        

\subsection{Resistivity with thermal conduction case}
\label{Res+TC_case}

In the right panels of Fig. \ref{fig:temperature_mass_density}, we display the results for the Res+TC case. For example, in panel (b), we identify the formation of a typical flare structure and its associated post-flare loops; however, in this case, the thermal conduction redistributes the plasma inside the flare loops, which makes the evaporation flow less visible than in the Res case. Furthermore, the maximum temperature reaches $\sim 10^{7}$ K, smaller than the maximum temperature in the Res case, as shown in panel (a). In panel (d), we do not observe any structured region on the top of loops, nor the increase of evaporation of the plasma inside it, as shown in panel (c). This feature could be related to the effect of thermal conduction along the magnetic field lines, which in this case would be high along the loops, redistributing the heat inside the loops and dominating the resistivity. Panel (f) shows the mass density at $t=25$ s, which is consistent with the behavior shown in the temperature maps of panels (b) and (d). In particular, inside the loops, the density is of the order of $10^{-15}$ gr cm$^{-3}$. Finally, in panel (h), the mass density at $t=50$ s shows a smooth distribution predominantly inside the loops.   

In panels (b), (d), (f), and (h) of Fig. \ref{fig:current_density_gas_pressure}, we show the z-component of current density $J_{z}$ in statA cm$^{-2}$ and plasma pressure in dyn cm$^{-2}$ for the Res+TC case. For instance, in panel (b), $J_{z}$ at $t=25$ s shows similarities to the obtained in the Res case. However, at $t=50$ s, as displayed in panel (d), we do not identify the substructure at the top of the post-flares as in the Res case. Instead, the CS becomes thin. In panels (f) and (h), the plasma pressure shows the same smooth behavior as $J_{z}$ at both times, $t=25$ and $t=50$ s. Notably, the regions with high plasma pressure ($\sim 18$ dyn cm$^{-2}$) are mainly around the post-flare loops. 

\begin{figure*}
    \centering
    \centerline{\Large \bf   
       \hspace{0.4\textwidth}  \color{black}{\Large{Res+TC}}
         \hfill}
          \centerline{\Large \bf   
      \hspace{0.275\textwidth}  \color{black}{(a)}
       \hspace{0.295\textwidth}  \color{black}{(b)}
         \hfill}
    \includegraphics[width=6.0cm, height=5.0cm]{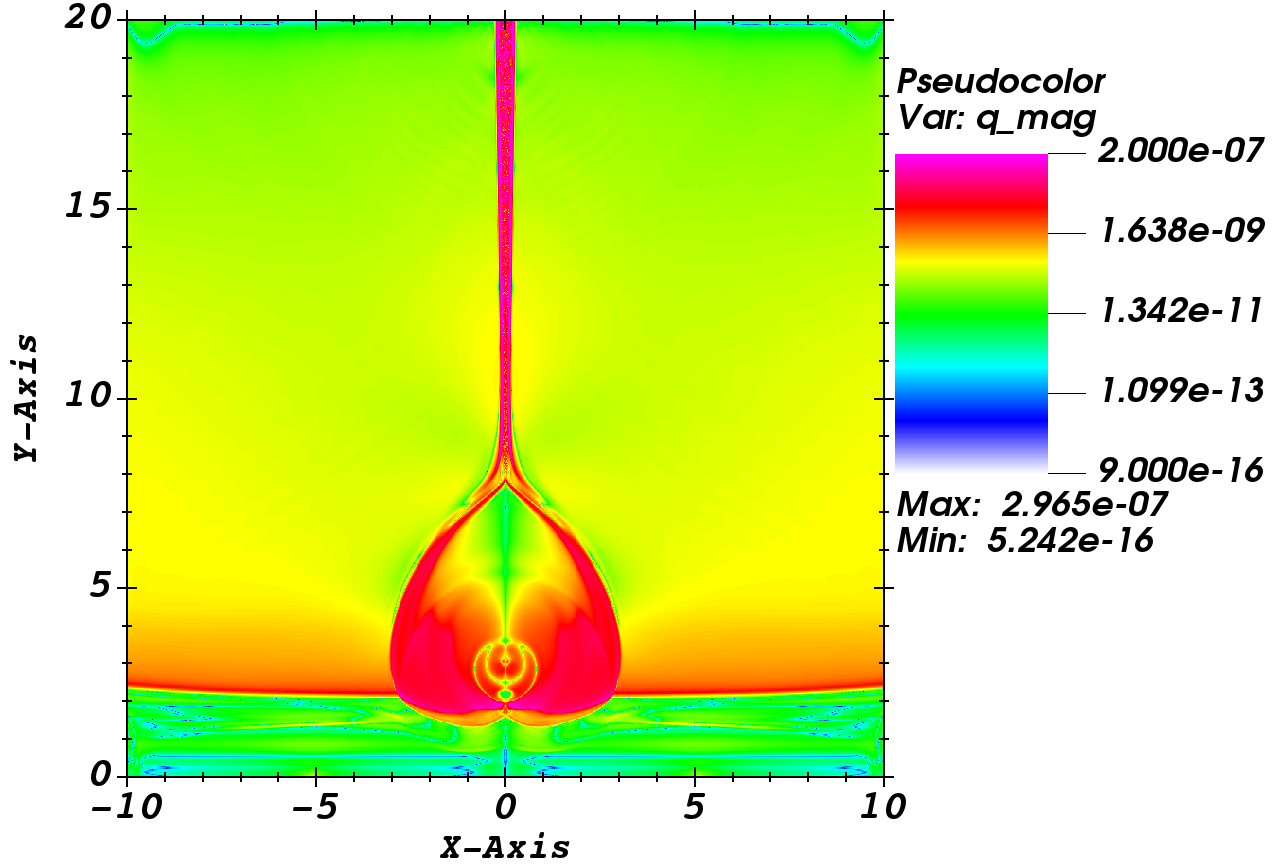}
     \includegraphics[width=6.0cm, height=5.0cm]{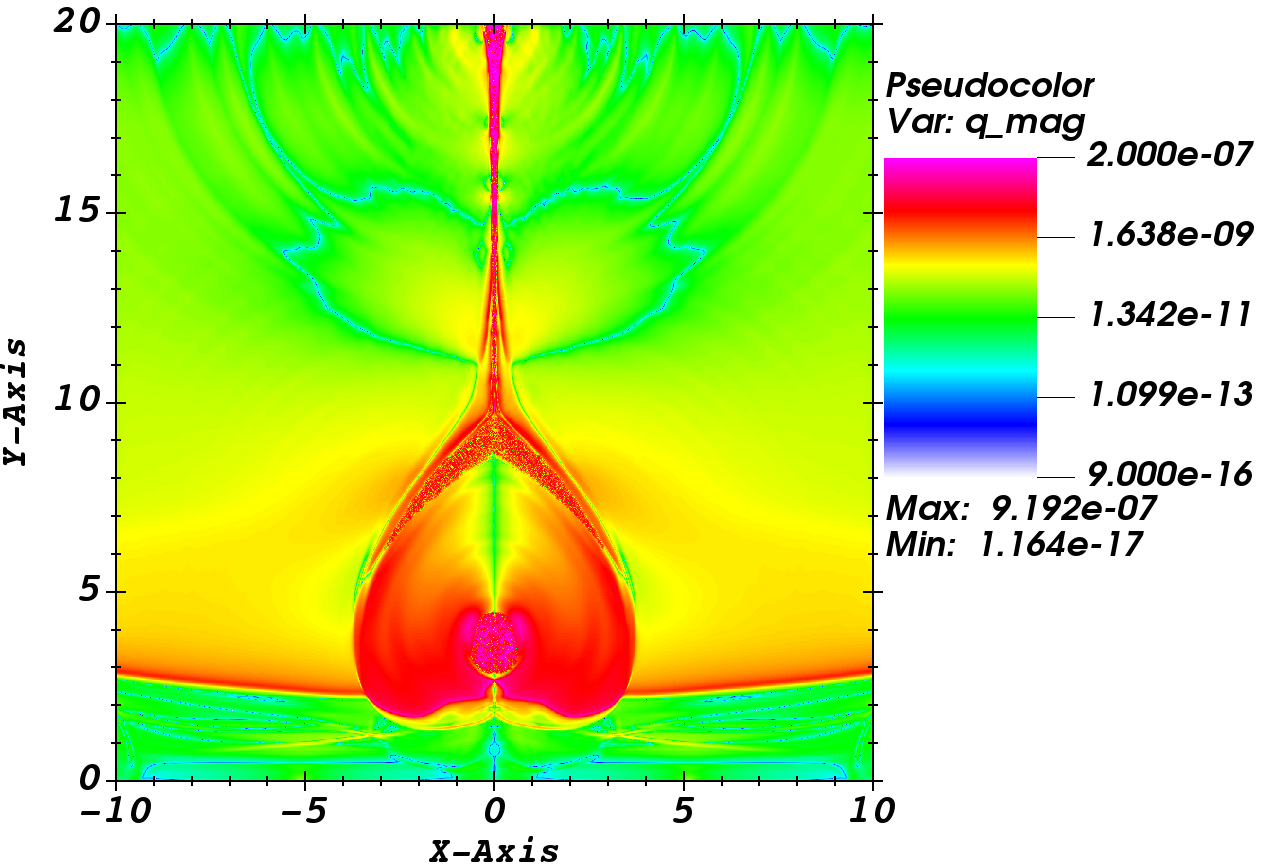}
       \caption{Maps of the local heat flux $|{\bf q}|$ in erg s$^{-1}$ cm$^{-2}$.}
    \label{fig:thermal_conduction_q}
\end{figure*}

To show how the thermal conduction behaves on the flare and post-flare loop structures, we estimate the magnitude of the heat flux $\textbf{q}$. In Fig. \ref{fig:thermal_conduction_q}, we display heat flux maps for two times $t=25$ and $t=50$ s. We see that the heat flux is high within the post-flare loops and the CS, which indicates that heat conduction could determine the thermal behavior of the post-flare loops. Therefore, this makes the structure smooth, as shown in panels (b), (d), (f), and (g) in the temperature and density maps of Fig. \ref{fig:temperature_mass_density}.

\begin{figure*}
    \centering
     \includegraphics[width=6cm, height=6cm]{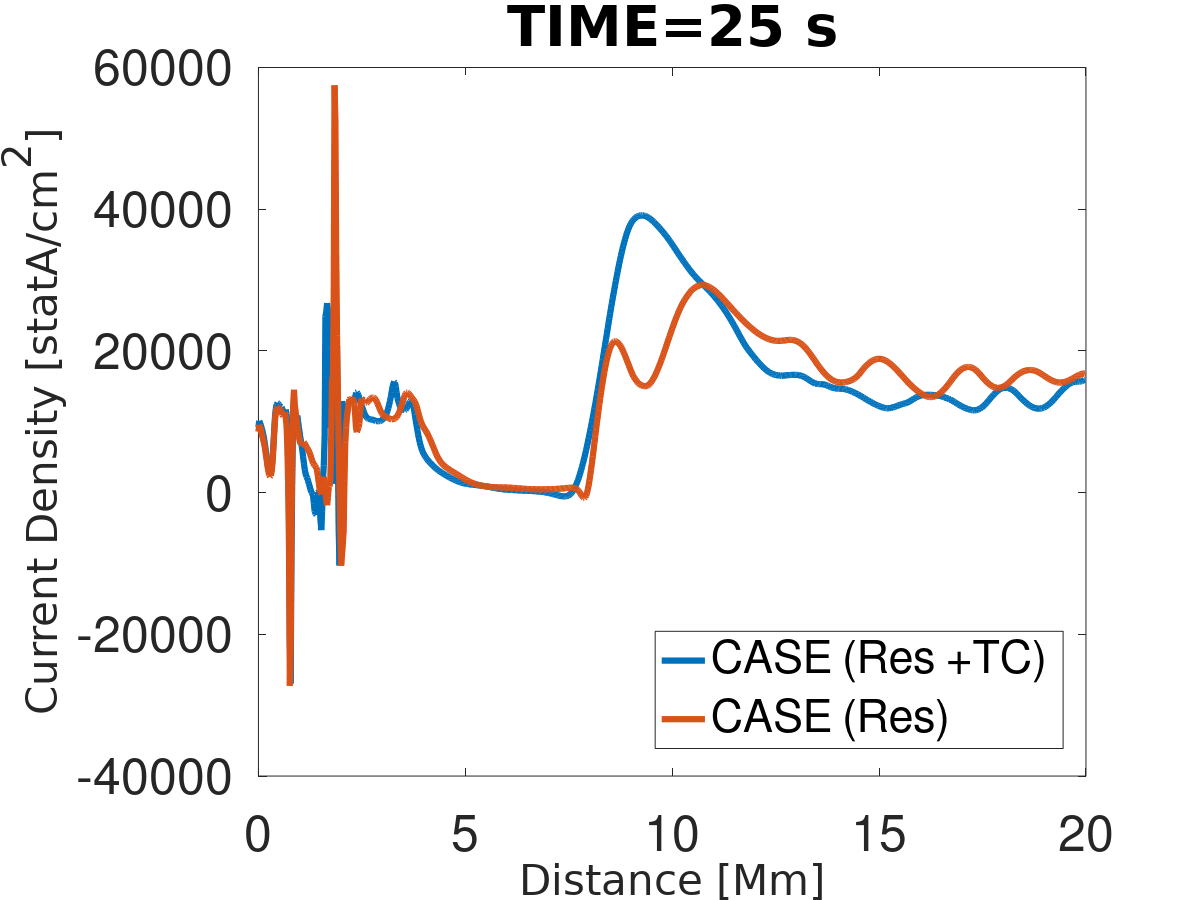}
   \includegraphics[width=6cm, height=6cm]{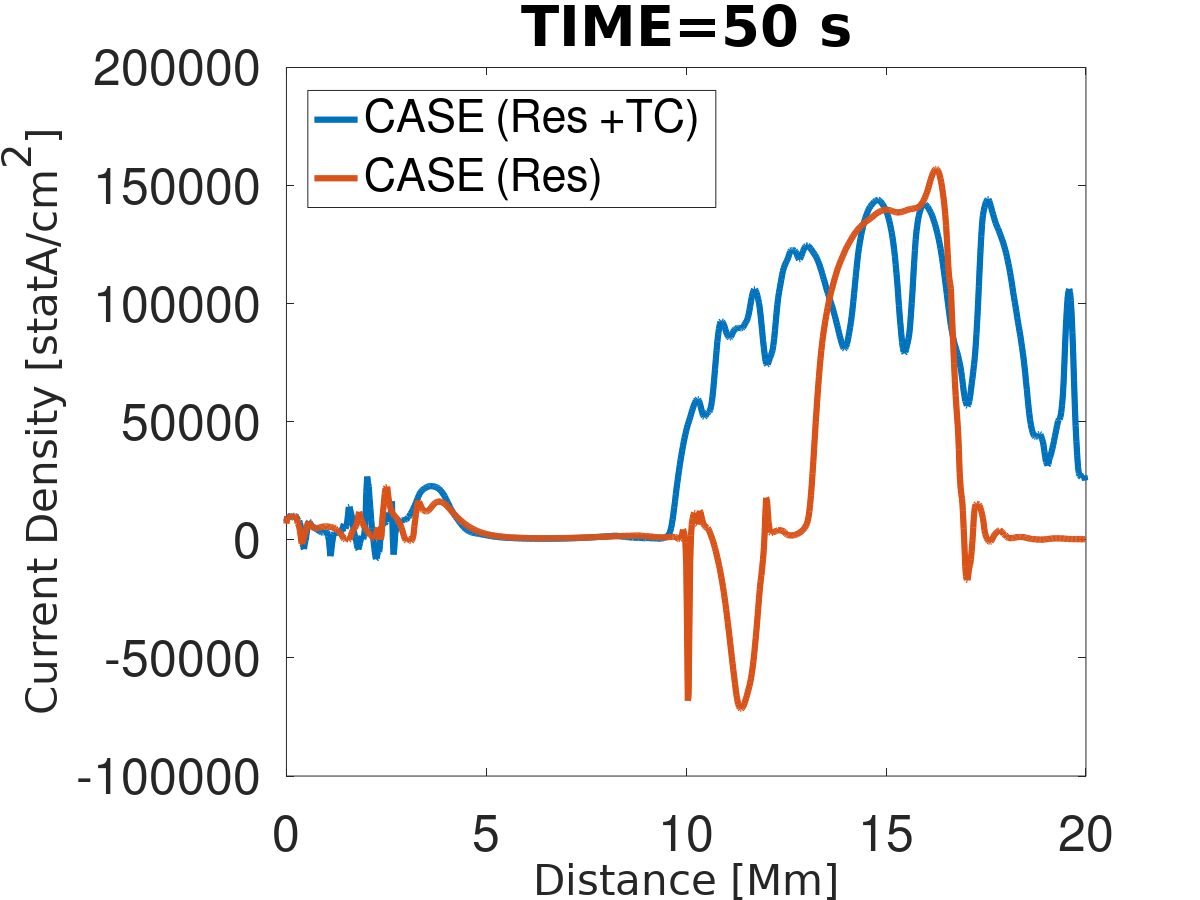}
        \includegraphics[width=6cm, height=6cm]{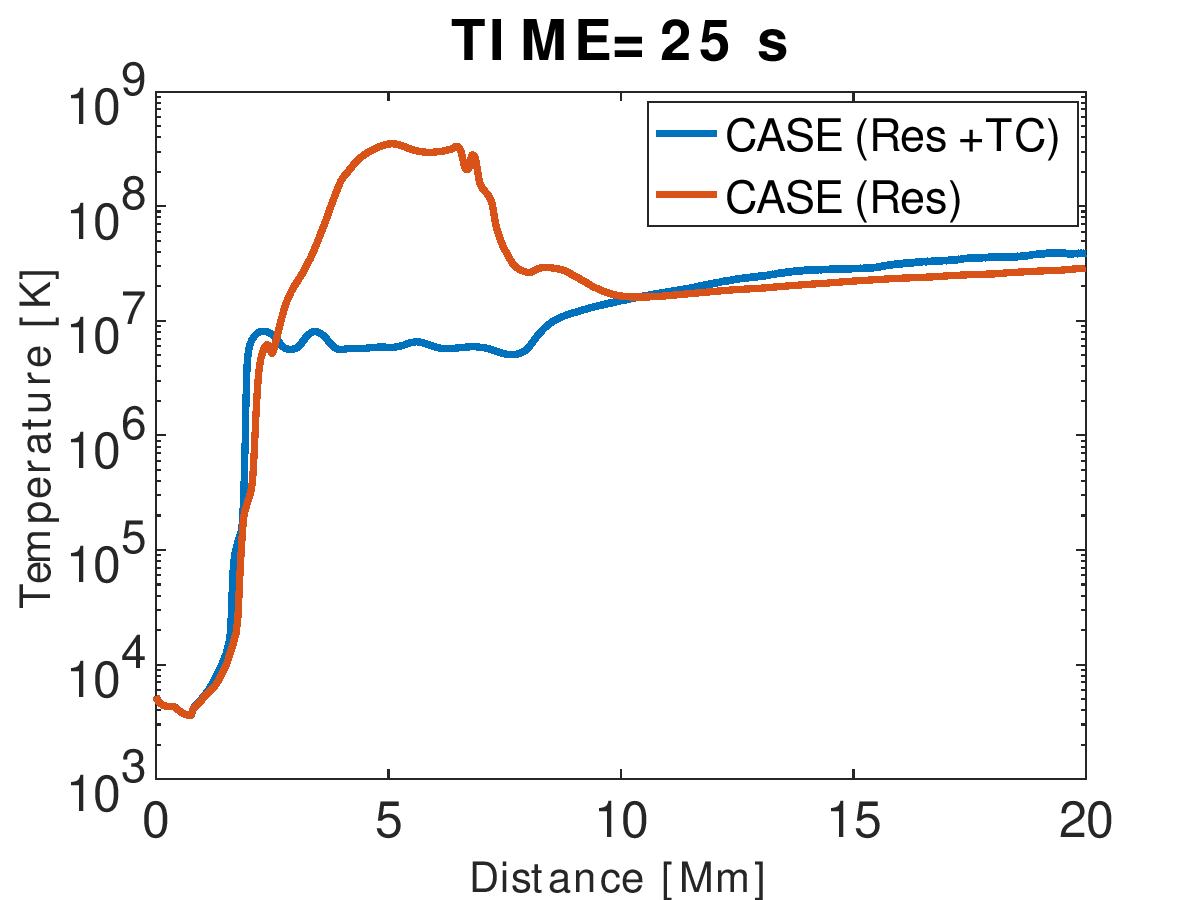}
         \includegraphics[width=6cm, height=6cm]{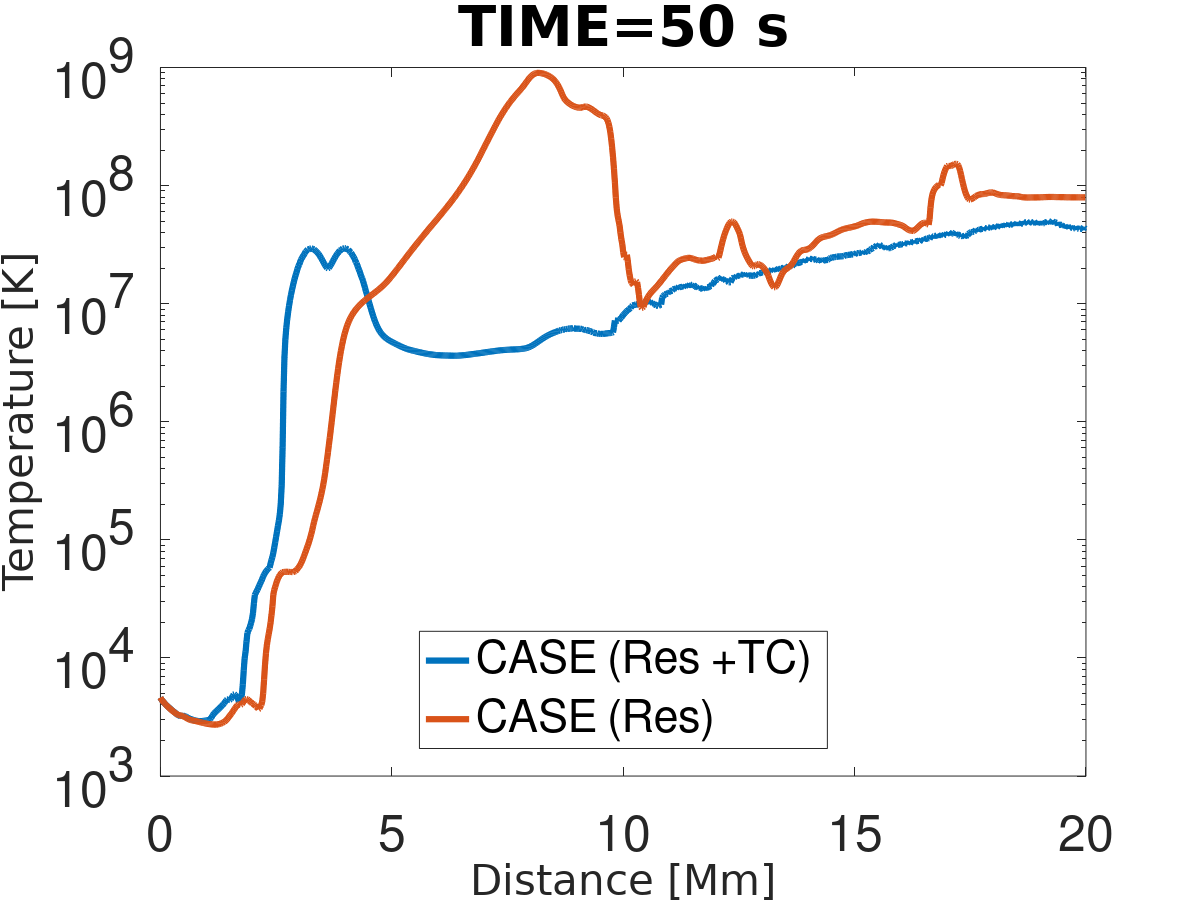}
    \caption{1D cuts along the vertical direction $y$ of the $z-$component of the current density $J_{z}$ in statA cm$^{-2}$ (top) and temperature in Kelvin (bottom) at times $t=25$ s (left panels) and $t=50$ s (right panels) the Res case (blue curves) and the Res+TC case (orange curves).}
    \label{fig:cross-cuts}
\end{figure*}

In Fig. \ref{fig:cross-cuts}, we show 1D cuts along the vertical distance $y$ and at $x=0$ Mm of $J_{z}$ in statA cm$^{-2}$ (top panels) and the temperature in Kelvin (bottom panels), at times $t=25, 50$ s. These plots help identify the current density and temperature variations with the vertical distance and highlight the behavior between these variables in the Res and Res+TC cases. For example, in the top-left panel, we show that $J_{z}$ at $t=25$ s behaves similarly in both scenarios; however, in the Res case, there is an abrupt growth at around the distance $y=3$ Mm, which is related to a high current density region, as shown in panel (a) of Fig. \ref{fig:current_density_gas_pressure}. On the other hand, at $t=50$ s, the current density variations are remarkably different for the Res+TC case in the interval of 10-20 Mm than the behavior in the Res case. These variations could be related to the formation of multiple small-scale magnetic islands along the CS, as we will analyze in detail in the following subsections. Finally, in the bottom panels, we show the variations of temperature as a function of the distance, which behave consistently with the results shown in Fig. \ref{fig:temperature_mass_density}. In particular, we observe a temperature growth at about $y\approx 3$ Mm, in both cases at the two times. Nevertheless, in the Res case, the plasma temperature reaches higher values ($\sim 10^{9}$ K). In contrast, in the Res+TC case, the temperature of the plasma maintains an order of magnitude of about $10^{7}$ K. We can interpret these results in terms of the effects that thermal conduction has over the plasma temperature, i.e., in the Res case, the heat continues to affect the fluid due to high dynamics. In contrast, in the Res+TC case, the heat is dissipated and redistributed, so the temperature does not rise considerably. 

\subsubsection{Instabilities}
\label{Instabilities}

In general, in a fluid the hydrodynamic instabilities such as Rayleigh-Taylor (RT) and Ritchmyer-Meshkov (RM) frequently appear when turbulence takes place. In particular, the RT instability occurs at an interface between different fluids when the lighter fluid is accelerated into the heavy. While the RM instability may be considered a particular case of RT instability, it develops when the acceleration provided is impulsive, resulting from a shock wave \citep{ZHOU2021132838}. 

\begin{figure*}
    \centering
    \centerline{\Large \bf   
      \hspace{0.26\textwidth}  \color{black}{\Large{Res}}
       \hspace{0.29\textwidth}  \color{black}{\Large{Res+TC}}
         \hfill}
          \centerline{\Large \bf   
      \hspace{0.17\textwidth}  \color{black}{(a)}
       \hspace{0.32\textwidth}  \color{black}{(b)}
         \hfill}
          \includegraphics[width=6.5cm, height=6.5cm]{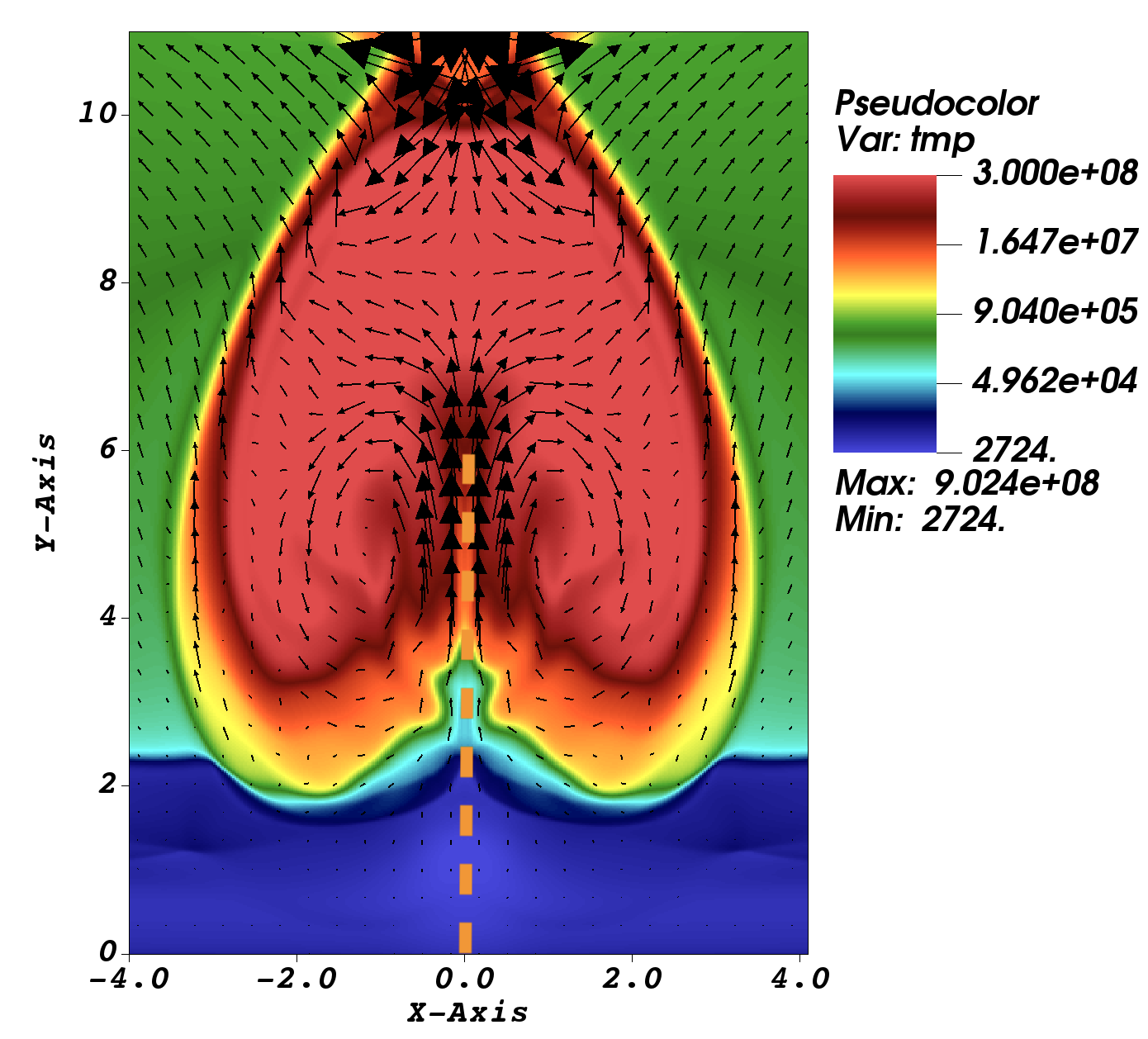}
     \includegraphics[width=6.5cm, height=6.5cm]{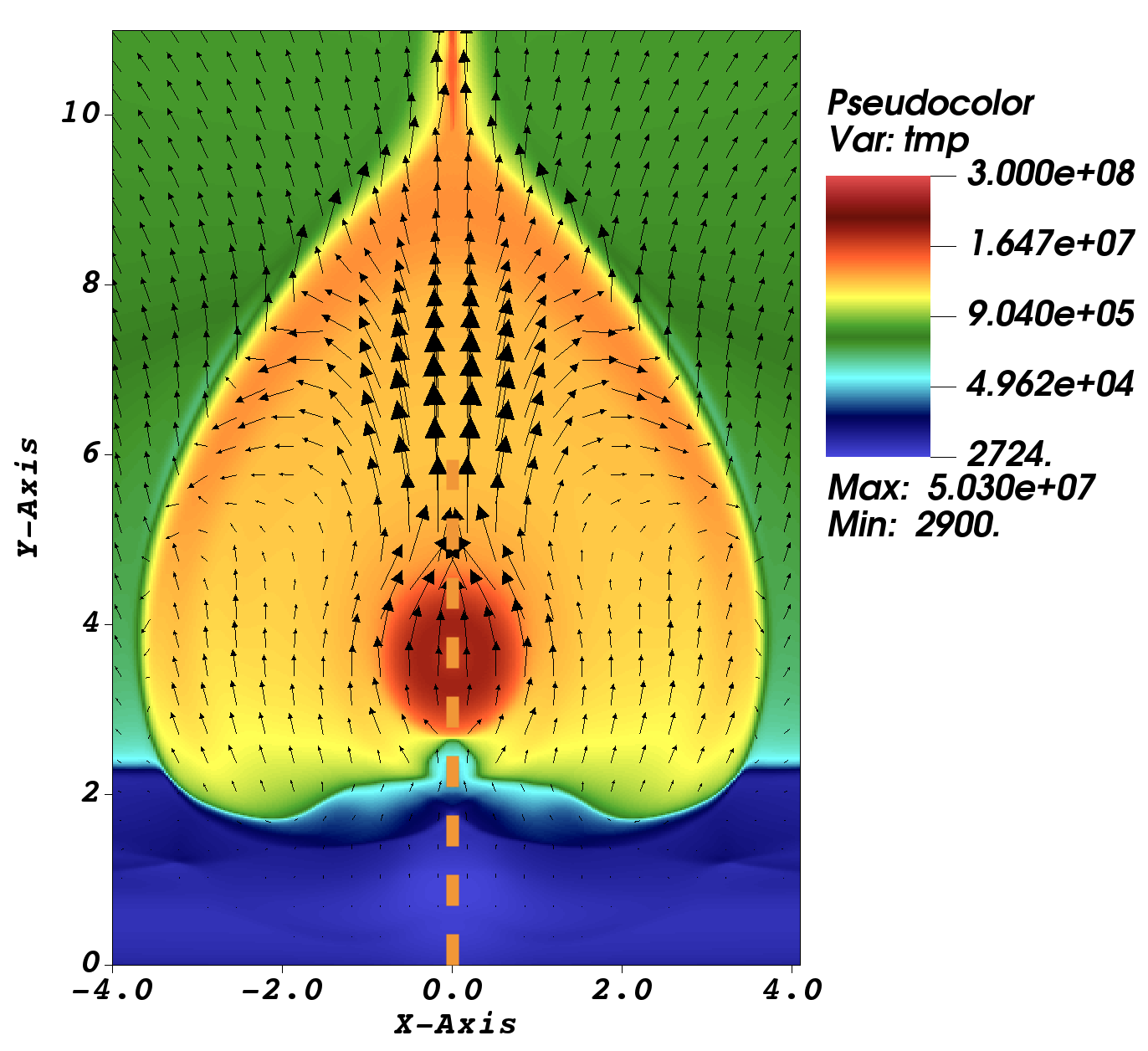}
      \centerline{\Large \bf   
      \hspace{0.175\textwidth}  \color{black}{(c)}
       \hspace{0.325\textwidth}  \color{black}{(d)}
         \hfill}
    \includegraphics[width=6.5cm, height=6.5cm]{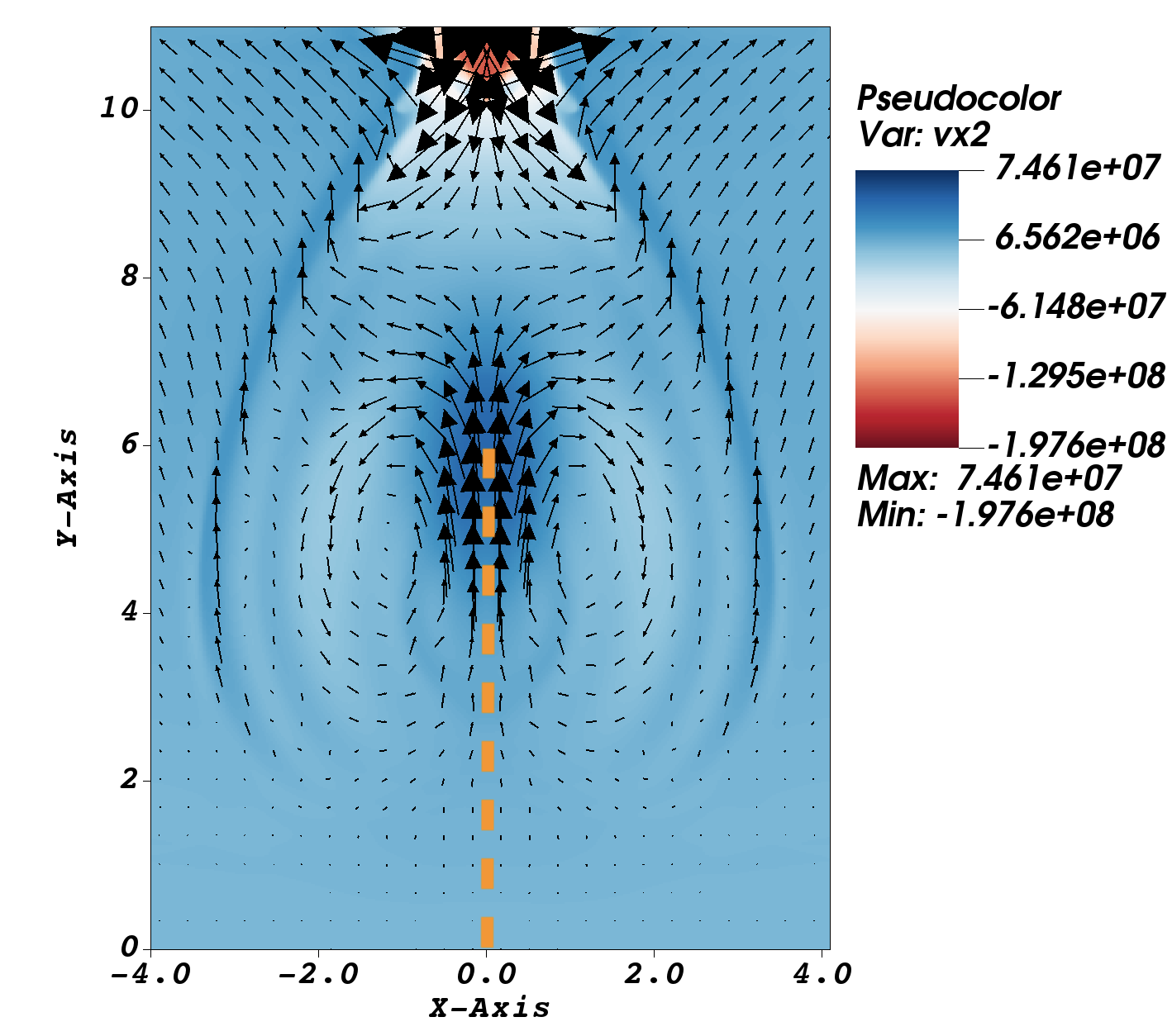}
     \includegraphics[width=6.5cm, height=6.5cm]{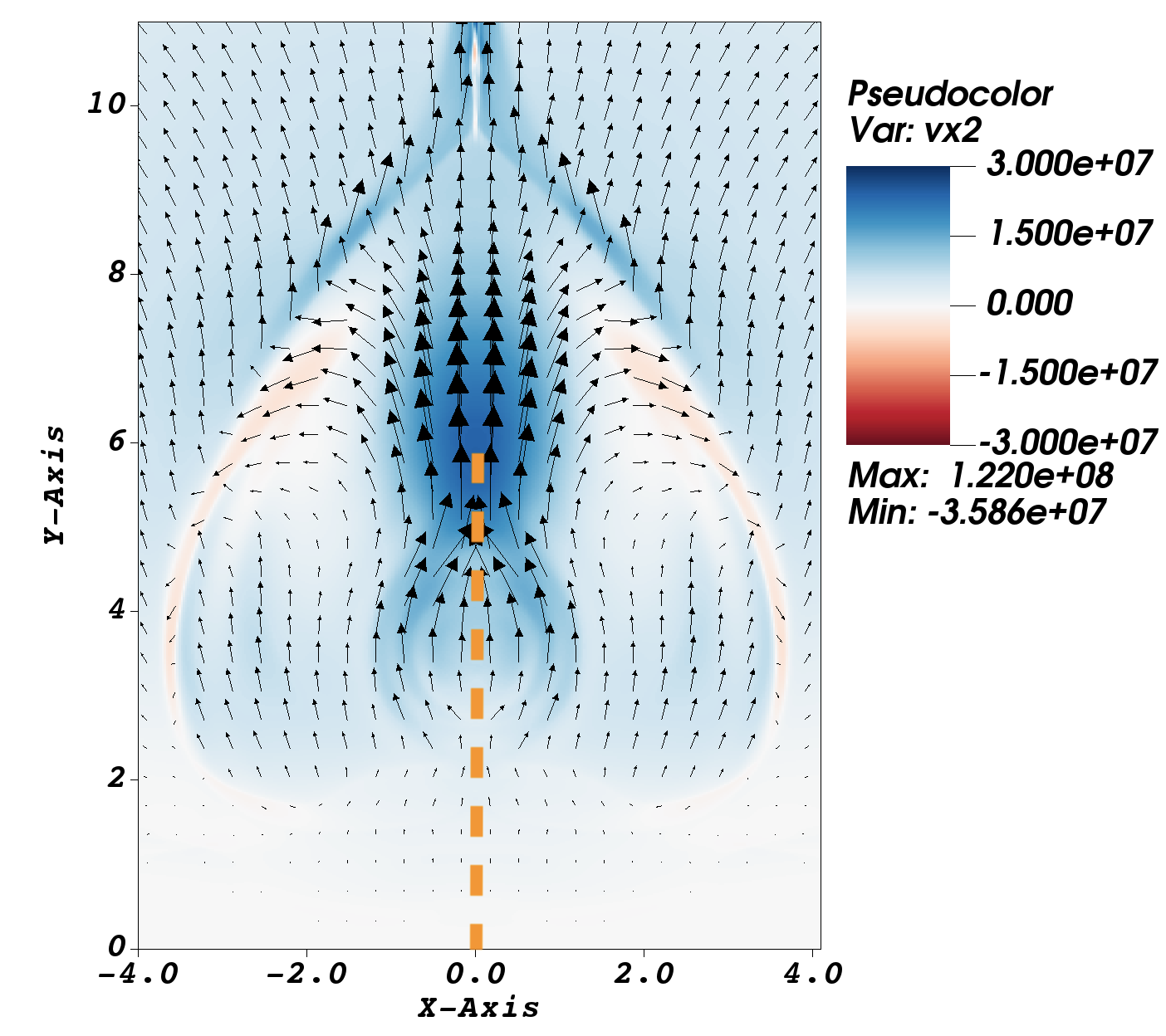}
     \caption{Zoom of the temperature in kelvin (top panels), and vertical velocity $v_{y}$ in cm s$^{-1}$ (bottom panels) overlap with vector velocity field at $t=50$ s, corresponding to the Res case (panels (a) and (c)), and the Res+TC case (panels (b) and (d)).}
    \label{fig:zoom_temp_vertical_vel}
\end{figure*}

Furthermore, in Fig. \ref{fig:zoom_temp_vertical_vel}, we show a zoom region of the temperature and vertical velocity overlap with the velocity vector field at time $t=50$ s, for the Res case (panels (a) and (c)) and for the Res+TC case (panels (b) and (d). In panels (a) and (c) of the figure, we identify plasma flowing downward and upward, shown in the velocity vector field; therefore, these flows could generate the 'spike' structures inside the post-flare loop regions. In contrast, in panels (b) and (d), we see that plasma moves predominantly upwards, and the post-flare loop looks homogeneous; this is consistent with the behavior of the heat flow that is high inside the flare loops, as schematized by the maps of $|{\bf q}|$ in Fig. \ref{fig:thermal_conduction_q}. In the four figures is an orange vertical dashed line, which represents the distance where the pressure, density, and velocity gradients are estimated for the instabilities analysis shown below.  

\begin{figure*}
    \centering
    \centerline{\Large \bf   
       \hspace{0.4\textwidth} \color{black}{\Large{Res case}}
         \hfill}
         \centerline{\Large \bf   
      \hspace{0.25\textwidth}  \color{black}{(a)}
       \hspace{0.36\textwidth}  \color{black}{(b)}
         \hfill}
     \includegraphics[width=7.0cm, height=5.0cm]{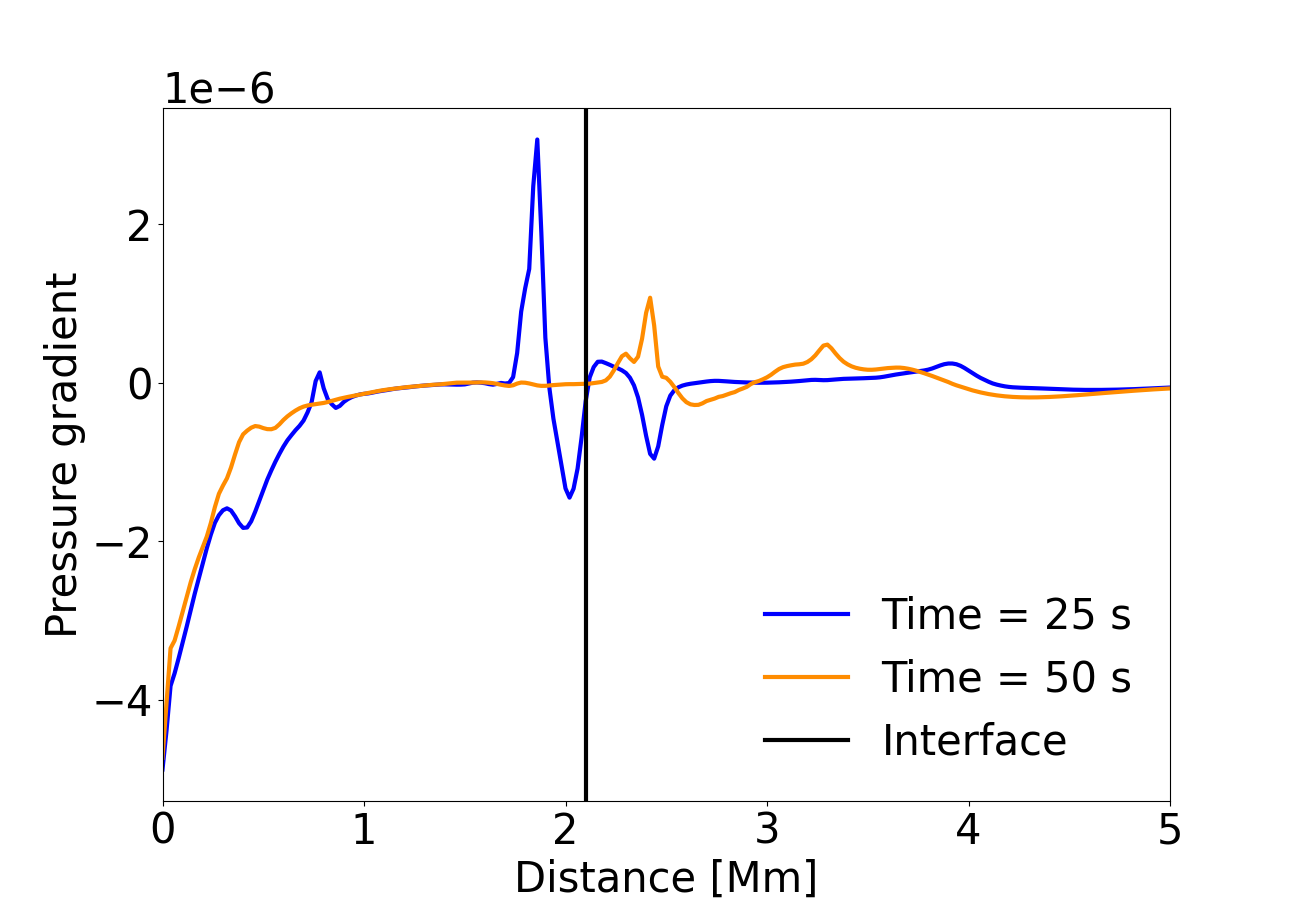}
     \includegraphics[width=7.0cm, height=5.0cm]{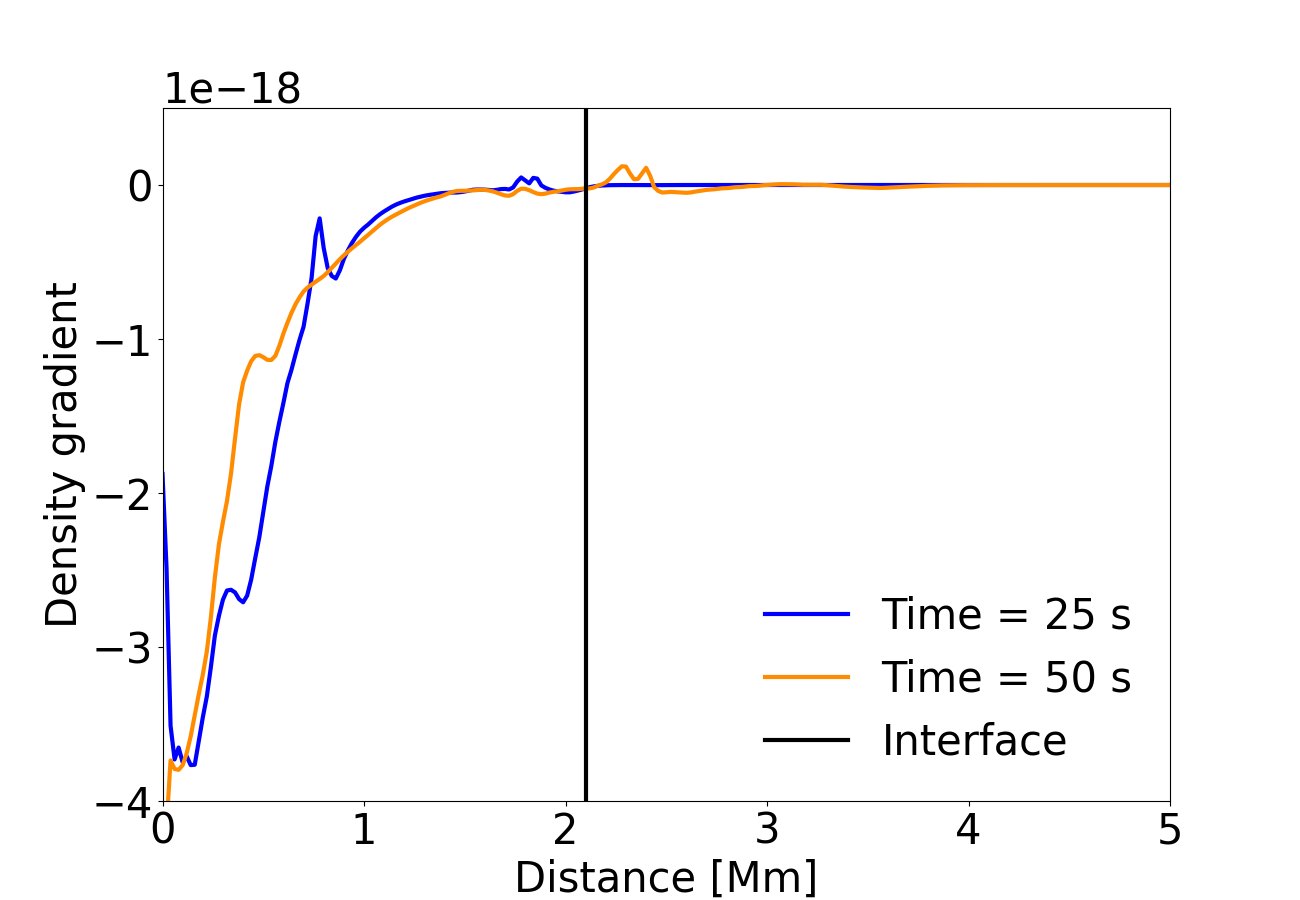}
      \centerline{\Large \bf   
      \hspace{0.25\textwidth}  \color{black}{(c)}
       \hspace{0.36\textwidth}  \color{black}{(d)}
         \hfill}
      \includegraphics[width=7.0cm, height=5.0cm]{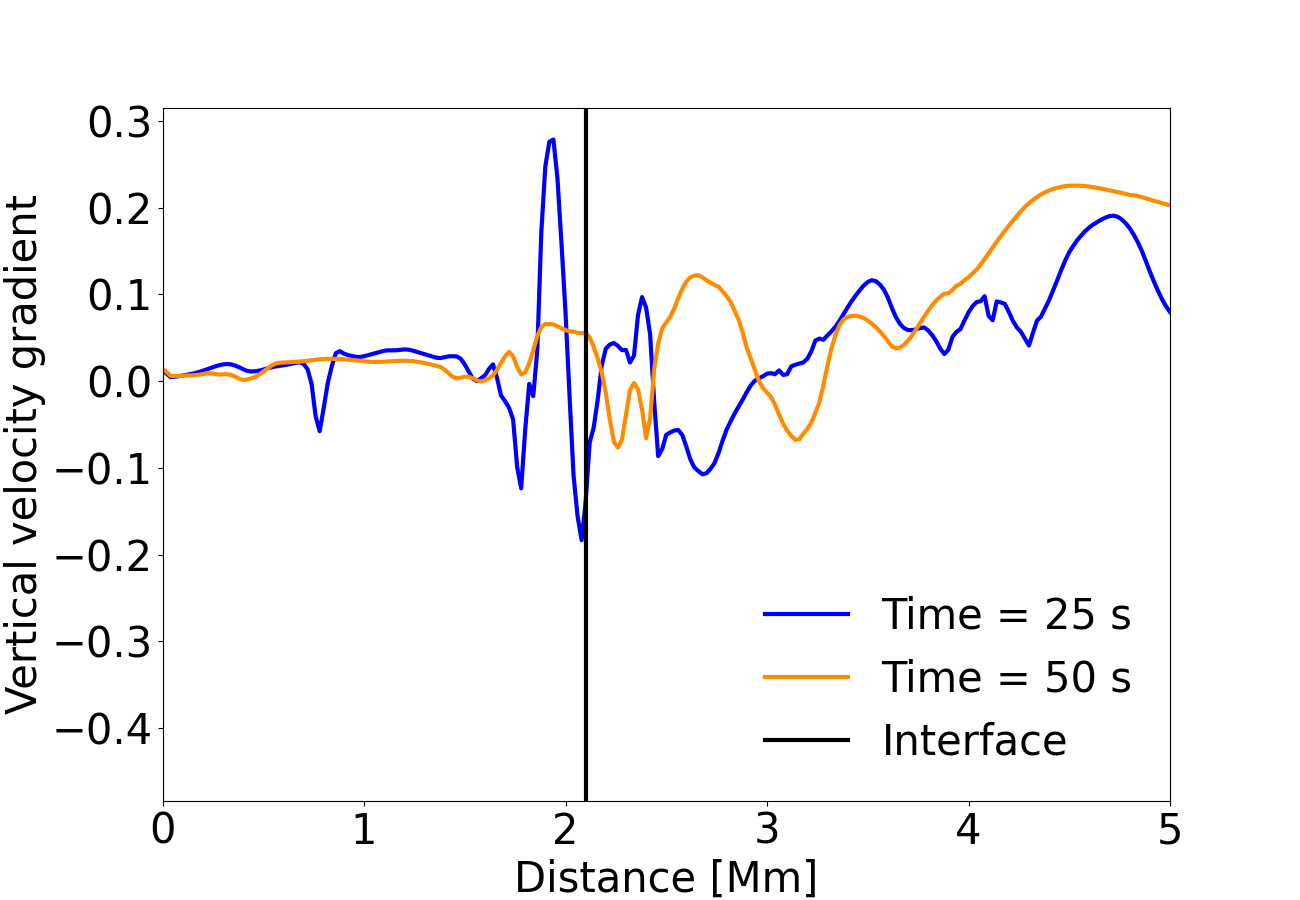}
       \includegraphics[width=7.0cm, height=5.0cm]{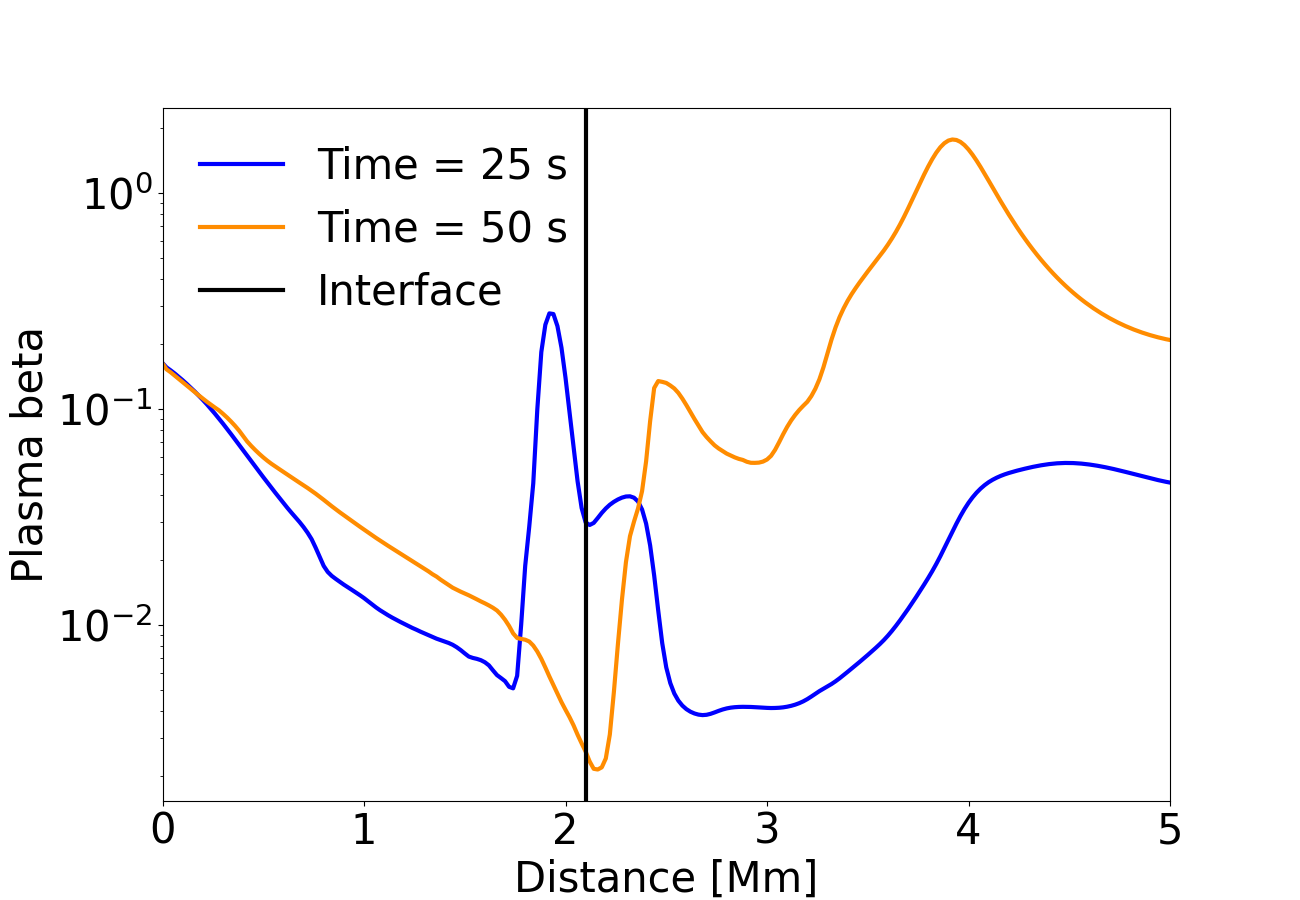}
       \centerline{\Large \bf   
       \hspace{0.4\textwidth} \color{black}{\Large{Res+TC case}}
         \hfill}
     \centerline{\Large \bf   
      \hspace{0.25\textwidth}  \color{black}{(e)}
       \hspace{0.36\textwidth}  \color{black}{(f)}
         \hfill}
     \includegraphics[width=7.0cm, height=5.0cm]{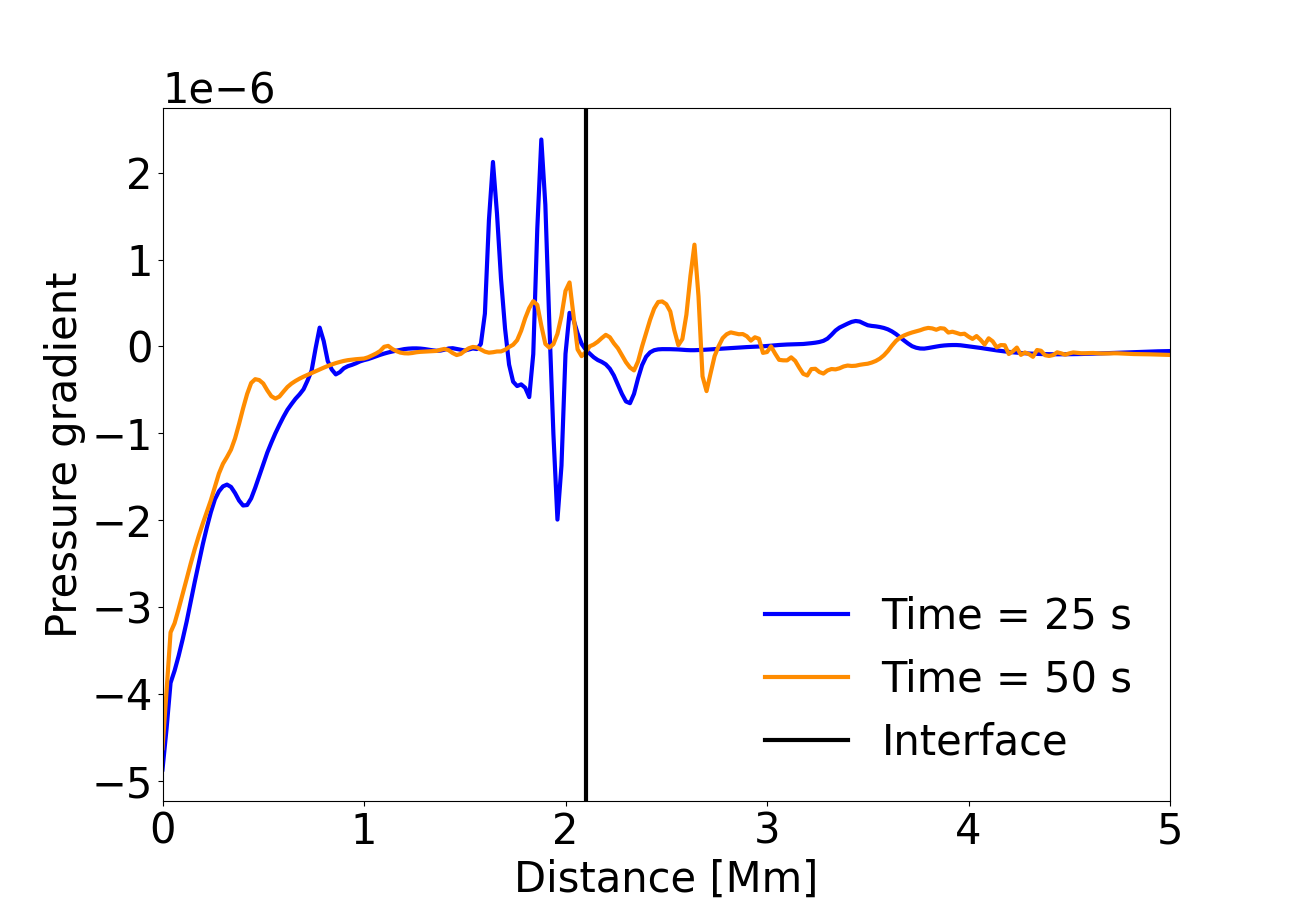}
     \includegraphics[width=7.0cm, height=5.0cm]{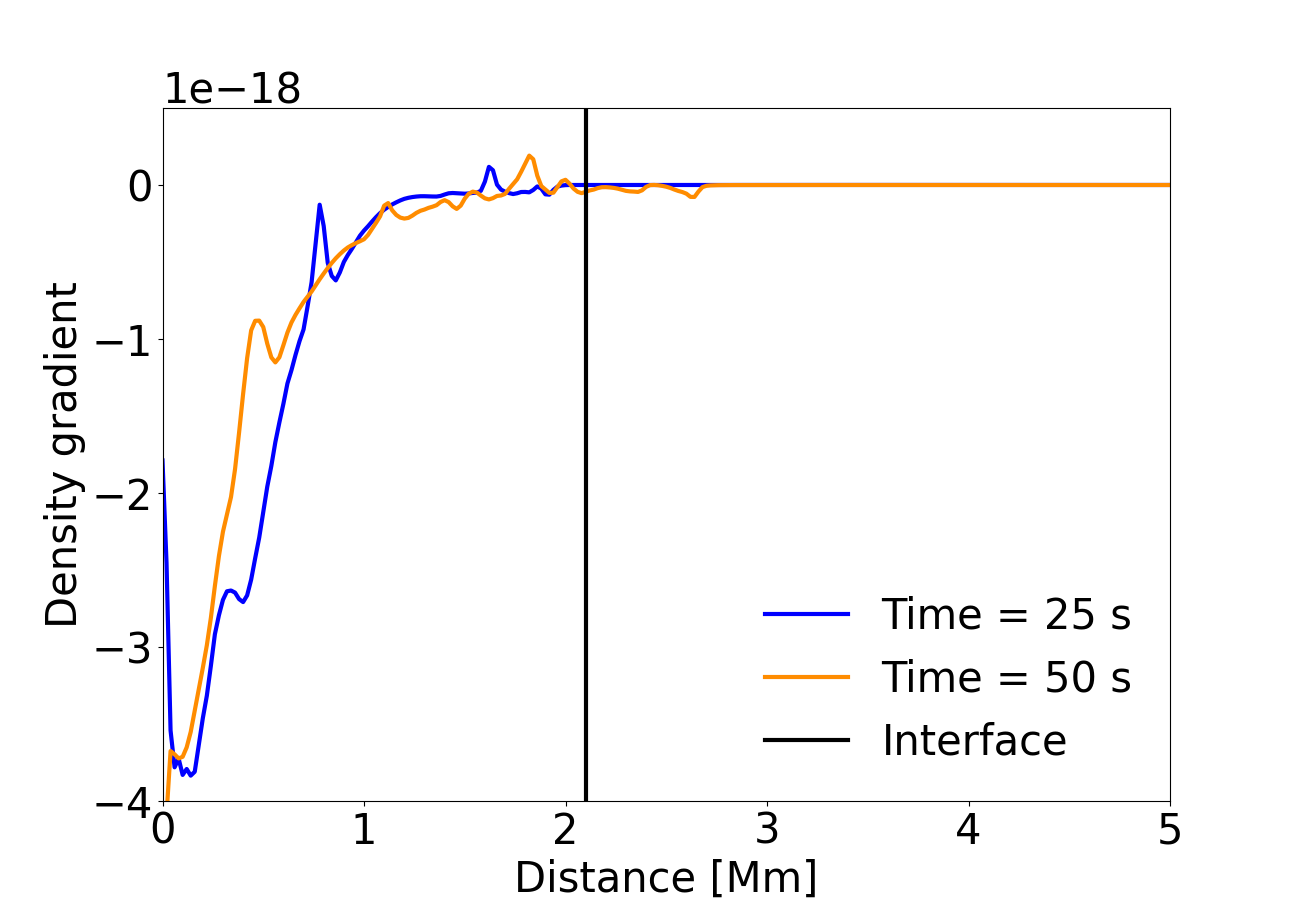}
     \centerline{\Large \bf   
      \hspace{0.25\textwidth}  \color{black}{(g)}
       \hspace{0.36\textwidth}  \color{black}{(h)}
         \hfill}
      \includegraphics[width=7.0cm, height=5.0cm]{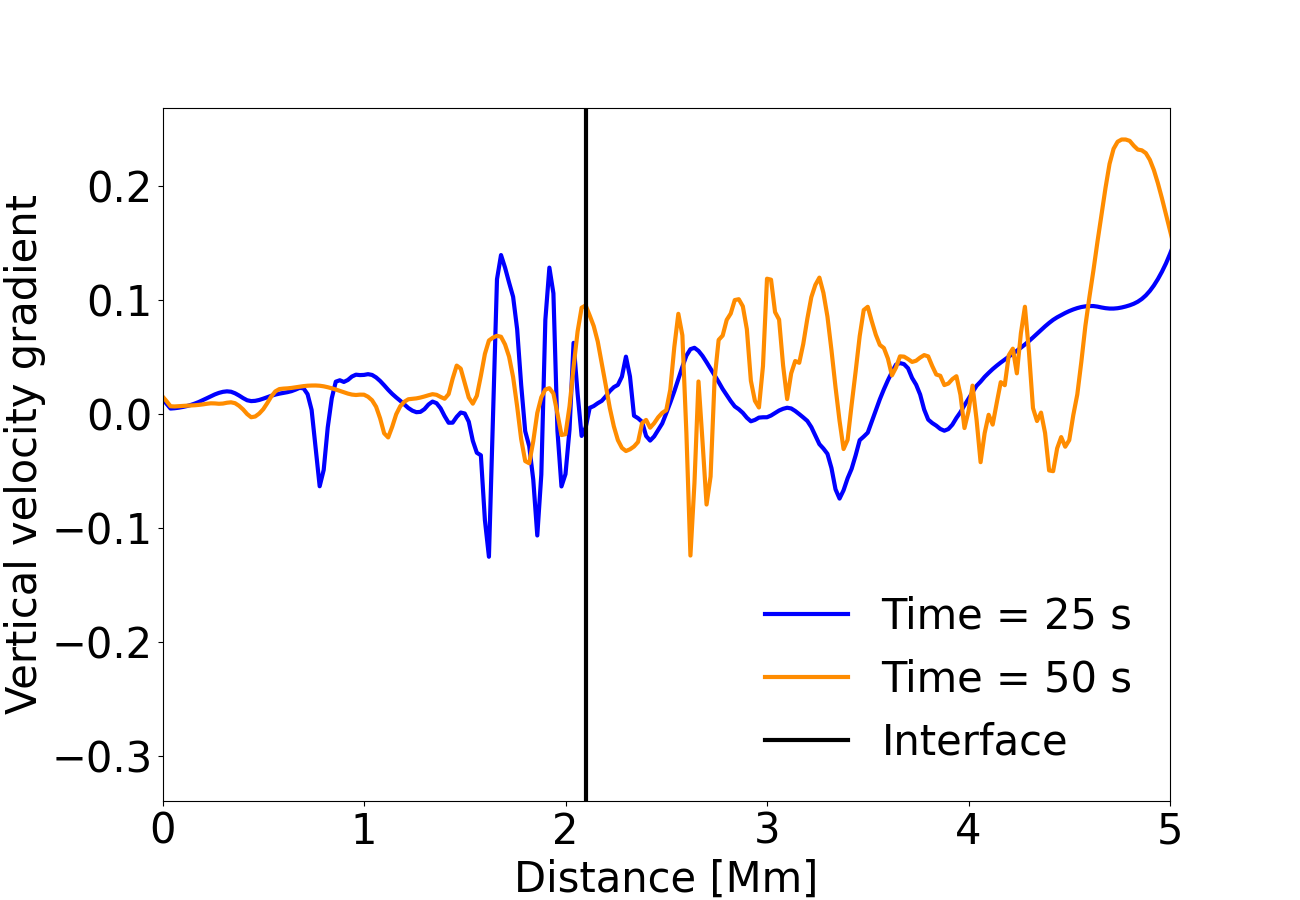}
      \includegraphics[width=7.0cm, height=5.0cm]{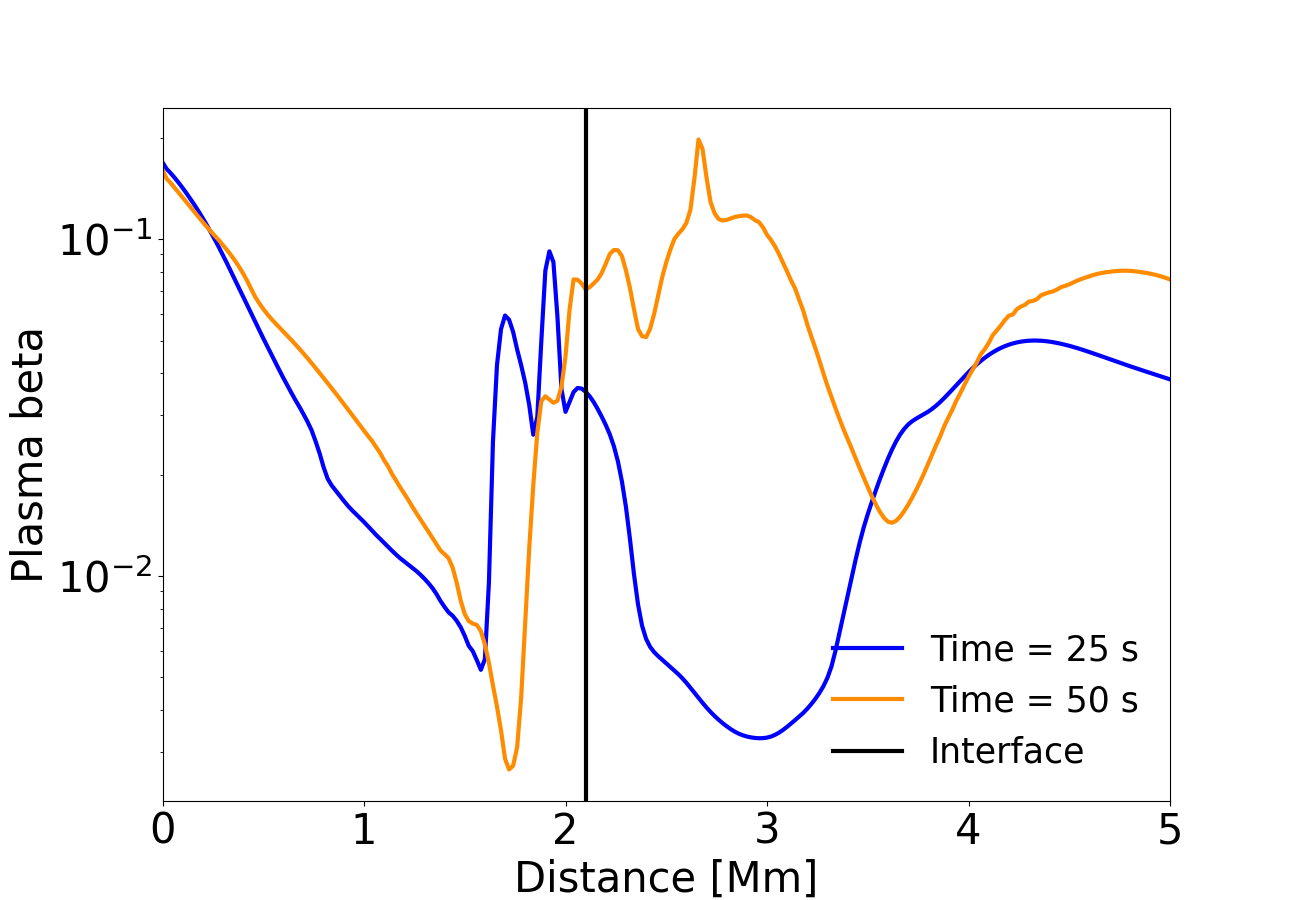}
      \caption{1D cuts of pressure gradient in dyn cm$^{-3}$, density gradient in gr cm$^{-4}$, vertical velocity ($v_{y}$) gradient in s$^{-1}$ and plasma $\beta$ along the vertical distance $y$ at $x=0$ Mm, corresponding to the Res case in panels (a), (b), (c),  and (d), and to the Res+TC case in panels (e), (f), (g) and (h) for times $t=25$ s and $t=50$ s. The black vertical dashed line represents the interface between the dense and tenuous plasma around the transition region at $y\sim$2.1 Mm.}
    \label{fig:cuts_for_instability}
\end{figure*}

To investigate the possible development of RTI/RMI, we estimate the variations of the pressure gradient, density gradient, vertical velocity gradient, and plasma $\beta$ along the orange vertical dashed line drawn in the panels of Fig. \ref{fig:zoom_temp_vertical_vel}, that is similar to the analysis performed by \cite{Shen_et_al_2022}. Notably, we estimate how the above-mentioned variable changes when crossing the interface around the transition region ($y=\sim$2.1 Mm). In Fig. \ref{fig:cuts_for_instability}, we show 1D cuts of pressure gradient in dyn cm$^{-3}$, density gradient in gr cm$^{-4}$, vertical velocity ($v_{y}$) gradient in s$^{-1}$ and plasma $\beta$, corresponding to the Res case in panels (a), (b), (c),  and (d), and to the Res+TC case in panels (e), (f), (g) and (h) for times $t=25$ s and $t=50$ s. 

The RMI generally occurs when the sign of the density and pressure gradients are opposite, i.e., $\nabla\rho\cdot\nabla p<0$, if a shock propagates from the light to the dense gas. Nevertheless, if a shock propagates from the heavy to the light gas, which is the scenario of our simulations, the RMI should occur when $\nabla\rho\cdot\nabla p>0$. However, in the Res case, we see that pressure and density gradients have the same signs just below the interface ($y=\sim$2.1 Mm) at $t=25$ s; therefore, the condition $\nabla\rho\cdot\nabla p>0$ it seems to be satisfied globally. If we analyze locally, let us say at $y=2$ Mm, there $\nabla p\approx-1.335\times10^{-6}$, and  $\nabla\rho\approx-4.736\times10^{-20}$, so the condition $\nabla\rho\cdot\nabla p>0$ it is satisfied. If we estimate the condition over the interface, at about $y=2.5$ Mm, we get $\nabla p\approx-2.993\times10^{-7}$, and $\nabla\rho\approx-4.326\times10^{-22}$, and the condition is also satisfied. If we go higher, i.e., at about $y=4.5$ Mm, $\nabla p\approx-8.741\times10^{-8}$, and $\nabla\rho\approx-3.753\times10^{-24}$, the condition remains. Now, if we estimate the condition for the same point, but at $t=50$ s, we obtain that at $y=2$ Mm, $\nabla p\approx-2.014\times10^{-8}$, and $\nabla\rho\approx-3.121\times10^{-20}$, so the condition $\nabla\rho\cdot\nabla p>0$ is present there. If we go over the interface, at about $y=2.5$ Mm, we get, $\nabla p \approx 6.095\times10^{-8}$, and $\nabla\rho\approx-4.730\times10^{-20}$, which does not satisfy the condition. At $y=4.5$ Mm, we get, $\nabla p\approx-1.659\times10^{-8}$, and $\nabla\rho\approx-1.716\times10^{-22}$, and again, $\nabla\rho\cdot\nabla p>0$. The same happen at about $y=5$ Mm, where $\nabla p\approx-7.299\times10^{-8}$, and $\nabla\rho\approx-7.308\times10^{-23}$. According to the above estimations, we identify that the condition $\nabla\rho\cdot\nabla p>0$ takes place for most of the points, so it is likely the development of an RMI for the Res case. 

The vertical velocity gradient (panel (c)) shows a positive value near the interface but higher up, and this gradient varies from negative to positive values, which means the presence of upward and downward flows. In panel (c), plasma $\beta$ at $t=50$ s shows an evident variation from a magnetically dominant regime ($\beta<1$) to a fluid-dominated regime ($\beta>1$). In particular, the RMI features appear in regions where $\beta\approx1$, i.e., where the fluid and magnetic pressure are approximately equal. The latter results are consistent with the found in \cite{Shen_et_al_2022}. On the other hand, in the Res+TC case, we obtain that for $t=25$ s, $\nabla p\approx-8.534\times10^{-8}$, and  $\nabla\rho\approx-8.263\times10^{-22}$ at $y=2$ Mm, while at $y=2.5$ Mm, $t=25$ s, $\nabla p\approx-3.094\times10^{-8}$, and  $\nabla\rho\approx-8.179\times10^{-24}$. At approximately $y=4.5$ Mm, $t=25$ s, $\nabla p\approx-8.978\times10^{-8}$, and $\nabla\rho\approx-1.189\times10^{-22}$. Finally, at $y=5$ Mm, $\nabla p\approx-5.298\times10^{-8}$, and $\nabla\rho\approx-1.189\times10^{-22}$, and the condition $\nabla\rho\cdot\nabla p>0$ satisfies for all the points. But, if we estimate for $t=50$ s, we obtain $\nabla p\approx-6.398\times10^{-7}$, and $\nabla\rho\approx3.380\times10^{-20}$ at $y=2$ Mm, while at $y=2.5$ Mm, $\nabla p\approx 4.872\times10^{-7}$, and  $\nabla\rho\approx-1.114\times10^{-20}$. At approximately $y=4.5$ Mm, $\nabla p\approx-7.993\times10^{-8}$, and $\nabla\rho\approx2.145\times10^{-22}$. Finally, at $y=5$ Mm, $\nabla p\approx-9.791\times10^{-8}$, and $\nabla\rho\approx-8.918\times10^{-23}$. In most of the points, we get $\nabla\rho\cdot\nabla p>0$. However, we do not see instabilities in all the maps for the Res+TC case; consequently, this analysis does not apply to this scenario. In panel (g), the vertical velocity gradient oscillates between positive and negative values below and over the interface up to about $y=2.5$ Mm. However, at a higher distance, the gradient is positive, which indicates a predominantly upward flow, consistent with the behavior shown in the 2D map of the vertical velocity in panel (f) of Fig. \ref{fig:zoom_temp_vertical_vel}. Finally, in panel (h), we note that plasma $\beta$ is always smaller than 1, which means that magnetic pressure dominates over fluid pressure, and therefore, the plasma is flowing in this for the two times $t=25,50$ s. This behavior is consistent with any observed RMI since there is no transition between magnetically-dominated and fluid-dominated regimes.  

According to the previous analysis, we can conclude that the internal structure of the loop region is related to an RMI in the Res case. In the Res+TC case, the thermal conductivity determines the plasma dynamics in the loop region, so the plasma moves through the loops and dissipates the heat, and therefore, any RMI develops.
 
\subsubsection{Shock development}
\label{Shock_development}

\begin{figure*}
    \centering
    \centerline{\Large \bf   
       \hspace{0.42\textwidth} \color{black}{\Large{Res case}}
         \hfill}
          \centerline{\Large \bf   
      \hspace{0.27\textwidth}  \color{black}{(a)}
       \hspace{0.32\textwidth}  \color{black}{(b)}
         \hfill}
    \includegraphics[width=6.5cm, height=6.5cm]{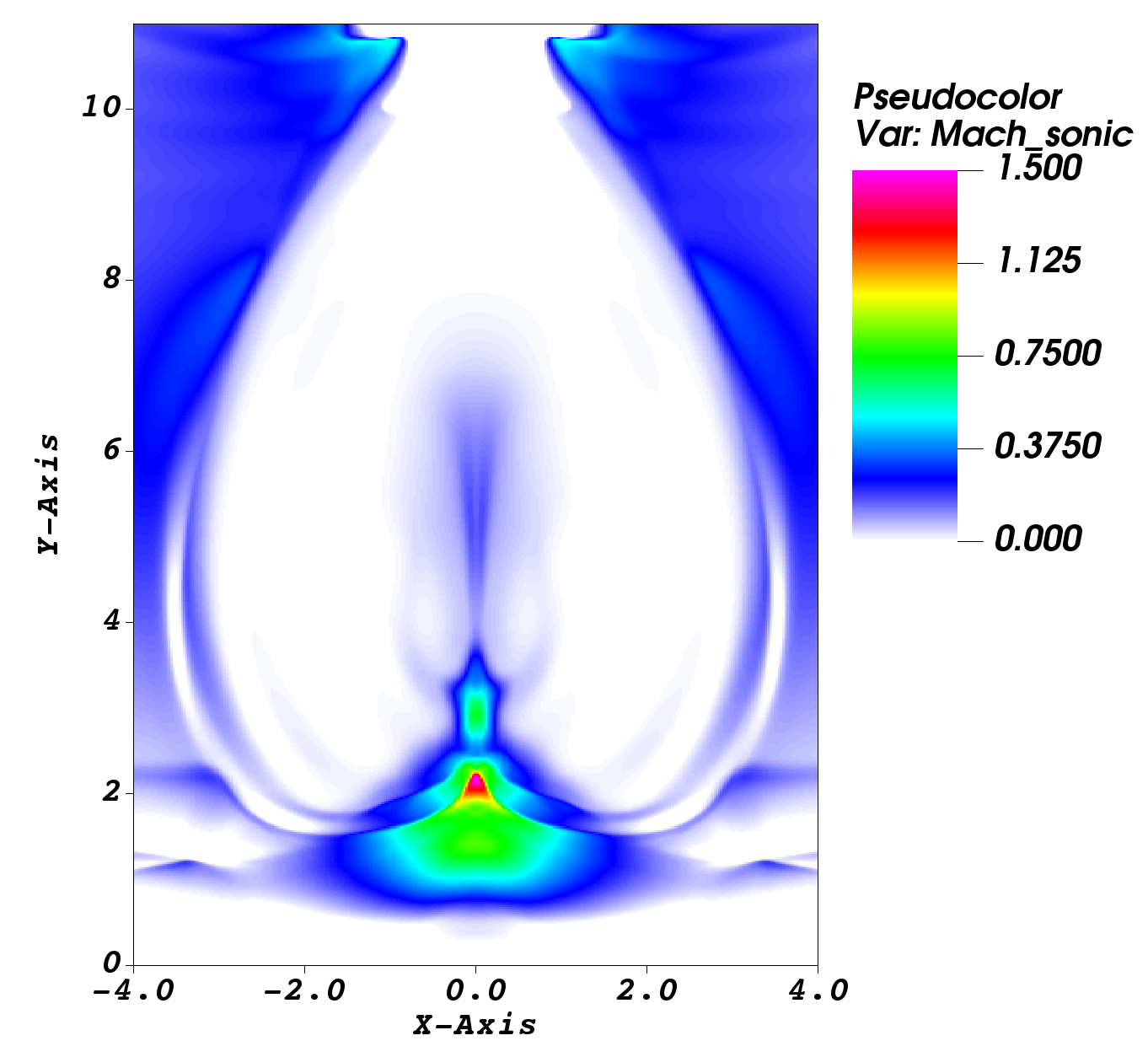}
    \includegraphics[width=6.5cm, height=6.5cm]{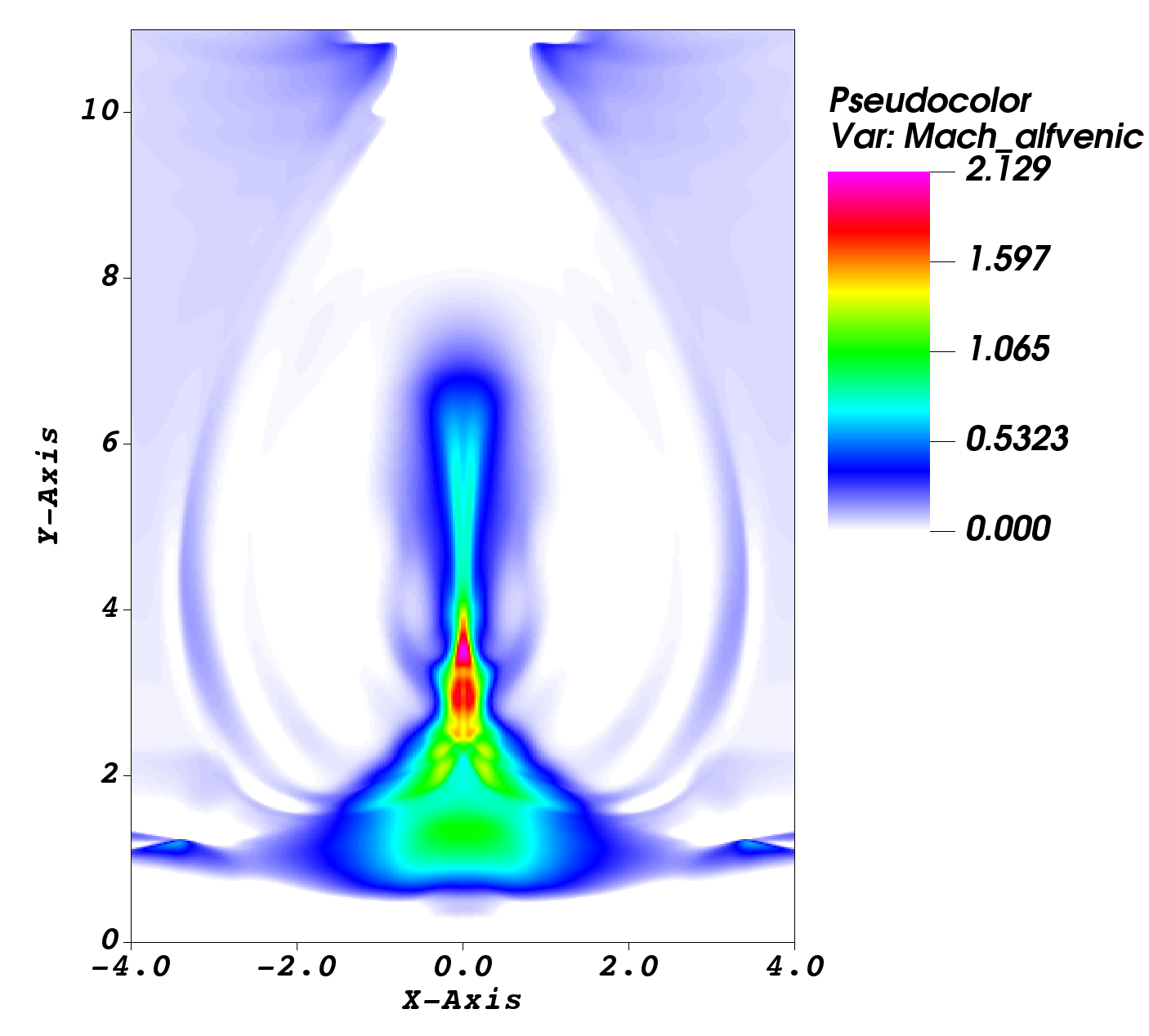}
    \centerline{\Large \bf   
       \hspace{0.4\textwidth} \color{black}{\Large{Res+TC case}}
         \hfill}
          \centerline{\Large \bf   
      \hspace{0.265\textwidth}  \color{black}{(c)}
       \hspace{0.32\textwidth}  \color{black}{(d)}
         \hfill}
    \includegraphics[width=6.5cm, height=6.5cm]{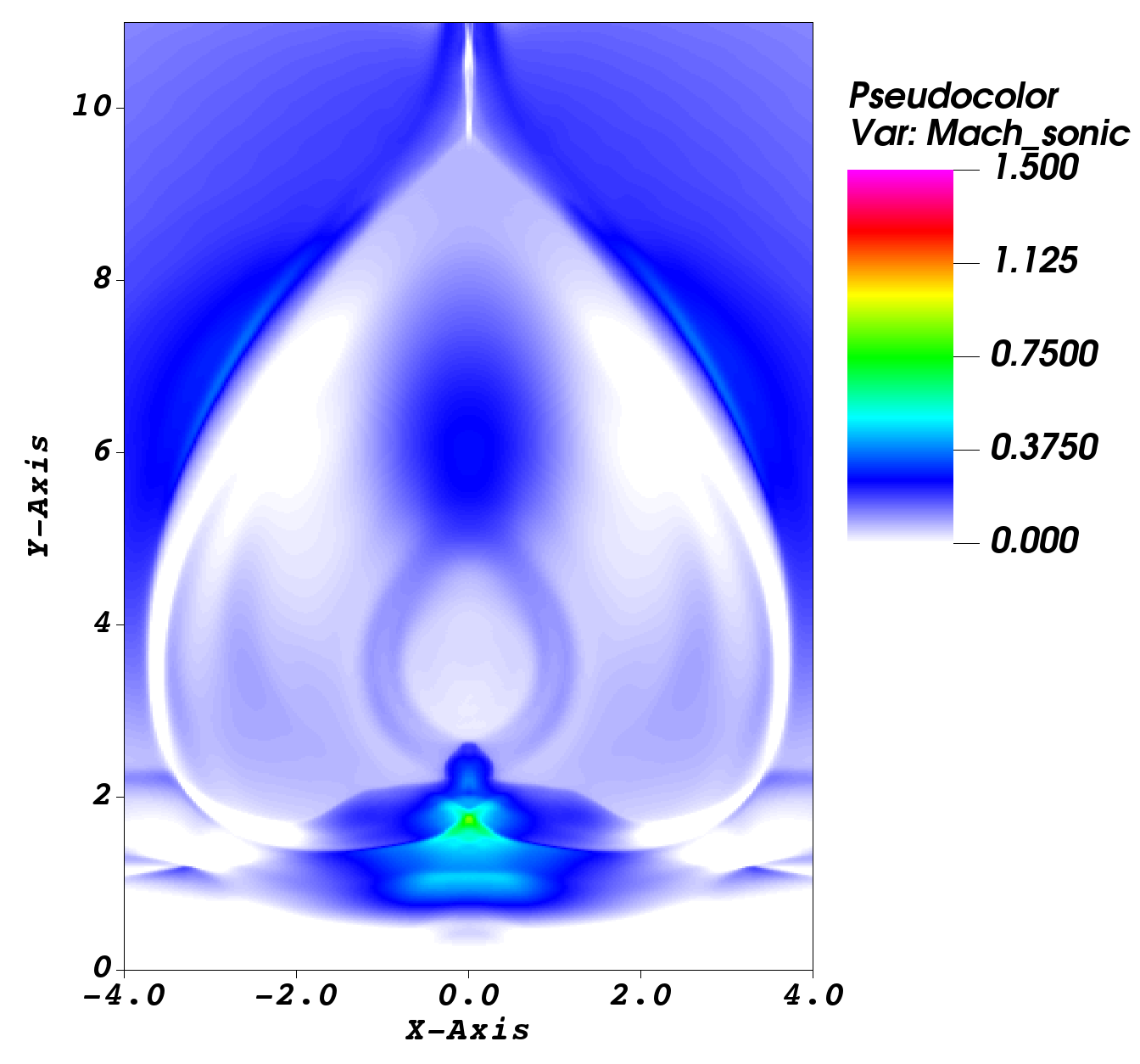}
    \includegraphics[width=6.5cm, height=6.5cm]{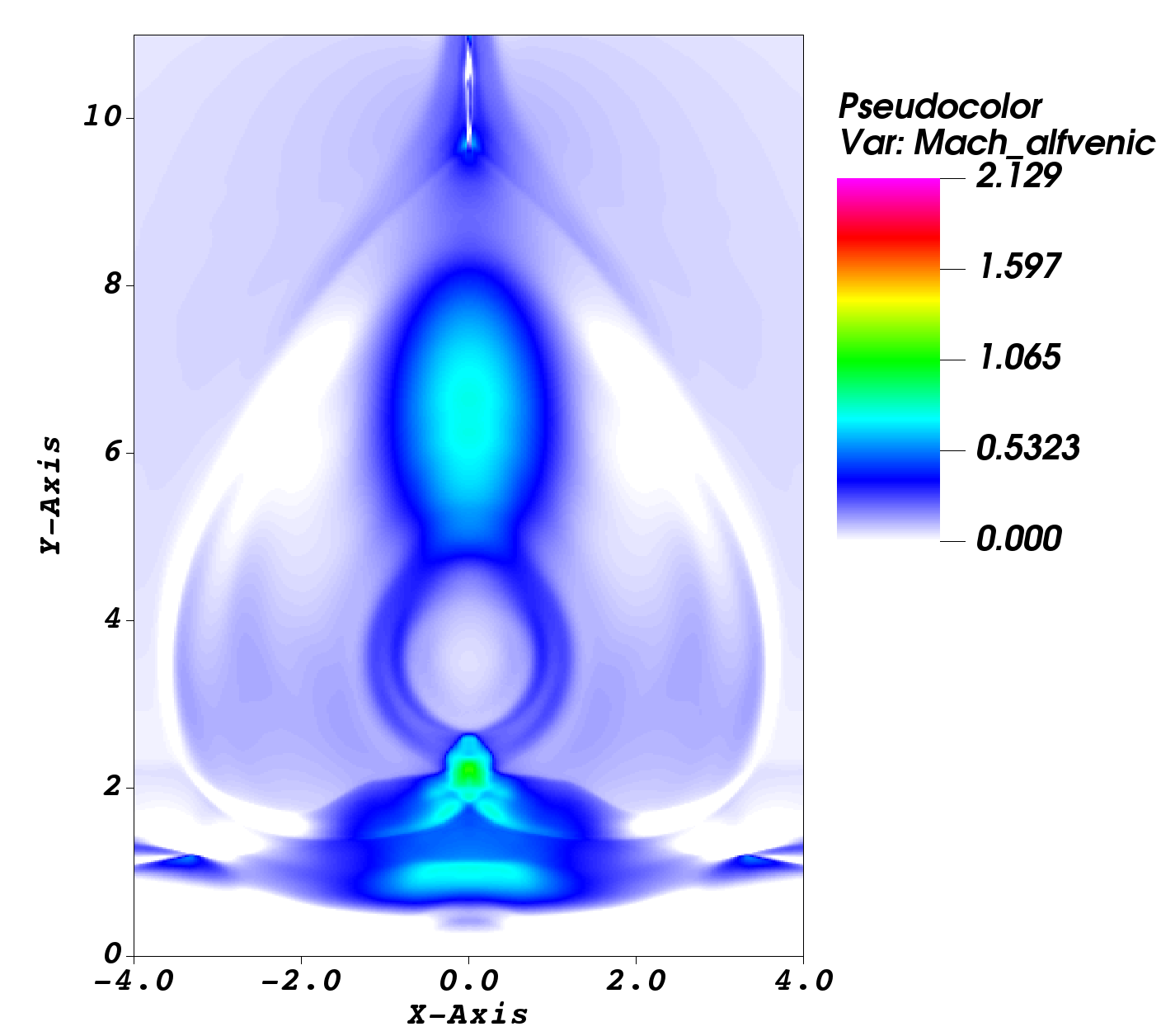}
       \caption{Sonic and Alfv\'enic Mach numbers, $M_{s}$ (panels (a) and (c)) and $M_{A}$ (panels (b) and (d)) correspond to the Res case (top panels) and to the Res+TC case (bottom panels) at $t=50$ s.}
    \label{fig:Mach_numbers}
\end{figure*}

To investigate if there are the physical conditions so that a shock could develop in the post-flare loops structures of any of the two simulation cases, we estimate the sonic Mach number, $M_{s}=V_{y}(C_{s}$ number, and the Alfv\'enic Mach number, $M_{A} = V_{y}/C_{A}$, where, $V_{y}$ is the local vertical velocity of the plasma, $C_{s}=\sqrt{\gamma p/\rho}$, is the local speed of sound, and $C_{A}=|{\bf B}|/\sqrt{4\pi\rho}$, is the local Alfv\'en speed. For example, in panel (a) of Fig. \ref{fig:Mach_numbers}, we display the sonic Mach number and the Alfv\'enic Mach number for the Res case at $t=50$ s. There, we see that the sonic Mach number is greater than one, i.e., the plasma is supersonic in the bottom regions of the loops, where the plasma propagates upwards. We also note in panel (b) that the Alfv\'enic Mach number is greater than one in regions where the plasma propagates upwards, which means that plasma flow is superAlfv\'enic, and as a consequence of the two later conditions, a shock develops in the Res case. On the other hand, in panels (c) and (d), we show the sonic and Alfv\'enic Mach numbers for the Res+TC case. In this scenario, it is discernible that plasma is neither supersonic nor superAlfv\'enic; therefore, the environment is unsuitable for the shock to develop. The results shown for the Res case in Fig. \ref{fig:Mach_numbers} are consistent with those obtained in \cite{Chen_et_al_2015} in the scenario of particle acceleration by a solar flare termination shock.  

\subsubsection{Magnetic islands formation}
\label{Mag_islands_formation}

\begin{figure*}
    \centering
    \centerline{\Large \bf   
      \hspace{0.375\textwidth}  \color{black}{\Large{Res+TC case}}
      \hfill}
      \centerline{\Large \bf   
      \hspace{0.21\textwidth}  \color{black}{(a)}
       \hspace{0.39\textwidth}  \color{black}{(b)}
         \hfill}
     \includegraphics[width=6.6cm, height=7.5cm]{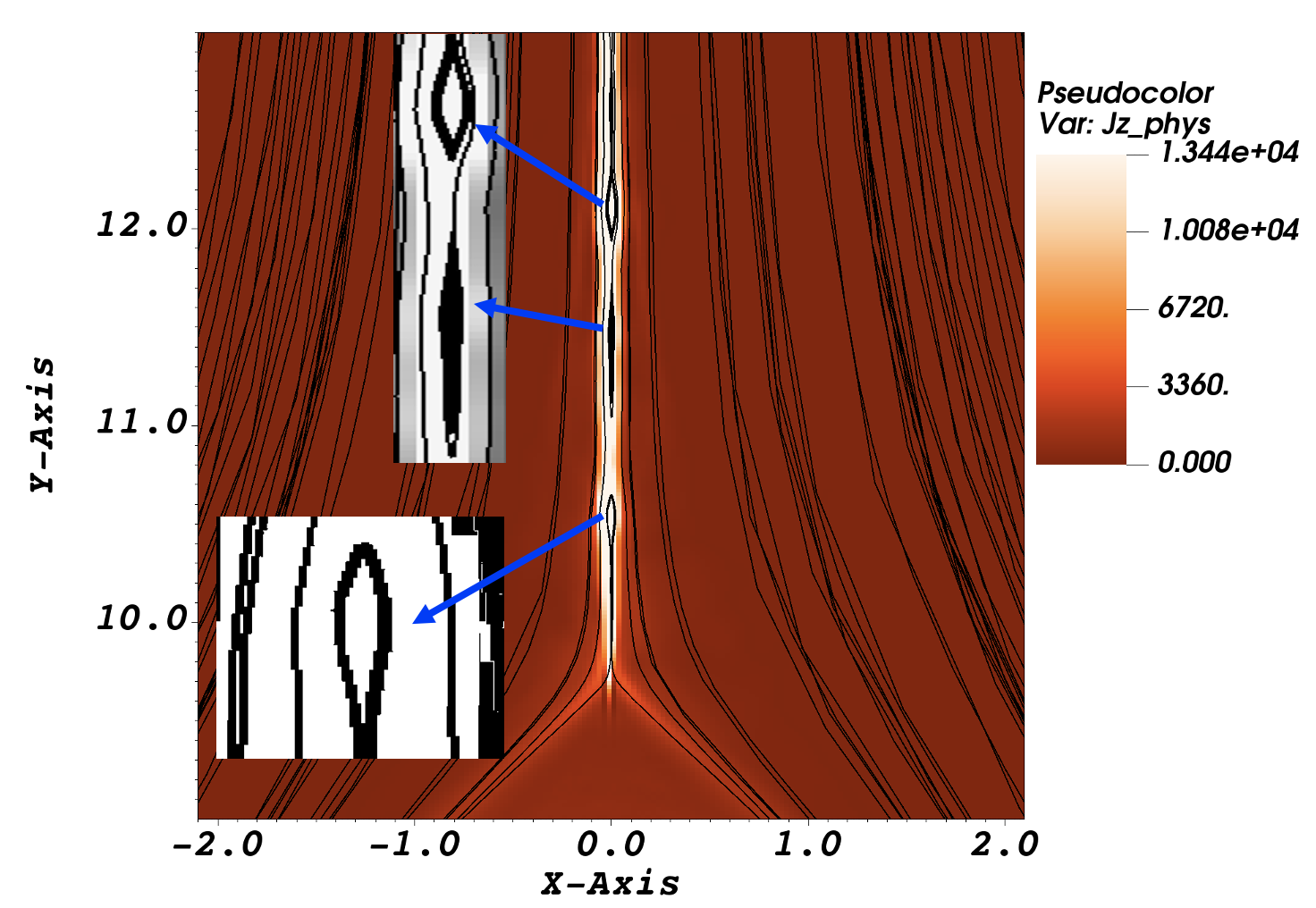}
     \includegraphics[width=9.0cm, height=7.5cm]{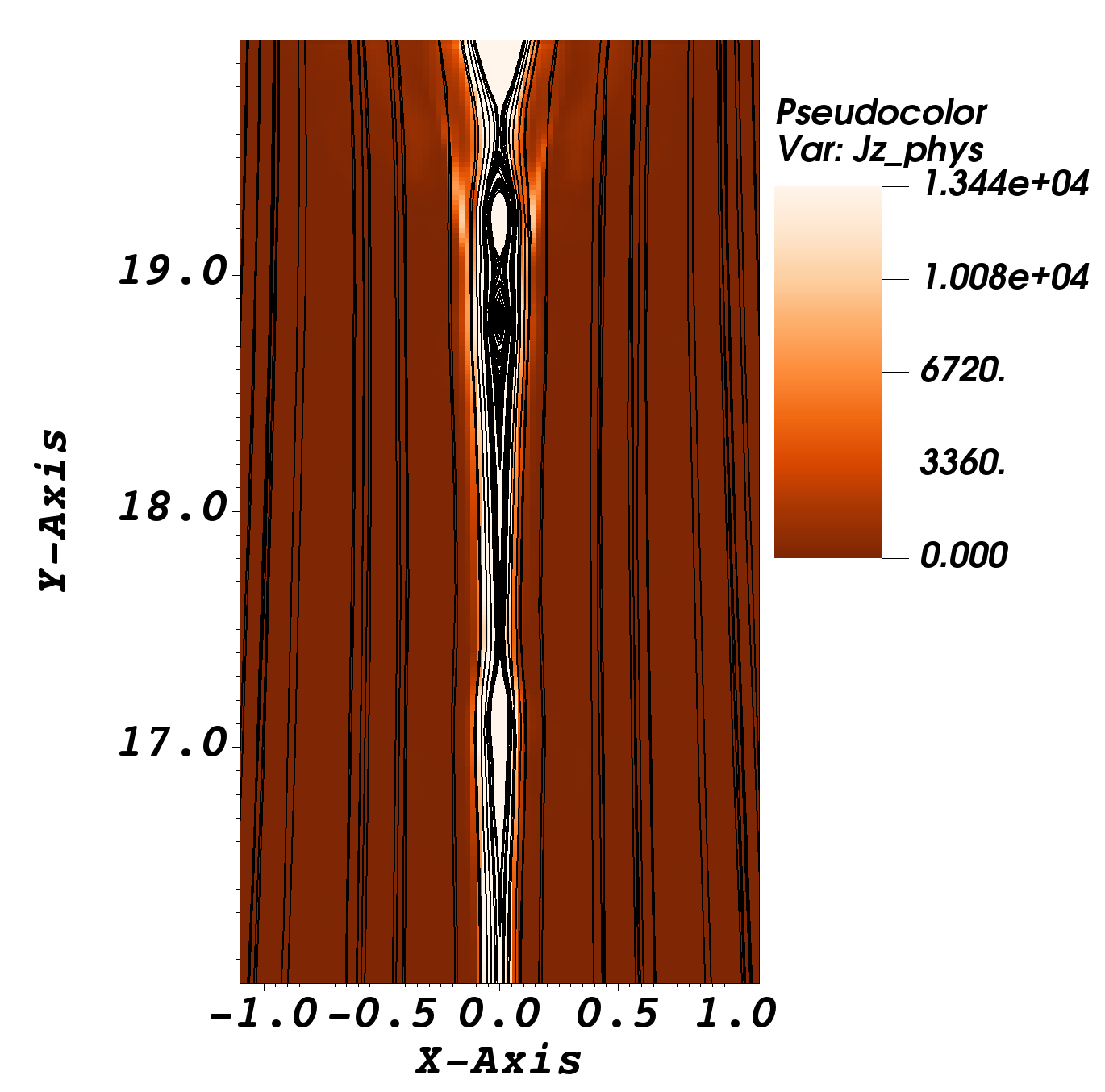}
     \caption{Zoom of the $z$ component of the current density in statA cm$^{-2}$ at $t=50$ s that schematize the magnetic islands corresponding to the Res+TC case.}
    \label{fig:islandR}
\end{figure*}

In the Res+TC case, we can zoom in, for example, the $J_{z}$ and observe the CS closely to identify multiple magnetic islands, as shown in Fig. \ref{fig:islandR}. In panel (a), we display a zoom region that covers $x\in[-2,2]$ Mm and $y\in[9,13]$ Mm, while in panel (b), we show a zoom region that covers $x\in[-1,1]$ Mm and $y\in[15,20]$ Mm. For instance, in panel (a), we note the development of three magnetic islands along the CS, i.e., in regions where $J_{z}$ is high. These structures are also observed along the CS at higher altitudes as schematized panels (b). In both panels, we can identify separatrices regions that indicate the presence of the magnetic islands. These regions represent the most recent reconnected field line that separates regions with magnetic fields of different topology; these magnetic fields may be anti-parallel, while the reconnected fields are found where the oppositely directed fields have been interconnected. Generally, the separatrices meet at the X-line, surrounded by a diffusion region, i.e., a high current density region, as shown in the small plots indicated by the blue arrows in panel (a). These magnetic islands can be formed due to a Tearing-instability in the non-linear regime. The tearing-instability is a linear stability of a CS of infinite \citep{10.1063/1.1706761}, which is characterized by making a system in equilibrium unstable due to the tearing mode, leading to the formation of X-points and plasmoids during the reconnection process. This instability is part of a basic resistive instability that is linked with a long-wave tearing mode, corresponding to the breakup of the layer along current-flow lines \citep[see, e.g.,][]{10.1063/1.1706761}. It can occur on timescales $\tau$, such that $\tau_{A}<\tau<\tau_{d}$. The latter times are the time last the instability to traverse the CS at the diffusion speed $v_{d}$ and the Alfvén speed $v_{A}$, respectively, and it has the effect of creating many small-scale magnetic islands in the CS \citep{Shen_et_al_2011}. Besides, these magnetic islands are likely to develop in solar flares, \citep[see, e.g.,][]{Kliem_et_al_2000, Wang_et_al_2021}. 

\begin{figure*}
    \centering
       \centerline{\Large \bf   
      \hspace{0.37\textwidth}  \color{black}{\Large{Res+TC case}}
      \hfill}
      \centerline{\Large \bf   
      \hspace{0.26\textwidth}  \color{black}{(a)}
       \hspace{0.32\textwidth}  \color{black}{(b)}
         \hfill}
     \includegraphics[height=8.5cm]{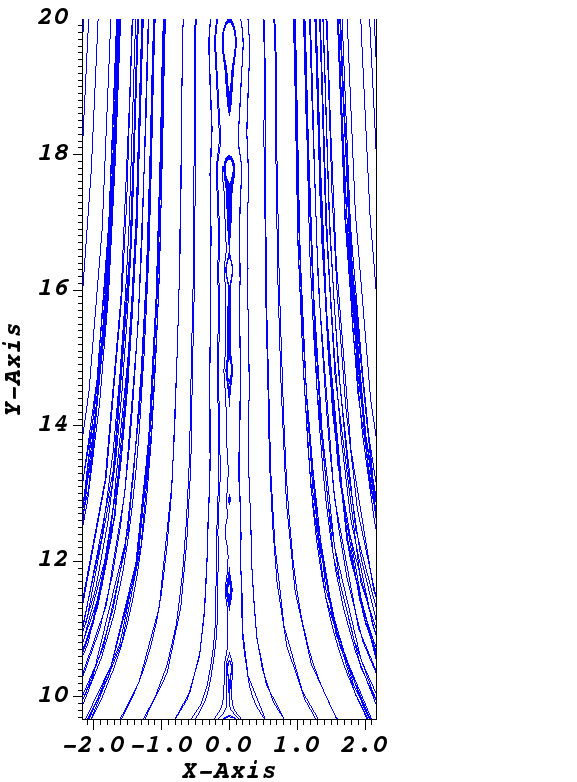}
     \includegraphics[height=8.5cm]{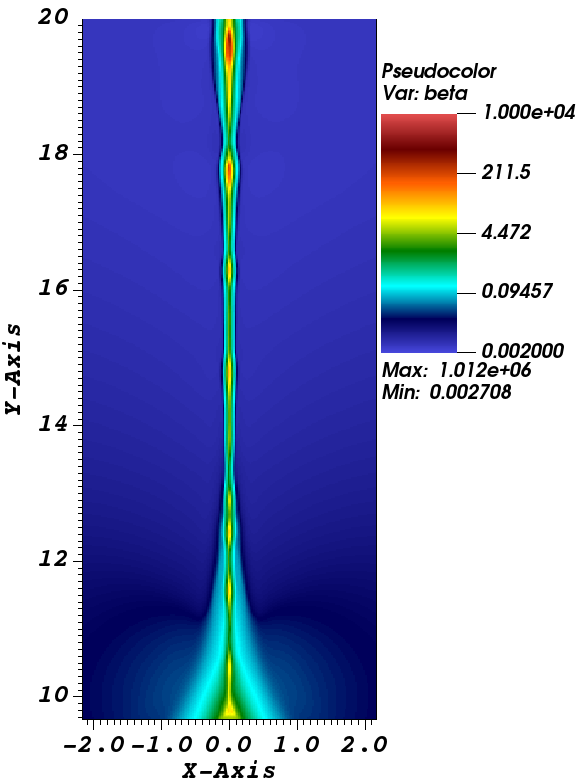}
     \caption{Zoom of the magnetic field lines (a) and the plasma $\beta$ (b) at time $t=51$ s for the Res+TC case.}
    \label{fig:field_lines_beta}
\end{figure*}

To complement Fig. \ref{fig:islandR} and go into more detail in the formation of the magnetic islands due to the Tearing-instability, in Fig. \ref{fig:field_lines_beta}, we show a zoom region that covers $x\in[-2,2]$ Mm and $y\in[9,20]$ Mm of the magnetic field lines (a) and the plasma $\beta$ (b) at time $t=51$ s. For example, in panel (a), we note the formation of various small-scale ($\sim 0.5$ Mm of width) magnetic islands, which can be related to regions of high ($\sim 10^{4}$) plasma $\beta$ values. We can also see that magnetic islands are souring by low ($\sim 10^{-1}$) plasma $\beta$ values, which means that overall magnetic islands are localized in a mixed region. The latter means that magnetic islands are more likely to form in regions where gas pressure dominates over magnetic pressure, i.e., in regions of high plasma $\beta$. This result makes sense since the magnetic islands form in X-points or O-points, where, by definition, the magnetic field is zero. Besides, it is interesting that magnetic islands are surrounded by regions where the magnetic pressure dominates over gas pressure ($\beta\sim10^{-1}$), which implies that the magnetic field confines the material that forms the islands. Specifically, in the Res+TC case, the thermal conduction helps distribute the plasma along the magnetic field lines, leading to the generation of a thinner CS compared to the Res case. 

\begin{figure*}
    \centering
    \centerline{\Large \bf   
      \hspace{0.28\textwidth}  \color{black}{(a)}
       \hspace{0.38\textwidth}  \color{black}{(b)}
         \hfill}
     \includegraphics[height=7cm]{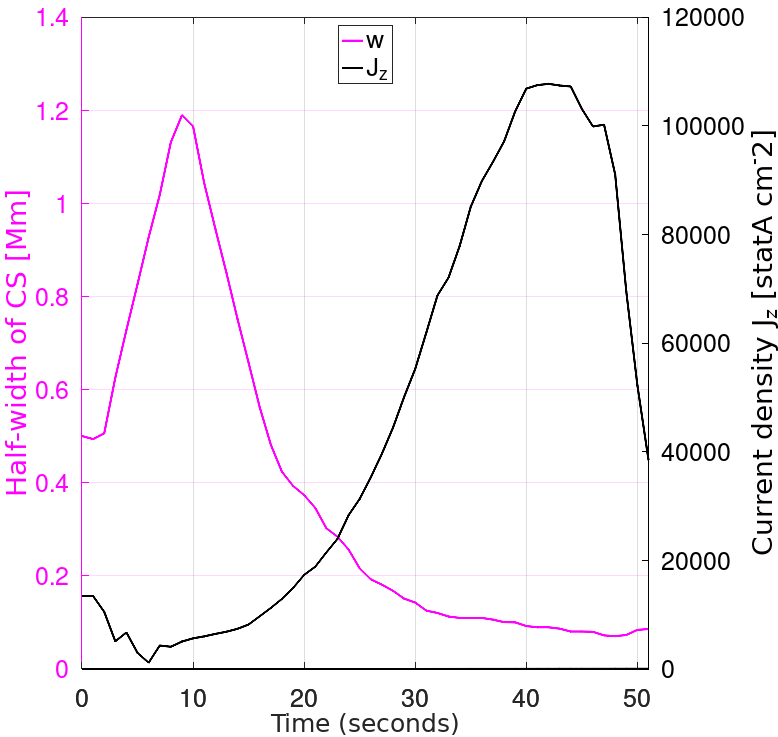}
      \includegraphics[height=7cm]{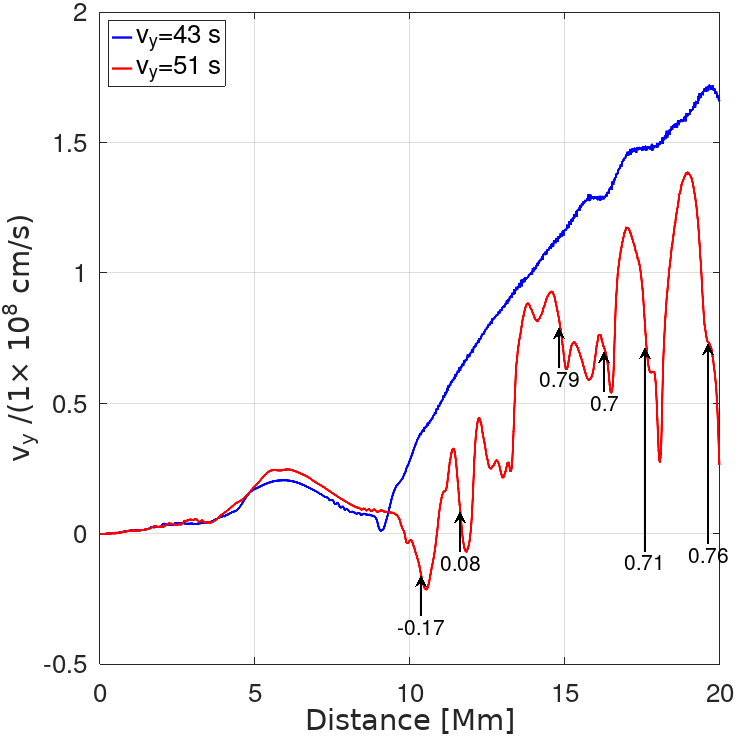}
     \caption{Variations of $w$ (magenta curve), and the $z-$ component of the current density $J_{z}$ (black curve) at $y=$10.4 Mm as functions of time (a) and the vertical velocity along the vertical distance $y$ at two times $t=43$ s and $t=51$ s (b) for the Res+TC case.}
    \label{fig:Location_magnetic_island}
\end{figure*}

Moreover, it is possible to indicate the evolution of the CS's width, which could help identify the formation of multiple X- and O-points in the CS, \citep[see, e.g.,][]{Shen_et_al_2011}. For instance, in Fig. \ref{fig:Location_magnetic_island}, we plot the half-width $(w)$ of the CS in Mm and the $z-$ component of the current density $J_{z}$ at the point $(x,y)=$(0,10.4) Mm as functions of time (a) and the vertical component of the plasma velocity along the $y-$axis at two representative times $t=43$ s and $t=51$ s (b) for the Res+TC case. Interestingly, the formation of magnetic islands is related to the width of the CS, and these structures are expected to form for small values of $w$ \citep[see, e.g., ][]{Shen_et_al_2011,mactaggart2020topics}. For instance, in Fig. \ref{fig:Location_magnetic_island}(a), we show the time evolution of the $w$ at the point  $(x,y)=(0,10.4)$ Mm represented by the magenta-colored curve. This figure shows that the CS's width reduces over time. Also, in Fig. \ref{fig:Location_magnetic_island}(a), we display the evolution of $J_{z}$ at the same point as $w$, which is shown by the Mm black curve. One can notice how $J_{z}$ increases with time until around $t\approx43$ s, and then it experiences a sudden decrease up to the final time of the simulation, which is $t=51$ s. This increase is related to the location of an O-type (neutral) point at $y=$ 10.4 Mm, which in turn triggers the magnetic island \citep[see, e.g., ][]{Shen_et_al_2011}. In this scenario, plasma pressure dominates over magnetic pressures inside the magnetic islands, as shown in the plasma $\beta$ map of Fig. \ref{fig:field_lines_beta}. Since magnetic effects are small, a sudden drop in current density is observed in the last seconds of the simulation. The appearance of magnetic islands in the last seconds of the simulation is also consistent with $w$ minimum values ($\sim$ 0.1 Mm). Besides, to identify the formation of six O-points along the CS, in Fig. \ref{fig:Location_magnetic_island}(b), we display the plasma velocity along the y-direction for two times $t=43$ s and $t=51$ s. According to our simulation results, these times cover a period before and after the formation of magnetic islands. Each arrow in Fig. \ref{fig:Location_magnetic_island}(b) indicates the location of each magnetic island and its respective vertical velocity, which is normalized with the velocity scaled factor $v_{0}=10^{8}$ cm s$^{-1}$. Before forming the magnetic islands, the plasma vertical velocity increases with time and does not show significant variations. For example, at $t=51$ s, which is the time when we observe the formation of magnetic islands, and from $10$ Mm onwards, the plasma vertical velocity fluctuates. These fluctuations indicate the presence of the magnetic islands, known as plasma blobs \citep[see, e.g.,][]{Shen_et_al_2011}. In Fig. \ref{fig:Location_magnetic_island}(b), we also specify the location and vertical velocities of the magnetic islands themselves. We see an island that moves downstream and is close to $y=10.4$ Mm with a vertical velocity of about $v_{y}=-0.17$. The other five magnetic islands move upstream at $y=11.5$ Mm, $y=14.8$ Mm, $y=16.3$ Mm, $y=17.8$ Mm, and $y=19.6$ Mm. They correspond to vertical velocities of approximately $v_{y}=0.08$, $v_{y}=0.79$, $v_{y}=0.7$, $v_{y}=0.71$, and $v_{y}=0.76$, respectively. The plasma vertical velocity around these islands is high, although for the magnetic island of approximately $y=$16.3 Mm, the vertical velocity of the plasma around and the island are similar. For other magnetic islands, the vertical velocity of the surrounding plasma is considerably higher, which is consistent with the found in, for example, \cite{Shen_et_al_2011}.

\begin{figure*}
    \centering
    \includegraphics[width=15cm]{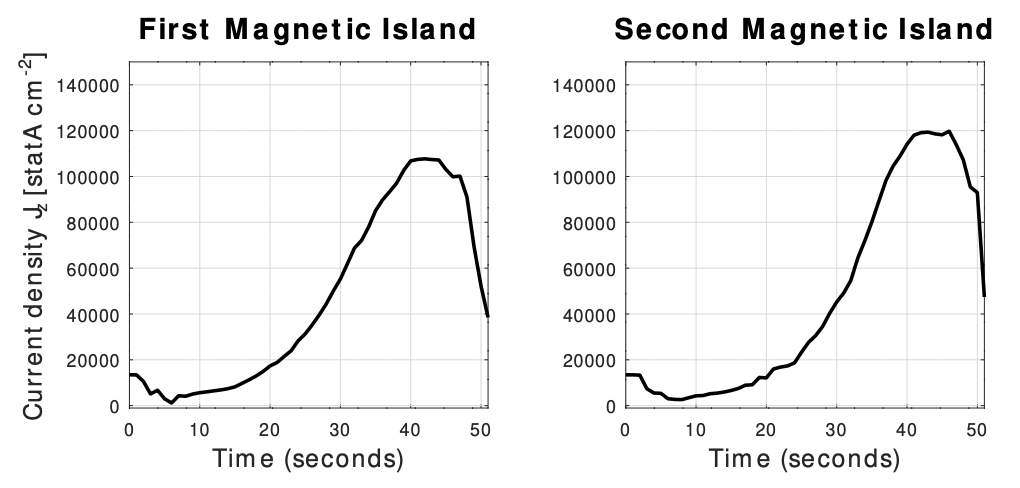}
    \includegraphics[width=8cm]{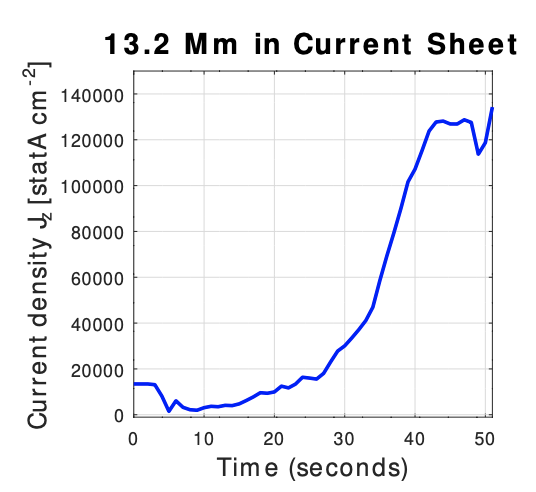}
     \caption{Temporal evolution of the $z-$component of the current density $J_{z}$ estimated at the locations of the first and second magnetic islands of Fig. \ref{fig:Variations_of_Jz_island} (black curves) and at $y=$13.2 Mm (blue curve).}
    \label{fig:Variations_of_Jz_island}
\end{figure*}

Additionally, to indicate how the variations of the vertical velocity shown in Fig. \ref{fig:Location_magnetic_island}(b) are related to the magnetic island formation, in Fig. \ref{fig:Variations_of_Jz_island}, we show the temporal evolution of the $z-$ component of the current density $J_{z}$ at the locations of the first and second magnetic island as schematized by the arrows of Fig. \ref{fig:Location_magnetic_island}(b). Essentially, these plots show that the behavior of $J_{z}$ in the two magnetic islands is similar, i.e., all reach their maximum value of around $t\approx43$ s and then suddenly decrease to the final of the simulation. The increase in $J_{z}$ is due to the appearance of magnetic islands. In the bottom of Fig. \ref{fig:Variations_of_Jz_island}, we show $J_{z}$ (colored in blue) calculated at $y=$13.2 Mm, where it is a point without magnetic island. Here, we note that the behavior of $J_{z}$ is different from the plots where magnetic islands form since, in the final times of the simulation, $J_{z}$ grows instead of suddenly decreasing.

\section{Concluding remarks}
\label{Conclusion}

In this paper, we study the formation and dynamics of solar flares, including the post-flare loops in a local region of the solar atmosphere, for two scenarios, according to the classical 2D model. In the Res case, localized resistivity produces more substructure around the post-flare loops, related to the plasma moving upwards and downwards. In addition, we identify the development of a complex structure, which could be associated with a magnetic island at the top of post-flare loops. On the other hand, in the Res+TC case, the whole solar flare structure looks smooth, including the post-flare loops and no apparent substructures are developed, which means that thermal conductivity redistributes the plasma homogeneously. However, we observe the formation of multiple small-scale magnetic islands from the top of the post-flare loops and along the CS in the vertical direction. Additionally, we include the case of an anomalous resistivity without thermal conduction. In that scenario, any post-flare loops and thin CS are formed. Instead, we see a hot plasma structure forming that becomes wider and moves upwards while magnetic field lines reconnect at the coronal level. This kind of plasma feature could be related to a plasmoid.
Furthermore, in the mass density maps, we identify the development of a small jet ($\sim 3$ Mm of height) moving upwards from the transition region. This feature is interesting since it could be related to the magnetic reconnection process at microscopic scales, which is triggered by anomalous resistivity, given a relevant result despite the limitations of the MHD to capture the microscopic mechanism associated with the magnetic reconnection process. However, it is clear that in the case with anomalous resistivity, we do not observe the formation of a typical solar flare and associated post-flare loops as indicated by the classical 2D model introduced by \cite{Yokoyama&Shibata_2001}.

The development of the multiple magnetic islands in the Res+TC case is related to the Tearing instability in the nonlinear regime. This instability is characterized by making a system in equilibrium unstable due to the tearing mode, leading to the formation of X-points and plasmoids during the reconnection process. This instability is part of a fundamental resistive instability linked with a long-wave tearing mode, corresponding to the breakup of the layer along current-flow lines. Besides, in the Res+TC case, the magnetic islands form along the CS in the vertical direction, consistent with the results obtained by \cite{Wang_et_al_2021}. Moreover, we observe that magnetic islands tend to form in regions with high plasma $\beta$ ($\sim 10^{4}$) but surrounded by low plasma $\beta$ regions ($\sim 10^{-2}$), which suggest that on those regions plasma pressure and magnetic pressure must contribute. That is, the magnetic pressure confines the plasma, and the plasma pressure makes the structure of the islands. In the magnetic islands, the current density tends to increase, and the CS's half-width becomes thinner when forming. Finally, when magnetic islands form, we identify the generation of plasma bobs in the vertical velocity, which appears due to the flow generated by the magnetic reconnection process.

In the case of the analysis for the instabilities performed to identify the development of RTI and RMI, we conclude that in the Res case, the internal spike-shaped structures are likely related to an RMI since the general physical conditions that state this instability are satisfied. In the Res+TC case, we show that any condition for RMI or RTI is satisfied since the post-flare loops look smooth. Although our simulation setup for the Res case is more straightforward than the one used in \cite{Shen_et_al_2022}, we identify desirable conditions for developing the RMI/RTI instabilities since we use only a localized resistivity profile. This result is crucial since it could indicate that a simple model with magnetic resistivity can solely produce such instabilities observed in underdense plasma downflows associated with magnetic reconnection in solar flares. 

A difference of the present study regarding, for example, \cite{Yokoyama&Shibata_2001, Takasao_et_al_2015, Ruan_et_al_2020, Shen_et_al_2022}, is the inclusion of a highly anisotropic thermal conductivity that smoothly varies between the classical and saturated thermal conduction regimes, which according to our knowledge, it had not been used in numerical simulations of solar flares, so far. Additionally, we adopt a model for the solar atmosphere that includes the minimum temperature of the chromosphere and the sharp gradient of the transition region, which is more realistic than the one adopted by the abovementioned works since they typically consider a simple tangent function to describe a smooth variation of temperature from the photosphere to the solar corona. Furthermore, in comparison to \cite{2024arXiv240107048M}, where the authors studied the formation and evolution of plasmoids in coronal CS, subjected to an external velocity perturbation under the condition of uniform resistivity, in our paper, we employed a localized resistivity that triggers the magnetic reconnection and leads the formation of the CS. Besides, \cite{2024arXiv240107048M} indicates that CS initially goes in thinning and Petschek-type magnetic reconnection followed by tearing instability and plasmoid formation. In contrast, one of the main objectives of our paper focuses on studying the associated small-scale structures of solar flares without entering into detail about the reconnection mechanism. However, both analyses could complement each other and engage in the context of the highly dynamic behavior observed along the CS in flare-like structures.

Finally, the results of this paper could help to continue persuading the dynamics of the formation of solar flares based on the standard model. In particular, it opens the gate to investigate if the anomalous resistivity itself could generate plasmoids and jets, which complete the picture of the magnetic reconnection process using the MHD approximations. Additionally, the current results could help better determine that solar flares can develop substructures even in simple scenarios where, for example, turbulence and particle acceleration are not considered.

\section*{Acknowledgements}

We gratefully acknowledge the constructive remarks of the referee that highly improved our manuscript. J.J.G.-A. acknowledges to "Consejo Nacional de Humanidades Ciencias y Tecnolog\'ias (CONAHCYT)" 319216 project "Modalidad: Paradigmas y Controversias de la Ciencia 2022." M.G.-S. is grateful to ``Programa de Becas para Estudios de Posgrado en México'' granted by CONAHCYT. 

\section*{Data Availability}

The data underlying this article will be shared on reasonable request to the corresponding author.




\bibliographystyle{mnras}
\bibliography{MHD_simulations_post-flare_loops_accepted_MNRAS_2024} 







\bsp	
\label{lastpage}
\end{document}